\documentclass{elsart}
\usepackage{graphicx}  

\begin{document}

\begin{frontmatter}

\title{Precision physics of simple atoms:\\
QED tests, nuclear structure\\
and fundamental constants}

\author{Savely G. Karshenboim}
\ead{sek@mpq.mpg.de}
\address{D. I. Mendeleev Institute for Metrology, 190005 St. Petersburg, Russia}
\address{Max-Planck-Institut f\"ur Quantenoptik, 85748 Garching, Germany}

\begin{abstract}
Quantum electrodynamics is the first successful and still the most
successful quantum field theory. Simple atoms, being essentially
QED systems, allow highly accurate theoretical predictions.
Because of their simple spectra, such atoms have been also
efficiently studied experimentally frequently offering the most
precisely measured quantities. Our review is devoted to comparison
of theory and experiment in the field of precision physics of
light simple atoms. In particular, we consider the Lamb shift in
the hydrogen atom, the hyperfine structure in hydrogen, deuterium,
helium-3 ion, muonium and positronium, as well as a number of
other transitions in positronium. Additionally to a spectrum of
unperturbed atoms, we consider annihilation decay of positronium
and the $g$ factor of bound particles in various two-body atoms.
Special attention is paid to the uncertainty of the QED
calculations due to the uncalculated higher-order corrections and
effects of the nuclear structure. We also discuss applications of
simple atoms to determination of several fundamental constants.
\end{abstract}

\end{frontmatter}

\newpage

\tableofcontents

\newpage

\section{Introduction}

Several atoms play basic roles in modern physics and, in fact,
very different roles. A unit of time, the second, is defined via
the hyperfine interval in the cesium atom, while the atomic mass
unit and the Avogadro number are defined via the mass of a carbon
atom. These two atoms are significant for our system of units, SI.
In addition, there are some favorite atomic systems where the
basic laws of Nature find their expression in the most transparent
way. These simple atoms, most of which consist of two bound
particles, have been crucial for our understanding of key moments
of modern physics and their study is still of essential interest
and importance.

The simplicity and harmony of the theory of bound systems have
been tempting and challenging for a while. Johannes Kepler
believed the Solar planetary system to be governed by a harmony of
discrete numbers via geometry, trying with the so-called {\em
Platonic} or {\em Regular solids}. He failed to verify that kind
of the harmony and discovered instead some regularities in the
planetary orbital motion known now as Kepler's laws. His discovery
became a milestone in the development of theory of gravitation and
mechanics and eventually led to the establishment of new mechanics
(classical mechanics).

Three centuries after the Kepler's time, a planetary model was
suggested for atoms. Meantime certain regularities in the spectrum
of the hydrogen atom were discovered. Those regularities, like the
Kepler's laws, led once again to the establishment of new
mechanics (quantum mechanics) and simultaneously realized the
Kepler's dream of the harmony of numbers governing the orbital
motion.

By now we know that a quantity describing a classical object can
be of an arbitrary value while in the quantum case only discrete
values are possible for some quantities. And this is how certain
integer numbers enter the basic equations of modern physics.

Working on a new mechanics, a new dynamics or a new model, one
used to try first to apply it to some `simple' objects. The
simplest object is a free particle. However, only a limited number
of properties of a free particle can be derived {\em ab initio\/}
and studied with a high accuracy. Study of simple atoms opens a
broad field for possible effects and applications. A two-body
atomic system such as the hydrogen atom is a natural object to
verify a new model or to test a new approach. Studies of the
properties of the hydrogen atom have already served to establish
the so-called `old quantum mechanics' (the Bohr theory), modern
nonrelativistic quantum mechanics, relativistic quantum mechanics
(based on the Dirac equation) and quantum electrodynamics (QED),
which was the first successful quantum field theory. Perhaps, we
should even say that QED is the only quantum field theory which is
successful for a really broad range of problems from atomic
spectra to scattering, from low energy, related to microwave
radiation, to high energy phenomena with hard annihilation and
bremsstrahlung, from nano- to giga- electronvolt.

\begin{figure}[hbtp]
\begin{center}
\includegraphics[width=\textwidth]{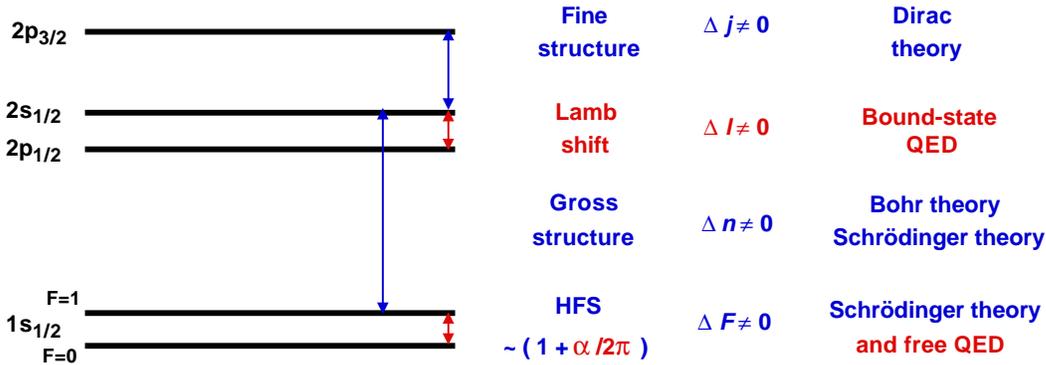}
\caption{Some low-lying levels in the hydrogen atom (not to
scale). The hydrogen $nl_j(F)$ levels are labelled by the values
of the principal quantum number $n$, orbital moment $l$, electron
angular momentum $j$ and atomic angular momentum ${\bf F}={\bf j}
+{\bf I}$, where $I$ is the nuclear spin. The gross structure
($\Delta n\neq0$) is well explained by the Bohr theory (so-called
`old quantum theory') and Schr\"odinger theory which also predicts
the hyperfine structure ($\Delta F\neq 0$). The fine structure
($\Delta j\neq 0$) is explained by the Dirac theory while the Lamb
shift ($\Delta l\neq 0$) is a result of bound state QED effects.
Meanwhile effects of quantum electrodynamics for free particles
are responsible for the $\alpha/2\pi$ anomaly in the hyperfine
structure. \label{00hydr}}
\end{center}
\end{figure}

Figure~\ref{00hydr} shows several crucial contributions to
hydrogen energy levels. We note here that one of reasons for
choosing a non-relativistic equation by Schr\"odinger over a
relativistic Klein-Gordon-Fock equation was an incorrect
description by the latter of the fine structure effects in the
latter. Another remark on importance of the hydrogen atom for QED
is that the anomalous magnetic moment of an electron was first
discovered by Rabi and his colleagues \cite{rabi} as an anomaly in
the hyperfine structure of hydrogen. Immediately that was
interpreted as a possible anomaly related to a free electron and
only afterwards was that confirmed by a direct experiment. A
historic overview of the `contribution' of the hydrogen atom to
modern physics can be found in \cite{rigden}.

One can expect that the simplest atoms are the easiest for a
theoretical prediction. That is true only in part. A simple atom
usually possesses a simple spectrum which has relatively simple
and transparent properties. The atomic energy levels are often
perturbed during measurements by various external factors such as
a residual magnetic or electric field. Because of the simplicity
of the spectra, there is a good chance to understand the influence
of those external factors on simple atoms. As examples of the
highest-precision experiments on simple atoms, we remind here that
the hyperfine interval in the ground state of hydrogen was the
most accurately measured physical quantity for a few decades and
today the $1s-2s$ interval in the hydrogen atom is among the most
precisely measured values.

Decade after decade, theorists and experimentalists investigated
simple atoms. As a result their theory is the most advanced atomic
theory and it has to compete with very sophisticated experiments.
The theory of simple atoms goes now far beyond non-relativistic
quantum mechanics with the reduced mass of an electron. One has
also to take into account relativistic effects, recoil effects,
quantum electrodynamics, effects of the nuclear spin and nuclear
structure.

In an early time of modern physics a list of simple atoms
consisted of hydrogen only and later deuterium, neutral helium and
helium ion were added. Now the list is much longer and quite
impressive. It also contains tritium, hydrogen-like ions of almost
all elements (and other few-electron ions). Artificial two-body
atoms are the easiest to produce and the list also includes
muonium and positronium, muonic atoms, pionic, kaonic and
antiprotonic atoms, exotic bound systems of two unstable particles
(such as pionium and $\pi\mu$-atoms) and antihydrogen.

Often accuracies of theory and experiment are not compatible.
However, there is a broad range of effects, for which theory and
experiment approach the same high level of accuracy. The study of
such effects forms a field called {\em precision tests of bound
state QED\/}, which is reviewed in part here.

A number of sources have contributed to uncertainty of such tests,
and the current accuracy of QED calculations for free particles
and two-body atoms is not a limiting factor for QED tests. The
accuracy of the tests is limited by one of the three other
sources:
\begin{itemize}
\item an experimental uncertainty;
\item an inaccuracy of taking into account effects of the strong
interactions;
\item an
uncertainty due to a determination of the fundamental constants.
\end{itemize}

The latter enters consideration because theory is not in a
position to give itself any quantitative  predictions. It provides
us with some expressions containing values of certain fundamental
constants, such as the Rydberg constant $R_\infty$, the fine
structure constant $\alpha$, the electron-to-proton mass ratio
etc. To make a prediction, one needs first to determine the
fundamental constants by extracting their values from some other
experiments. Thus, theory serves as a bridge between different
experiments. That makes the determination of fundamental physical
constants to be another important part of precision physics of
simple atoms.

The contemporary situation with the QED uncertainty being below
the total uncertainty of any QED test is a result of significant
theoretical progress for the two last decades. Twenty or thirty
years ago the QED uncertainty was often the dominant one.

The {\em precision tests of QED\/} form a {\em
multidisciplinary\/} field involving atomic, nuclear and particle
physics, laser spectroscopy, frequency metrology, accelerator
physics, advanced quantum mechanics, quantum field theory etc.
Those tests are rather not to test QED itself, but to check the
overall consistency of the results and methods from different
fields and in part to search for possible new physics beyond the
Standard Model. In doing more and more advanced QED calculations,
we also need to verify our approaches to deal with infrared and
ultraviolet divergences, renormalization and bound state problem
for few-body atoms in the case of high order terms of perturbation
theory.

As already mentioned, the simplest object to test any theory is a
free particle. A study with free leptons provides us with a
possibility of testing the QED Lagrangian. The most accurate data
for a free particle are related to the anomalous magnetic moments
of an electron and a muon. In the former case the limitation of
the accuracy of a comparison of theory versus experiment is due to
a determination of the fine structure constant $\alpha$ while the
latter QED test is limited by effects of strong interactions
entering via hadronic intermediate states for the vacuum
polarization effects and by the experimental accuracy. The QED
theory of the anomalous magnetic moment is quite advanced
including many hundreds of multiloop diagrams (up to the four-loop
level for the electron and five-loop level for the muon). That
differs from a so called {\em bound state QED\/}, a quantum
electrodynamics theory of simple atoms, which deals mainly with
relatively few one-loop and two-loop diagrams, but those are not
for free particles, but for the particles bound by the Coulomb
field. Three-loop contributions are rather below the uncertainty
of most of the bound QED calculations and if necessary can be
calculated neglecting binding effects.

These two QED theories, the free QED and the bound state QED, are
very different in their approaches, problems and applications and
it is worth to consider their tests separately. The bound state
problem makes all calculations more complicated. Intending to
perform a calculation within the free QED, one can in principle
find needed basic expressions in a number of textbooks. On the
contrary, the bound state QED is not a well-established theory and
there are no published common prescriptions for the relativistic
quantum bound problem. It involves different effective approaches
to solve the two-body bound problem.

Precision tests of the bound state QED offer a number of different
options:
\begin{itemize}
\item The approaches for the bound problem can be essentially
checked with the low-$Z$ two-body atomic systems like hydrogen and
deuterium, neutral helium and helium ions, muonium, positronium,
etc. At low value of the nuclear charge $Z$ the binding energy is
of order of $(Z\alpha)^2 mc^2$ and it is essentially smaller than
the energy related to the rest mass $mc^2$ of the orbiting
particle. That is a regime of a weak coupling in contrast to the
high-$Z$ physics. The weak coupling allows efficient use of a
perturbation expansion over the Coulomb strength $Z\alpha$. Many
contributions of higher-order in $Z\alpha$ are calculated
diagrammatically and thus are closely related to other QED
diagrams (e.g., for scattering theory).
\item Studies of high-$Z$ ions are related to a strong coupling
regime, however, it is unlikely to provide us with more
information on bound state QED because of substantial
contributions due to the nuclear structure. Such an investigation
is rather useful for testing different nuclear models. However, in
some particular cases, atomic systems with not too high $Z$ can
give some important information on higher order terms of the QED
$Z \alpha$ expansion. Importance of particular atomic systems also
depends on the energy interval under study. For instance, the
hyperfine structure (HFS) interval depends more strongly on the
nuclear-structure effects than the Lamb shift, and the related
calculations involve more details of the nuclear structure. The
accuracy of the nuclear-finite-size contribution is often claimed
to be very high, however, their uncertainty is customarily not
well verified. It is often estimated from a variation of the
result for an energy shift after application of certain models of
the charge distribution while the charge radius is fixed. However,
to claim such a calculation as an {\em ab initio\/} result, one
has to reconsider first the original scattering and spectroscopy
data on the nuclear form factor and related nuclear theory and to
check how strongly they constrain the shape of the distribution in
general and its parameters and afterwards to question how much the
nuclear-size correction varies within allowed shapes and
parameters. That may be done only on a nucleus-by-nucleus basis.
Lack of such a consideration in study of high-$Z$ study reduces
importance of the bound-state QED calculations, which are in such
a case one more confirmation of applicability of the
phenomenological model of the charge distribution.
\item Studies of few-electron atoms involve electron-electron
interactions. In the case of high $Z$ ions, the electron-electron
interaction is proportional to $\alpha$, while the interaction
between an electron and the nucleus is proportional to $Z\alpha$.
If the ion is highly charged with only few electrons remaining,
the electron-electron interaction can be treated as a
perturbation. As a result, at high $Z$ the electron must be
treated relativistically, i.e., no expansion over $Z\alpha$ can be
used, however, the treatment of the electron-electron interaction
as a perturbation leads to the $1/Z$ expansion. In the case of
light atoms, electrons are essentially nonrelativistic but the
electron-electron interaction is compatible with the
nucleus-electron interaction. The few-electron atoms (like, e.g.,
the neutral helium atom or heavy lithium-like ions) is the only
case when the uncertainty of the QED calculations used to be a
limiting factor for a comparison of theory versus experiment.
\item There are some other two-body atoms under investigation.
They contain a muon or a hadron as an orbiting particle. The orbit
in a muonic atom is much closer to the nucleus than in the case of
a conventional atom with an orbiting electron, and the muonic
energy levels are much more sensitive to the nuclear structure. In
the case of a hadronic atom, the orbit lies even lower than in a
muonic atom, while the interaction of the orbiting particle and
the nucleus is dominated by the strong interaction, in contrast to
the electromagnetic interaction in conventional (i.e., electronic)
and muonic atoms. Different exotic or muonic atoms offer a unique
opportunity to study certain particle properties by spectroscopic
means with high precision.
\end{itemize}

Frequently high-$Z$ spectroscopy is quoted as a QED test {\em at a
strong field\/}. However, that is not exactly the case. A value of
macroscopic meaning, such as the electric field strength {\bf E},
has not much sense inside an atom. Other details are more
important. For example, the strength of the field can be
characterized by the average potential energy
\begin{equation}\label{eaver}
\langle U \rangle \sim \frac{(Z\alpha)^2 m c^2} {n^2}\,,
\end{equation}
which increases with the nuclear charge $Z$ and the mass of the
orbiting particle $m$ and decreases with the principal quantum
number $n$. The strongest field is related to high-$Z$ atoms with
an orbiting particle, heavier than electron, at the ground state.
Muonic atoms have been studied for a while and with $m_\mu\simeq
207m_e$ they offer a test at a field stronger than in electronic
atoms and at shorter distances
\begin{equation}\label{raver}
\langle r \rangle \sim \frac{\hbar}{Z\alpha m c}n^2\;.
\end{equation}
The distance $\langle r \rangle$ (or a related characteristic
value of the momentum transfer $\langle q \rangle \sim
\hbar/\langle r \rangle$) is another important characteristic of
the electric field inside an atom.

We also note that a characteristic value of the potential $\langle
U \rangle \sim Z\alpha/\langle r\rangle$, the distance $\langle r
\rangle$ and the strength of the  field $\vert {\bf E}\vert\sim
\langle U \rangle/\langle r\rangle$ strongly depends on a
particular quantity under study. What we present above in Eqs.
(\ref{eaver}) and (\ref{raver}) is related to the leading
contributions to atomic energy levels. Higher order corrections,
such as QED corrections, may involve various short-distance
effects with characteristic distance $\langle r \rangle \sim
{\hbar}/m c$ and momentum transfer of about $mc$. They correspond
to higher field than long-distance effects.

The case of {\em strong fields\/} at short distances in which a
characteristic momentum transfer is higher than $2m_ec$ leads to
an enhancement of polarization effects of the electron-positron
vacuum. That makes the physics of muonic atoms quite different
from that of conventional atoms. High-$Z$ ions offer another
opportunity
--- a {\em strong coupling\/} regime with the binding energy
comparable to $mc^2$ which implies a completely relativistic
consideration. The strong-coupling regime is very different from
perturbative weak-coupling one. The calculations are more
complicated. They may be of interest for study of methods to be
applied in the case of strong interactions. Some of the high-$Z$
results are important also for high-accuracy low-$Z$ physics.
However, one has to remember that the strong coupling is rather
not a problem of QED, but a problem of its application. In this
sense, the stronger the coupling, the less `simple' the system. To
study pure QED problems in more detail we should prefer a
weak-coupling situation.

In muonic atoms the coupling constant and other parameters can
take quite different values. While for $Z=1,2$ the states of
interest are $n=1,2$, the principal quantum number $n$ for medium
$Z$ may be higher than 1 or 2, which used to be common for the
high-$Z$ experiments with conventional atoms.

However, in both cases (muonic/exotic atoms and high-$Z$ ions),
understanding the nuclear properties is needed for any theoretical
predictions. For instance, let us look at studies of the $1s$ Lamb
shift in hydrogen-like uranium. Recently, the experimental result
was improved \cite{gumb}. The experimental uncertainty allows to
check one-loop radiative corrections, but not two-loop effects
calculated few years ago \cite{numtwoloop1}. Those two-loop
corrections are an object of intensive theoretical study and are
of great interest (see Sects.~5, 7 and 13). The
finite-nuclear-size uncertainty for the Lamb shift in U$^{91+}$ is
estimated at a level approximately tenfold below the experimental
uncertainty. However, the uncertainty of the result was obtained
\cite{franosch} (see also \cite{mohr_pr,yesh_u}) by comparison of
two distributions of the nuclear charge, which were the homogenous
spherical distribution and the Fermi distribution. The value of
the mean square radius was fixed as $\langle r^2
\rangle^{1/2}=5.8604(23)\,$fm  for both distributions. However,
this result was obtained in Ref.~\cite{zumbro} from muonic uranium
spectroscopy suggesting a modified Fermi distribution, which is
different from both distributions applied in \cite{mohr_pr}. It
was stated \cite{zumbro} that the uncertainty presented there was
of pure statistical nature, while the model-dependence had not
been studied and related systematic error was not included.
Apparently, the characteristic atomic momentum in muonic uranium
is much higher than in the conventional hydrogen-like uranium ion
and muonic spectra are substantially more sensitive to the
nuclear-shape effects than conventional ones. If one expects that
a comparison of the homogenous spherical distribution and the
Fermi distribution leads to a plausible estimation of the
finite-nuclear-size effects (which should be verified), that
should be applied first to muonic atoms to learn the systematic
error of the mean square radius. Then, with a value of the radius
related to each distribution, one should calculate the energy
levels. That should substantially increase uncertainty. This
example shows that how fragile QED tests with high-$Z$ ions can be
and how much work should be additionally done to develop them.

A purpose of this paper is to give a brief review on {\em
precision\/} physics of simple atoms related to the {\em
accurate\/} tests of quantum electrodynamics for bound states and
the determination of fundamental constants. Because of that, we
focus our considerations mainly on light hydrogen-like atoms
(hydrogen, deuterium, helium ion, muonium, positronium) and some
medium $Z$ ions, where the nuclear structure and hadronic effects
are not too significant and allow a crucial test of advanced QED
theory with high order contributions.

We distinguish here QED and bound state QED, which example is an
application of QED to the simplest atoms. Studying less simple
atoms we deal not with just bound state QED, but with its
realization for not-too-simple atoms. The additional problem may
be related to strong field, strong coupling, crucial effects due
to the nuclear structure, electron-electron interaction in
few-electron atoms etc. Definitely, a number of investigations of
less simple atoms are of interest. However, dealing with them is
to go beyond the simplest bound state QED.

We note, that the light hydrogen-like atoms are the most
attractive from a theoretical point of view. They involve neither
electron-electron interactions, nor strong-coupling effects or so.
We consider in the next section, what is the most favorite choice
for experimental accuracy options in testing bound state QED for
hydrogen-like atoms. That is related to the light atoms. They are
also favorite in principle for theoretical accuracy, being the
simplest atomic systems. That does not mean that study of other
atoms are out of interest. First of all, what is important is not
just an atomic system, but a certain transition there. As one can
see in this review, certain transitions, or combinations of
certain quantities, related to different transitions, may offer
various theoretical or experimental advantages.

Because of the simplicity of simple atoms and multidisciplinary
nature of the {\em precision tests of the bound state QED\/} we
have tried to review the results as simply and briefly as possible
in order to make the paper readable by non-experts. Detailed
reference information on crucial theoretical contributions is
collected in tables.

While considering QED, there is always the problem of selecting
units. From a theoretical point of view one should like to
simplify equations by applying relativistic units in which
$\hbar=c=1$, or using the atomic units. Meanwhile, from an
experimental point of view the equations should be expressed in
units convenient for measurements. In our paper we choose a
compromise. All results are presented in the units of the SI.
However, we present, when possible, most of the results using
natural constants as a kind of units. For example, to define the
so-called Fermi energy which is a result of the non-relativistic
interactions of the electron and nuclear magnetic moments, we
write for hydrogen
\[
E_F= - \frac{8\pi \alpha}{3}\frac{\hbar^3}{ m_e^2
c}\,\frac{1}{1+a_e}\, \frac{\langle \mbox{\boldmath$\mu$}_e \cdot
\mbox{\boldmath$\mu$}_p\rangle}{\mu_B^2 } \,\big\vert
\Psi_{1s}({\bf r}=0) \big\vert^2\;,
\]
and thus the proton and electron magnetic moments are explicitly
expressed in units of the Bohr magneton. The other factors do not
directly contain electron charge $e$, but only the fine structure
constant $\alpha$, which does not depend on a choice of
`macroscopic' units (in which it may be defined as
$\alpha=e^2/\hbar c$, $\alpha=e^2/4\pi\hbar c$,
$\alpha=e^2/4\pi\epsilon_0\hbar c$ depending on the definition of
the unit for the electric charge). Still, here we make an
exception for numerical values of atomic energy, which are always
expressed in frequency units, i.e., in terms of $E/h$. This is
because it is widely preferred to write equations for energy,
while the actually measured quantities are the transition
frequencies, spectroscopic linewidths and decay rates. The most
frequently used notations are explained in Appendix~\ref{s:not}.

More details on physics of hydrogen-like atoms can be found in:
\begin{itemize}
\item various basic questions in books \cite{00BS,00S};
\item an overall review with an extended comparison of theory and
experiments related to sixties and early seventies in
\cite{lautrup};
\item minireviews on particular questions in
\cite{00S,KinoQED,ibook,newbook};
\item review on theory of light hydrogen-like atoms in
\cite{SYQED,report};
\item original results presented at {\em Hydrogen Atom\/}
conferences and on International conferences of {\em Precision
Physics of Simple Atomic Systems\/} (PSAS) in
\cite{00H1,ibook,newbook,icjp,newcjp}.
\end{itemize}
The books \cite{ibook,newbook} published in series {\em Lecture
Notes in Physics}, volumes 570 and 627, are also available
on-line. The recent PSAS conference on simple atoms took place
in early August 2004 in Brazil as a satellite meeting to the
International Conference on Atomic Physics (ICAP). The coming PSAS
meeting is scheduled for June 2006 in Venice.

A few problems related to our paper are not presented here in
detail.
\begin{itemize}
\item We consider the fundamental constants only in connection to
simple atoms and QED. More detail on fundamental constants can be
found in \cite{codata1986,codata,00Mohr,newcodata,UFN}.
\item Heavy few-electron ions, which are of a great interest for
study of application of bound state QED to strong-coupling and
few-electron systems, are reviewed in, e.g.,
\cite{mohr_pr,00Myer,00Sto}.
\item Recent progress with exotic and muonic atoms is presented in
detail in \cite{00Neme,00Yam,00Mar,00Yama}. Study of such atoms
are not of a big interest because of QED, on contrary, they
deliver us a crucial information on other part of physics, namely,
particle and nuclear physics.
\end{itemize}

Most of this review was ready before new results on the
fundamental constants \cite{newcodata} became available and
through out the paper we compare the QED-related results on the
fundamental constants with a previous set of the recommended
constants \cite{codata}. We also note that a substantial part of
QED results under review appeared between the publications of two
recommended sets (their deadlines for collecting the input data
were 1998 and 2002) and most of recent results have been
accommodated in \cite{newcodata}.

We start our review with an introductory discussion of spectra of
simple atoms and related basic QED phenomena and next consider QED
tests with hydrogen (the Lamb shift and hyperfine structure) and
other light atoms. We discuss studies of pure leptonic atoms such
as muonium and positronium. In addition to spectra we consider the
magnetic moments of bound particles in various two-body atomic
systems. The concluding sections of the paper are devoted to
fundamental constants and problems of the bound state QED.

\section{Spectrum of simple atoms and main QED phenomena}
Let us discuss the spectrum of simple two-body atoms in more
detail. The gross structure of atomic levels in a hydrogen-like
atom comes from the Schr\"odinger equation with the Coulomb
potential and the result is well known:
\begin{equation}\label{SchC}
E_n=-\frac{(Z\alpha)^2m_Rc^2}{2n^2}\;,
\end{equation}
where $Z$ is the nuclear charge in units of the proton charge,
$m_R$ is the reduced mass of the atomic orbiting particle (mostly,
an electron)
\begin{equation}\label{mred}
m_R = \frac{Mm}{M+m}\;.
\end{equation}
Here, $m$ and $M$ are masses of the orbiting particle and the nucleus.

There are a number of various corrections:
\begin{itemize}
\item relativistic corrections (one can find them from the Dirac
equation);
\item hyperfine structure (due to the nuclear magnetic moment);
\item recoil corrections;
\item radiative (QED) corrections;
\item nuclear-structure corrections.
\end{itemize}
A structure of levels with the same value of the principal quantum
number $n$ is a kind of signature of any atomic system. For most
of the precision applications the substructure of interest is
related to $n=2$. The corrections decrease with a value of the
principal quantum number as $n^{-3}$ or faster. The only exception
is the Uehling correction for muonic and exotic atoms which scales
as $n^{-2}$ for medium and high $Z$.

\begin{table}[hbtp]
\begin{center}
\begin{tabular}{|l|c|c|c|}
\hline
Contribution & Hydrogen-like & Positronium & Hydrogen-like \\
& electronic atom & & muonic atom \\
\hline
Schr\"odinger contributions &  & & \\
- with $M=\infty$ & 1 & 1 & 1\\
- with $m_R$ (correction) & $m/M$ & 1 & $m/M$ \\
Relativistic corrections &&&\\
- Dirac equation & $(Z\alpha)^2$ & $\alpha^2$ & $(Z\alpha)^2$ \\
- Two-body effects & $(Z\alpha)^2m/M$ & $\alpha^2$ & $(Z\alpha)^2m/M$ \\
Quantum electrodynamics&&&\\
- Self energy & $ \alpha(Z\alpha)^2\ln(Z\alpha)$  &$\alpha^3\ln\alpha$   &$\alpha(Z\alpha)^2\ln(Z\alpha)$   \\
- Radiative width & $\alpha(Z\alpha)^2$  &$\alpha^3$   &$\alpha(Z\alpha)^2$   \\
- Vacuum polarization&$\alpha(Z\alpha)^2$&$\alpha^3$&$\alpha\ln(Z\alpha m/m_e)$\\
- Annihilation&&&\\
-- virtual&$-$&$\alpha^2$&$-$\\
-- real & $-$ & $\alpha^3$& $-$ \\
Nuclear effects&&&\\
- Magnetic moment (HFS) & $(Z\alpha)^2m/M$ & $\alpha^2$& $(Z\alpha)^2m/M$ \\
 & or $\alpha(Z\alpha)m/m_p$ & &or $\alpha(Z\alpha)m/m_p$ \\
- Charge distribution & $(Z\alpha mc R_N/\hbar)^2$ & $-$ & $(Z\alpha mc R_N/\hbar)^2$ \\
\hline
\end{tabular}
\vspace{5mm} \caption{Various contributions to the energy levels.
The results are in units of $(Z \alpha)^2 m c^2$, where $m$ is the
mass of the orbiting particle. Here: $M$ is the nuclear mass and
$m_p$ is the proton mass which enters equations if one measure the
nuclear magnetic moment in units of the nuclear magneton. A
contribution of the nuclear magnetic moment, i.e., the hyperfine
structure, appears if the nuclear spin is not zero. $R_N$ stands
for the nuclear (charge) radius.\label{Tmainqed}}
\end{center}
\end{table}

We summarize the most important corrections to the levels
determined by Eq.~(\ref{SchC}) in Table~\ref{Tmainqed}. The main
phenomena essentially contributing to the structure of the energy
levels in hydrogen-like atoms are listed below.
\begin{itemize}
\item {\bf Non-relativistic consideration}
\begin{itemize}
\item[-] The dominant contribution is determined by the
Schr\"odinger equation for a particle with mass $m$ bound by a
central Coulomb field. The result has order $(Z\alpha)^2 mc^2$ and
we discuss all other corrections in units of $(Z\alpha)^2 mc^2$.
\item[-] The leading non-relativistic correction is already
incorporated into Eq.~(\ref{SchC}) by the introduction of the
reduced mass (\ref{mred}). The correction has a fractional order
of $m/M$, and the result, presented in Eq.~(\ref{SchC}), is
already corrected. This correction is responsible, for example,
for the {\em isotopic shift\/} of hydrogen and deuterium gross
structure. In electronic atoms, the correction has order
$m_e/M\simeq 5\cdot 10^{-4}/A$, where $A$ is the atomic number
(i.e., the atomic mass in the universal atomic mass units). The
isotopic shift between atoms with atomic numbers $A$ and
$A^\prime=A+\Delta A$ involves for heavy elements an additional
small factor of $\Delta A/A$. In muonic atoms the correction is of
order of $0.1/A$ and in positronium the effect reduces the result
in the leading order by a factor of two.
 \end{itemize}
The result (\ref{SchC}) depends on a principal quantum number $n$
only. The structure of levels due to the non-relativistic
approximation does not have a single common notation. It used to
be named as the {\em gross structure\/}, the main structure, the
Rydberg structure, the basic structure etc.
\item {\bf Relativistic effects}
\begin{itemize}
\item[-] The leading relativistic corrections are determined by a
solution of the Dirac equation
\begin{equation}
E_D = mc^2 \times F_D(nl_j)\;,
\end{equation}
where
\begin{equation}\label{defFD}
F_D(nl_j)=
\biggl(1+\frac{(Z\alpha)^2}{\left(n-j-1/2 +
\sqrt{(j+1/2)^2-(Z\alpha)^2}\right)^2} \biggr)^{-1/2}
\end{equation}
and thus
\begin{equation}
E_D = mc^2 \times\left[1-\frac{(Z\alpha)^2}{2n^2}-
\frac{(Z\alpha)^4}{2n^3}\left( \frac{1}{j+1/2} - \frac{3}{4n}
\right) +\dots\right]\;.
\end{equation}
The first term above is for the rest energy, the next is the
leading non-relativistic term and the third term (of order of
$(Z\alpha)^4mc^2$ in absolute units, or $(Z \alpha)^2$ in units of
$(Z\alpha)^2mc^2$ which we apply in this section) is the
relativistic correction. This relativistic correction splits
levels with the same $n$, but different values of the angular
momentum $j$. However, the result does not depend on the orbital
momentum $l$. In particular, these  relativistic effects are
responsible for the {\em fine structure\/} in the hydrogen atom.
They split the $2p_{3/2}$ and $2p_{1/2}$ states, but the energy of
the $2s_{1/2}$ and $2p_{1/2}$ is still the same.
\item[-] The Dirac equation itself is an equation ignoring the
nuclear motion. Both the nonrelativistic reduced-mass effects and
the leading relativistic recoil correction, which contributed
 in order $(Z\alpha)^2m/M$, can be expressed in terms of the Dirac energy (see, e.g., \cite{SYQED})
\begin{equation}\label{FDirRec}
\Delta E_{\rm rec} = mc^2 \times\left[
\frac{m_R-m}{m}\Bigl(F_D-1\Bigr)
-\frac{m_R^3}{2Mm^2}\Bigl(F_D-1\Bigr)^2 \right] \;.
\end{equation}
\end{itemize}
We note that a simple substitution $m\to m_R$ reproduces the
correct results for the leading terms for both gross and fine
structure since only the first term in (\ref{FDirRec}) contributes
there. However, the substitution leads to an incorrect result for
the leading relativistic correction for, e.g., the $1s$ state,
where the second term in (\ref{FDirRec}) is also important.
\item Contributions involving the {\bf nuclear structure effects}
appear already in the non-relativistic approximation.
\begin{itemize}
\item[-] One of the main effects is the hyperfine structure, a
result of interaction of the nuclear spin $I$ with the angular
momentum of the orbiting particle $j$. The energy of the levels
depends on the quantum number $F$ (where ${\bf F} = {\bf j} + {\bf
I}$). The result is of order of $(Z\alpha)^2m/M$ if we use the
notation of the particle physics, where considers a nucleus as a
particle of a charge $Ze$ and mass $M$ and measures its magnetic
moment in units of its own magneton $Ze\hbar/2M$. Here $e$ is the
proton charge. This is the case of positronium and muonium.

In the case of a compound nucleus (deuterium, tritium, helium-3
etc.) another convention is in use. The magnetic moment is
measured in units of the nuclear magneton $e\hbar/2m_p$, where
$m_p$ is the proton mass. In such a case one usually speaks about
a hyperfine structure of order of $\alpha(Z\alpha)m/m_p$.

Indeed, these two notations are related to the same order of
magnitude since $Z/M \simeq Z/(A\,m_p)\sim 1/m_p$ with $Z \leq A
<3 Z$. In the case of the hydrogen atom these two notations are
completely identical.

The hyperfine structure has the same parametrical order for states
with any $l$ and $j$, however, the result for the $s$ states is
usually considerably larger numerically.

A value of the nuclear magnetic moment is a consequence of its
nuclear structure and in conventional atoms (hydrogen, deuterium
etc.) cannot be calculated {\em ab initio\/} in contrast to
muonium and positronium.
\item[-] In the leading order, effects of a distribution of the
nuclear charge (so-called {\em finite-nuclear-size effects\/})
shift energy levels of the $s$ states only and the correction has
order $(Z\alpha mc R_N/\hbar)^2$, where $R_N$ is a characteristic
nuclear radius.
\end{itemize}
\item {\bf Quantum Electrodynamics} offers a variety of effects
and their hierarchy depends on a kind of atom.
\begin{itemize}
\item[-] The leading QED effect in the hydrogen atom (and other
conventional atoms) is the self energy contribution (see
Fig.~\ref{fi:se}) which leads to a correction of order
$\alpha(Z\alpha)^2\ln(1/(Z\alpha))$. It splits the levels with the
same $j$, but different values of $l$. An example is a splitting
of $2s_{1/2}-2p_{1/2}$ (the Lamb shift).
\begin{figure}[hbtp] 
\begin{center}
{\includegraphics[width=0.35\textwidth]{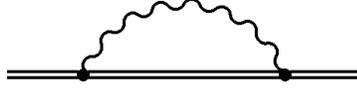}}
\end{center}
\vspace{10pt}
\caption{The self-energy diagram which is
responsible for the dominant contribution to the splitting of
$2s_{1/2}-2p_{1/2}$ (the Lamb shift) in the hydrogen atom. The
doubled electron line is related to the Green function of an
electron in the external Coulomb field.\label{fi:se} }
\end{figure}
\item[-] The vertex diagram in presence of the magnetic field is
chiefly responsible for a correction to the magnetic moment (i.e.,
for the anomalous magnetic moment of an electron) and other
effects are of higher order. As we already mentioned, the
anomalous magnetic moment of an electron was first detected as an
anomaly in the hyperfine structure of atomic hydrogen.
\item[-] An electron vacuum polarization (see Fig.~\ref{fi:vp}) in
the hydrogen atom is responsible for a small fraction of the Lamb
shift. In a general case, the correction, also called the {\em
Uehling correction\/}, has order $\alpha F(Z\alpha m/m_e)$, where
\begin{equation}
F(x) \sim \left\{
\begin{array}{ll}
x^2\,,~~& x\ll 1\,;\\
1\,,~~& x\sim 1\,;\\
\ln{x}\,,~~& x\gg 1\,.\\
\end{array}
\right.
\end{equation}

In the conventional atoms, where $m_e=m$ and $x=Z\alpha$, the
Uehling contribution does not change the hierarchy of the
intervals. In muonic atoms, where $x\simeq 1.5\, Z$, the Uehling
correction can produce $2s-2p$ splittings bigger than
$2p_{3/2}-2p_{1/2}$. The hierarchy depends on $Z$.
\begin{figure}[hbtp] 
\begin{center}
{\includegraphics[width=0.15\textwidth]{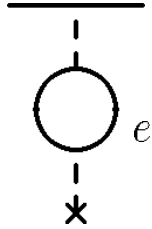}}
\end{center}
\vspace{10pt} \caption{The vacuum polarization diagram (the
Uehling contribution). In the light muonic atoms it leads to the
dominant contribution to the splitting of $2s-2p$ (the Lamb shift)
at low and medium $Z$.\label{fi:vp}}
\end{figure}
\item[-] In the case of positronium the two effects above are not
so important as the virtual annihilation of the bound electron and
positron into a single photon (see Fig.~\ref{fi:an}). The
one-photon annihilation splits two hyperfine levels related to the
$S$ states: the $^3S_1$ state (triplet) is shifted while the
$^1S_0$ state (singlet) is not affected.
\begin{figure}[hbtp] 
\begin{center}
{\includegraphics[width=0.25\textwidth]{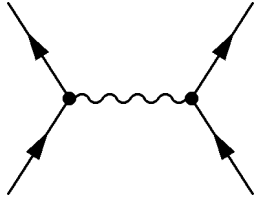}}
\end{center}
\vspace{10pt} \caption{The one-photon annihilation diagram. The
real annihilation into a single photon is not possible for
kinematic reasons, while the virtual one is possible and it shifts
the energy levels of $s$ states with total spin 1.\label{fi:an}}
\end{figure}
\item[-] The QED effects are also responsible for decay of atomic
states (see Fig.~\ref{fi:imse}). The radiative decay line width is
basically $\alpha(Z\alpha)^2$. This result is related to the
one-photon electric-dipole transition (so called $E1$ transition)
which is a dominant decay mode for all levels in light
hydrogen-like atoms except the $2s$ and $1s$ states. The upper
hyperfine component of the $1s$ states can decay via a
magnetic-dipole ($M1$) transition with an extremely small line
width while the metastable $2s$ state decays via a two-photon
electric-dipole ($2E1$) transition with the line width of order of
$\alpha^2(Z\alpha)^4$.
\begin{figure}[hbtp] 
\begin{center}
{\includegraphics[width=0.35\textwidth]{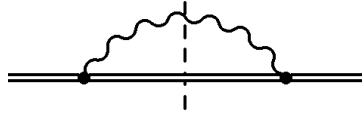}}
\end{center}
\vspace{10pt} \caption{The imaginary part of the self energy
diagram is the energy width of the level and thus it is equal to
the probability of its radiative decay via emission of a single
photon.\label{fi:imse}}
\end{figure}
\item[-] In the case of positronium the line width of the $S$
states\footnote{We use the low case letters ($s$, $p$ etc.) for
single-electron atoms, while the capital letters are used for
atoms with a few electrons. Other indexes for single-electron
atoms are related to the electron properties, while a total
angular momentum $F$ is indicated in the parentheses: e.g.,
$1s_{1/2}(F=1)$. In the case of positronium, since the masses of
electron and nucleus are the same, we use capital letters
indicating that we deal with the total atomic momentum related to
two particles and the subscripts and superscripts are related to
properties of an atom as a whole.} ($n=1,2$) depends on their
total angular momentum (spin). The triplet states ($^3S_1$) are
long living decaying via the three-photon annihilation, while the
two-photon annihilation is a dominant mode for the short-living
singlet states ($^1S_0$). The widths are $\alpha^4$ and $\alpha^3$
respectively. Higher $S$ states additionally to annihilation modes
can also decay radiatively into $P$ states (one photon $E1$
transition) with a related width of order of $\alpha^3$. In the
case of non-$S$ states ($P$, $D$ etc.) the radiative decay
dominates.
\end{itemize}
\end{itemize}

\begin{figure}[hbtp] 
\begin{center}
{\includegraphics[width=\textwidth]{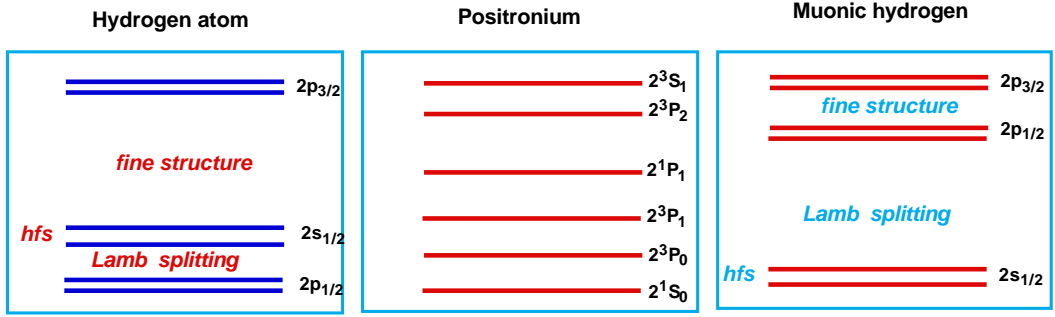}}
\end{center}
\vspace{10pt}
\caption{Scheme of the lowest excited levels ($n=2$) in different simple atoms (not to scale).\label{ficap1}}
\end{figure}

In any case the contribution in Eq.~(\ref{SchC}) dominates for an
interval between levels with $\Delta n\neq 0$. However, the
structure of levels with the same value of $n$ depends on details
of the atom. In Fig.~\ref{ficap1} we present three different basic
spectra of the structure at $n=2$.
\begin{itemize}
\item The first one is realized in `normal' (electronic)
hydrogen-like atoms (hydrogen, deuterium, helium ions etc.). The
muonium spectrum is of the same kind. The largest splitting, of
order $(Z\alpha)^4m_e c^2$, is the fine structure (i.e., a
splitting between levels with a different value of the electron
angular momentum $j$), the Lamb shift arising from the electron
self-energy effects is of order
$\alpha(Z\alpha)^4m_ec^2\ln\big(1/(Z\alpha)\big)$ and it splits
the levels with the same $j$ and different values of the electron
orbital momentum $l$. Some nuclei are spinless (like, e.g.,
$^4$He), while others have a non-zero spin ($I=1/2$ in hydrogen,
muonium, helium-3 and $I=1$ in deuterium etc.). In the latter
case, the interaction with the nuclear spin splits levels with the
same electronic quantum number. The splitting is of order
$(Z\alpha)^4m_e^2c^2/M$ or $\alpha(Z\alpha)^3m_e^2 c^2/m_p$, where
$M$ is the nuclear mass, and the structure depends on the value of
the nuclear spin. The scheme in Fig.~\ref{ficap1} is for a nuclear
spin of $1/2$. A difference for higher $I$ is that the states with
$j\geq 1$ (such as $2p_{3/2}$ with $j=3/2$) are split in three or
more subcomponents. In a general case, the number of hyperfine
components is the smaller number among $2j+1$ and $2I+1$.
\item The structure of levels in positronium and muonic atoms is
different because other QED effects enter consideration. For
positronium, an important feature is a real (into two and three
photons) and virtual (into one photon) annihilation. The former is
responsible for the decay of the $s$-states, while the latter
shifts triplet $s$-levels (1$^3S_1$ and 2$^3S_1$ in particular).
The shift is of the order of $\alpha^4m_ec^2$. Contributions of
the same order arise from relativistic effects for the both, the
electron and the nucleus (namely, positron), and from hyperfine
interactions. As a result, the structure of the positronium levels
at $n=2$ has no apparent hierarchy (Fig.~\ref{ficap1}).
\item Another situation is for the spectrum of the muonic atoms. A
difference to the hydrogen case comes from a contribution due to
the vacuum polarization effect (the Uehling potential). Effects of
electronic vacuum polarization shift all levels to the order of
$\alpha(Z\alpha)^2m_\mu c^2$. This shift is a nonrelativistic one
and it splits $2s$ and $2p$ levels, but in does not depend on $j$
and cannot split $2p_{3/2}$  and $2p_{1/2}$. The fine and
hyperfine structures are of the same form as for the normal atoms
(i.e., $(Z \alpha)^4m_\mu c^2$ and $(Z\alpha)^4m_\mu^2c^2 /M$,
respectively), but the latter is much more important than in a
conventional atom since $m_\mu/m_p\sim 1/9$. At the case of low
$Z$ the Lamb shift induced by the Uehling potential is the
dominant correction to the energy levels.
\end{itemize}
Table~\ref{Tmainqed} shows $Z$ dependence of various crucial
contributions to energy of bound states in two-body atoms and to
their decay width. Since the radiative linewidth increases with
$Z$, it is much easier to work with light atoms which provide a
longer lifetime of excited states. We note, however, that
importance of QED effects also increases with $Z$ in such a way
that a ratio of a correction to width, which is of interest, rather favors
heavy ions with high $Z$. Meanwhile there is a number of other
reasons which make low-$Z$ atoms more attractive.
\begin{itemize}
\item Studies of neutral atoms and ions involve quite different
methods and as a result, some quantities are better determined for
neutral two-body atoms (like, e.g., the hyperfine interval in the
ground state) while others lead to a higher accuracy for ions
(like, e.g., the hyperfine interval for the excited $2s$ state in
helium-ion $^3$He$^+$). Indeed, all neutral two-body atoms have a
low value of the nuclear charge, namely, $Z=1$.
\item Precision QED tests are realized in atomic systems which can
be calculated {\em ab initio\/}. Such a theory can be in principle
developed for any QED effects, but not for the nuclear structure.
Importance of the nuclear structure effects increases with $Z$. In
the leading non-relativistic approximation (see
Table~\ref{Tmainqed}) the related nuclear-spin-independent
contribution to the energy is proportional to $Z^4$ and a squared
value of the nuclear charge radius which also increases with $Z$.
In the case of the nuclear-spin-dependent contributions (hyperfine
effects) the related correction is proportional in the
non-relativistic approximation to $Z^5$ and to a characteristic
value of the nuclear size. In the case of high $Z$ the
non-relativistic approximation still leads to a correct order of
magnitude, however, the dependence on the nuclear charge and
nuclear radii becomes more complicated.

As a consequence, studying hyperfine interactions one has either
to study the nucleon-free atoms (muonium, positronium) or try to
cancel the nuclear effects between a few measured quantities. That
is possible because low-$Z$ atoms allow to present the nuclear
effects in terms of a few effective parameters (such as the
nuclear charge radius, nuclear magnetic radius etc.). On contrary,
for highly charged ions we rather need to apply certain models of
distribution of their charge and magnetic moment and thus the QED
tests with high-$Z$ ions become model-dependent, in comparison
with model-independent calculations for low-$Z$.
\item The Rydberg constant is related to an ultraviolet part of
spectrum. Thus, at low $Z$ ($Z=1,2$) there is a good chance to
study some transitions in gross structure with lasers while for
higher $Z$ laser spectroscopy can be used only for the fine
structure and the Lamb shift as well as for muonic atoms. Strong
dependence of the transition frequency on $Z$ allows only a few
opportunities for a spectrum of two-body atoms with $Z\neq 1$ and/or
$m\neq m_e$.
\item In contrast, a number of transitions in neutral atoms $Z=1$
are suited for laser spectroscopy. That is not a surprise. The
characteristic atomic and molecular energies we meet in our life
are of the same order as in hydrogen since they are mostly related
to certain single-electron excitations in a Coulomb field of an
effective charge of $Z_{\rm eff}\sim 1$. Since we are forced to
meet them in our common-day, industrial and scientific life the
most advanced radiation sources and useful tools were developed
for these regions (optical with some inclusion of infrared and
ultraviolet domains). Their development has a history of a few
centuries of making lenses, mirrors, prisms etc.
\item As we noted, the ratio of the crucial QED corrections of
interest (e.g., $\alpha^2(Z\alpha)^6mc^2$ contributions to the
Lamb shift) and the lifetime (which scales for the $E1$
transitions such as $2p\to 1s$ as $\alpha^5(Z\alpha)^4mc^2$)
clearly favors higher values of $Z$. However, that is not a whole
story. First, there is a metastable $2s$ state in hydrogen-like
atoms which lifetime scales as $\alpha^2(Z\alpha)^6mc^2$ and use
of this level is crucial for a number of experiments. Because of
the longer lifetime and a proper value of the $1s-2s$ interval it
is much easier to deal with this level at low $Z$. In the hydrogen
atom the $2p$ lifetime is about 10\% of the Lamb splitting
$2s_{1/2}-2p_{1/2}$ while a much smaller value of the natural $2s$
lifetime cannot be even seen from the line shape of the two-photon
$1s-2s$ excitation in precision spectroscopy experiments because
of various perturbations which significantly shorten the $2s$
lifetime (which is still much longer than that of the $2p$ state).
The narrow $1s-2s$ transition was successfully used in a number of
hydrogen experiments while there is no access to this narrow
transition for high $Z$. Secondly, the lifetime of the E1 width
sharply depends not only on a value of the nuclear charge $Z$ but
also on the principal quantum number $n$. With such an
$n$-dependence which can be roughly approximated as $1/n^3$ one
can realize that the $2p$ transition is one of the broadest. The
high-$Z$ experiments used to deal with $2s-2p$ and $1s-2p$
intervals and thus the involvement of the $2p$ state is
unavoidable. On contrary, the recent success in hydrogen
spectroscopy is related to high $n$ levels studying $2s-nd$ and
$2s-ns$ at $n=8-12$. Use of the metastable $2s$ level and highly
excited levels in hydrogen compensates all disadvantages of the
correction-to-width ratio for low $Z$ and makes the low-$Z$ case
most favorable for the bound state QED tests.
\end{itemize}
Since experimental and theoretical methods favor the light atoms,
we focus our consideration of particular QED tests on hydrogen and
other low $Z$ atoms and touch other atoms only in case if they
provide any results competitive with $Z=1$ atomic systems.

\section{Optical measurements in hydrogen atom and determination
of the Rydberg constant}

\begin{figure}[hbtp] 
\begin{center}
{\includegraphics[width=.7\textwidth]{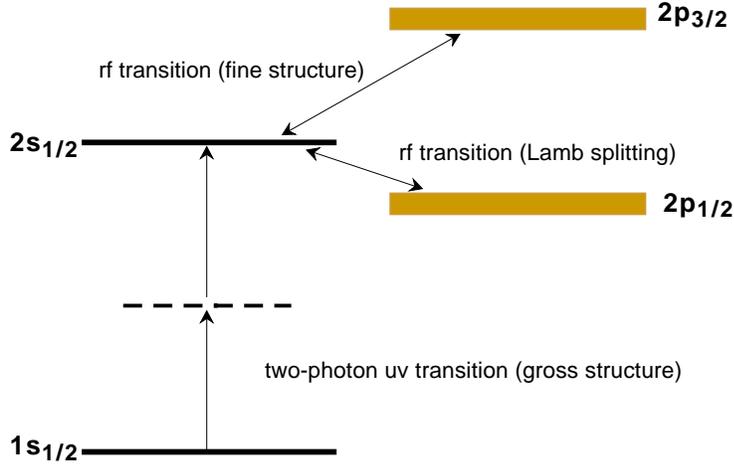}}
\end{center}
\vspace{10pt} \caption{An example of different transitions in the
hydrogen atom (not to scale). Transitions within the fine
structure and Lamb splitting are in the microwave range, while the
$1s-2s$ two-photon transition lies in the ultraviolet domain. The
Lamb splitting is the difference of the Lamb shifts of $2s_{1/2}$
and $2p_{1/2}$ levels. The hyperfine structure is neglected. The
$2p$ linewidth is approximately 10\% of the Lamb shift, which
itself is approximately a tenth part of the fine structure
splitting.\label{figHyd}}
\end{figure}

About fifty years ago it was discovered that in contrast to the
spectrum predicted by the Dirac equation, there are some effects
in hydrogen atom which split the $2s_{1/2}$ and $2p_{1/2}$ levels.
Their splitting known now as the Lamb shift (see
Fig.~\ref{figHyd}) was successfully explained by quantum
electrodynamics. The QED effects lead to a tiny shift of energy
levels and for thirty years this shift was studied by means of
microwave spectroscopy (see, e.g., \cite{hinds,pipkin}) measuring
either directly the $2s_{1/2}-2p_{1/2}$ splitting or the larger
$2p_{3/2}-2s_{1/2}$ splitting (fine structure) where the QED
effects are responsible for approximately 10\% of the interval.

A recent success of two-photon Doppler-free spectroscopy
\cite{twophot} opens another way to study QED effects directed to
high-resolution spectroscopy of gross-structure transitions. The
energy of such a transition between levels with different values
of the principal quantum number $n$ is mainly determined by the
Coulomb-Schr\"odinger formula in Eq.~(\ref{SchC}). However, an
unprecedently high accuracy achieved by to date allows to study
various tiny perturbations to Eq.~(\ref{SchC}), caused by effects
of relativistic physics, quantum electrodynamics and nuclear
structure. All studied transitions are related to the ultraviolet
part of the spectrum.

For any interpretation of hydrogenic lines in terms of QED effects
one has to determine a value of the Rydberg constant
\begin{equation}
\label{Ryd}
R_\infty = \frac{ \alpha^2m c}{2h}\;.
\end{equation}
One more problem for the interpretation of the optical
measurements is the involvement of few levels, significantly
affected by the QED effects. In contrast to radiofrequency
measurements, where the $2s-2p$ splitting was under study, the
optical measurements have been performed with several transitions
involving a number of states ($1s$, $2s$, $3s$ etc.). It has to be
noted that the Lamb shift for levels with $l\neq 0$ has a
relatively simple theory, while theoretical calculations for the
$s$ states lead to several serious complications. The problem has
been solved by introducing an auxiliary specific difference
\cite{del1,del2}
\begin{equation}\label{DefDelta}
\Delta(n)=E_{L}(1s)-n^3E_{L}(ns)\;,
\end{equation}
for which theory is significantly simpler and more clear than for
each of the $s$ states separately.

Combining theoretical results for the difference \cite{del2,rp}
with a measured frequency of two or more transitions one can
extract a value of the Rydberg constant and the Lamb shift in the
hydrogen atom.

\begin{figure}[hbtp]
\begin{center}
\includegraphics[width=.7\textwidth]{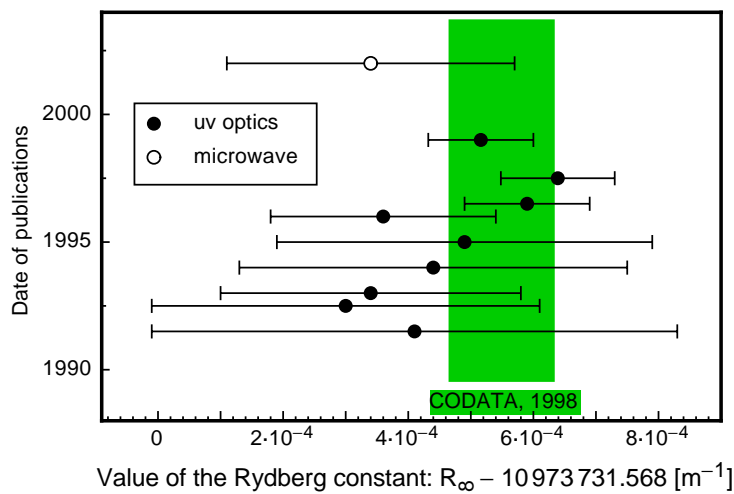}
\end{center}
\caption{Progress in determination of the Rydberg constant by
means of two-photon Doppler-free spectroscopy of hydrogen and
deuterium. The label {\em CODATA
} stands for the recommended
value of the Rydberg constant $R_\infty (1998)$ \cite{codata} from
Eq.~(\ref{ry98}). The most recent original value is a preliminary
result from MIT obtained by microwave means
\protect\cite{MIT_Ry}.\label{figRy}}
\end{figure}

The recent progress in determination of the Rydberg constant is
presented in Fig.~\ref{figRy} (see Refs.~\cite{twophot,codata} for
references). The progress is also clearly seen from the
recommended CODATA values of 1986 \cite{codata1986} and 1998
\cite{codata}:
\[
R_\infty (1986) = 10\,973\,731.534(13)\;{\rm m}^{-1}
\]
and
\begin{equation}\label{ry98}
R_\infty (1998) = 10\,973\,731.568\;549(83)\;{\rm m}^{-1}\,.
\end{equation}
The former value was derived from one-photon transitions (Balmer
series) and was `slightly' improved later (by a factor of 4.5),
but all further progress that led to the 1998's value (improvement
by more than 30 times) was a result of the study of two-photon
transitions in hydrogen and deuterium. Figure~\ref{figRy} shows a
comparison of several recently published values for the Rydberg
constant.

As mentioned before, the other result, which can be extracted from
the optical spectroscopy of the hydrogen atom, is the $1s$ ($2s$)
Lamb shift. However, studies of the Rydberg constant and the Lamb
shift, which are related quantities, are still not the same. While
the optical measurements deliver us the most accurate data on both
values (and in fact the data are strongly correlated), the
microwave experiments still supply us with various less precise
data on both quantities and the microwave data on the Lamb shift
and the Rydberg constant are completely independent from each
other.

The only example of a microwave determination of the Rydberg
constant is an MIT experiment \cite{MIT_Ry}, which dealt with
highly excited hydrogen levels. The transitions under study were
between $n=30$ and $n=27$ for high $l$ and because of such high
values of $n$ and $l$ the levels were not very sensitive to the
QED effects and thus the derived value of the Rydberg constant is
`Lamb-shift-free'. The preliminary result \cite{MIT_Ry}
\[
R_\infty=10\,973\,731.568\,34(23)\;{\rm m}^{-1}
\]
agrees with other values (see Fig.~\ref{figRy}). All other
important related microwave results in the field have been
achieved for either the Lamb shift or the fine structure.

\section{The Lamb shift in the hydrogen atom}

\begin{figure}[hbtp]
\begin{center}
\includegraphics[width=.7\textwidth]{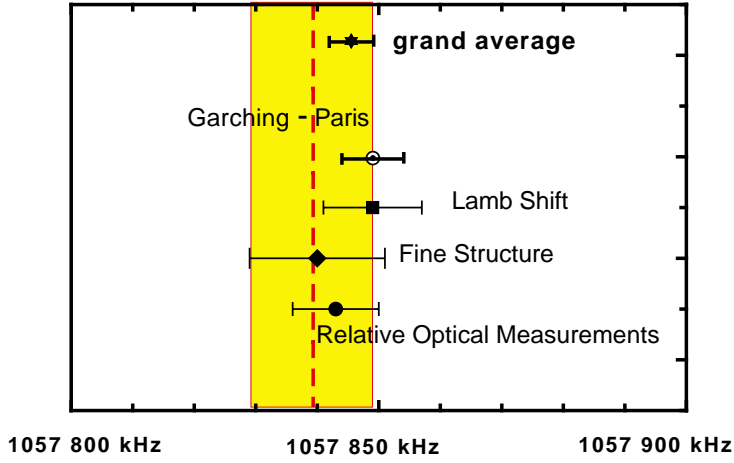}
\end{center}
\caption{Measurement of the Lamb shift in the hydrogen atom. The
most accurate experimental result comes from a comparison of the
$1s-2s$ interval measured at MPQ (Garching) \cite{mpq} and the
$2s-ns/d$ intervals at LKB (Paris) \cite{paris}, where $n=8,10,12$
(see also \cite{twophot} for detail). Three more results are shown
for the average values extracted from direct {\em Lamb shift\/}
measurements, measurements of the {\em fine structure\/} and a
comparison of {\em two optical\/} transitions within a {\em
single\/} experiment (i.e., a relative optical measurement). The
filled part is for theory. Theory and evaluation of the
experimental data are presented according to
Ref.~\cite{rp}.\label{figLamb}}
\end{figure}

After the progress of the last decade, the optical determination
of the Lamb shift \cite{twophot} superseded traditional
microwave measurements \cite{hinds,pipkin}, but still there is a
number of compatible microwave results among the published data.
The experimental results are summarized in Fig.~\ref{figLamb}.

\begin{figure}[hbtp]
\begin{center}
\includegraphics[width=.7\textwidth]{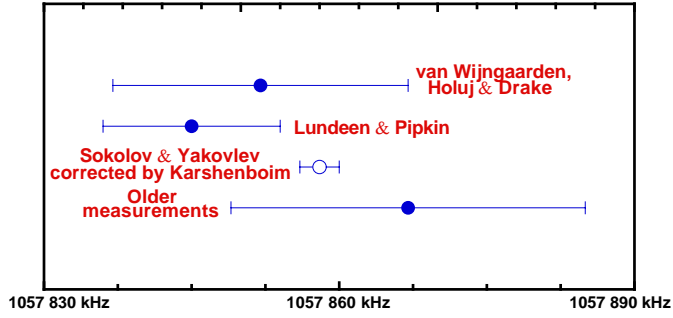}
\end{center}
\caption{Direct measurement of the Lamb shift in the hydrogen
atom. The most recent result was obtained in
\cite{Lundeen81,Wijn98}, while the older results \cite{oldLS} are
averaged. The result of Sokolov and Yakovlev \cite{Sokolov} has
been corrected according to \cite{PS98}.\label{f:icap3}}
\end{figure}

\begin{figure}[hbtp]
\begin{center}
\includegraphics[width=.7\textwidth]{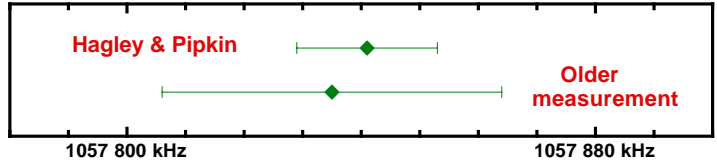}
\end{center}
\caption{An indirect determination of the Lamb shift
($2s_{1/2}-2p_{1/2}$) in atomic hydrogen via a study of the fine
structure $2p_{3/2}-2s_{1/2}$. The most recent result was obtained
in \cite{Hagley}, the older results \cite{oldFS} are
averaged.\label{f:icap2}}
\end{figure}

To discuss the state of the art in study of the Lamb shift in
detail let us start with the experimental data. A number of
different intervals in the hydrogen atom have been precisely
studied for about a century. Bound state QED and in part even QED
itself were established essentially in order to explain the Lamb
shift and the fine and hyperfine structure in the hydrogen atom as
well as in the deuterium atom and the helium ion\footnote{The Lamb
shift is sometimes named after Lamb and Retherford, referring to
their famous papers on the Lamb shift in the hydrogen atom
\cite{LambRe}. However, that is not quite correct, since at
approximately the same time a result on the Lamb shift in the
helium ion was presented in a paper by Skinner and Lamb
\cite{LambSki}.}. The corresponding transitions lie in the
microwave part of the spectrum. In the last decades, progress in
radiofrequency experiments with hydrogen and other light atoms was
rather slow. The results of over than thirty years related to the
Lamb shift in hydrogen are presented in Fig.~\ref{f:icap3}
(measured directly) and Fig.~\ref{f:icap2} (the Lamb shift deduced
from the measured fine structure interval $2p_{3/2}-2s_{1/2}$).
The most recent results are presented separately, while the older
experiments are averaged. The highest accuracy was claimed in
\cite{Sokolov} and we correct here their value according to
\cite{del1,PS98}. Nevertheless, the result is not included in the
Lamb shift average value in Fig.~\ref{figLamb} (the {\em LS\/}
value) because of unclear status of its uncertainty, which is
discussed in part below in this section.

To reach a value of the Lamb shift from the fine structure
($2p_{3/2}-2s_{1/2}$) measurement we need to use a value of the
$2p_{3/2}-2p_{1/2}$ splitting which was found theoretically
\begin{eqnarray}\label{L2pj}
\Delta E_L(2p_{j}) &=&\Delta E_L(2p)
+\frac{(Z\alpha)^4\,mc^2}{8}\times\left\{a_e \cdot A^a_{40}
\left(\frac{m_R}{m}\right)^2 +C_R\,\left(\frac{m}{M}\right)^2
\right.\nonumber\\
&+& \left. \frac{\alpha(Z\alpha)^2}{\pi}\,
\left(A_{61}\ln{\frac{1}{(Z\alpha)^2}}
+A^{VP}_{60}+G_{SE}(Z\alpha)\right) \right\}\nonumber\\
&+&\mbox{\rm higher order terms}\,,
\end{eqnarray}
where we introduce a $j$-independent part of the Lamb shift of the $2p$ state
\begin{eqnarray}\label{L2pp}
\Delta E_L(2p) &=&\frac{\alpha(Z\alpha)^4}{8\pi}\,mc^2\,
\left(\frac{m_R}{m}\right)^3\,\left\{-\frac{4}{3} \ln{k_0(2p)}
\times
\left(1+\frac{Z\,m}{M}\right)^2
\right.
\nonumber\\
&+&\left.\frac{\alpha (Z\alpha)^2}{\pi}\ln^2{\frac{1}{(Z\alpha)^2}}\,B_{62}
+\frac{Z\,m}{M}
\left(C_{\rm Rec}+\frac{\pi}{3}(Z\alpha)\right)\right\}\nonumber\\
&+&\mbox{\rm higher order terms}\;.
\end{eqnarray}
The state-dependent coefficients $A^a_{40}$, $A_{61}$,
$A^{VP}_{60}$, $C_{\rm Rec}$, $C_R$ and $B_{62}$ and the function
$G_{SE}(Z\alpha)$ are well known (see in Table~\ref{t:2pl}); $a_e$
stands for the anomalous magnetic moment of the electron and
$\ln{k_0(2p)}=-0.030\,016\,709\dots$ is the Bethe logarithm.

\begin{table}[hbtp]
\begin{center}
\begin{tabular}{lccc}
\hline
Coefficients & $\Delta E_L(2p_{1/2})$ & $\Delta E_L(2p_{3/2})$ & $\Delta(2)$\\
\hline
$A^a_{40}$ & $-\frac{1}{3}$ & $\frac{1}{6}$ & 0 \\
$A_{61}$ & $\frac{2}{9}+\frac{7}{20}$& $\frac{1}{10}+\frac{2}{9}$& $4\ln2-\frac{197}{60}$\\
$A^{\rm VP}_{60}$ & $-\frac{9}{140}$ & $-\frac{1}{70}$ & $\frac{4}{15}\ln2 + \frac{1}{140}$\\
$G_{\rm SE}$ &$-0.973\,5(2)$ & $-0.486\,5(2)$ & $0.894\;91(10)$\\
$B_{62}$ & $\frac{1}{9}$ & $\frac{1}{9}$& $\frac{16}{9}\ln2-\frac{7}{3}$\\
$C_{R}$ & $\frac{1}{6}-C_{\rm mix}\times\delta_{F1}$ & $-\frac{1}{12}+C_{\rm mix}\times\delta_{F1}$ & 0 \\
$C_{\rm Rec}$  & $-\frac{7}{18}$ & $-\frac{7}{18}$ & $\frac{14}{3}\left(\ln2-\frac{3}{4}\right)$\\
\hline
\end{tabular}
\end{center}
\caption{Coefficients for higher-order contributions to the
hydrogen Lamb shift $\Delta E_L(2p)$ and specific difference
$\Delta(2)$ (see Eq.~(\ref{DefDelta})). Most of the coefficients
are universal and are the same for, e.g., deuterium, with one
exception: coefficient $C_{R}$ depends on the nuclear spin, it is
presented for hydrogen ($I=1/2$); coefficient $C_{\rm mix}=
(g_p)^2/162$ is due to mixture of the $2p_{1/2}(F=1)$ and
$2p_{3/2}(F=1)$ states by the hyperfine interaction and was
considered in \protect\cite{pars} (see, also \cite{pach_mix}) and
$F$ is the complete angular momentum and its value for the
$2p_{1/2}$ state in hydrogen can be either 0 or 1, while for the
$2p_{3/2}$ state it can take value of 1 or 2.\label{t:2pl}}
\end{table}

The {\em direct\/} measurements of the Lamb splitting (i.e., the
interval between $2s_{1/2}$ and $2p_{1/2}$) need in principle no
auxiliary QED calculations. An exception is the result of
experiment \cite{Sokolov} where the originally measured quantity
was a product of the $2s_{1/2}-2p_{1/2}$ splitting and the
lifetime of the $2p_{1/2}$ state. The measurement was claimed to
be the most accurate determination of the Lamb shift (the assigned
uncertainty was about 2 ppm). The result \cite{Sokolov} has been
corrected due to a recalculation of the lifetime \cite{del1,PS98},
where the leading radiative correction was taken into account
within a logarithmic approximation
\begin{eqnarray}\label{tau2p}
\tau^{-1}(2p_{1/2})&=&\frac{2^{9}}{3^8}\,Z^4\alpha^3\,
\Bigl(2\pi\cdot
cR_\infty\Bigr)\frac{m_R}{m}\times\left(1+(Z-1)\frac{m}{M}\right)^2
\times\nonumber\\
&~&
\Biggl\{1+ \big(Z\alpha\big)^2\,\ln{\left(\frac{9}{8}\right)} \nonumber\\
&+& \frac{\alpha\big(Z\alpha\big)^2}{\pi}\,
\left[\big(-2.622\dots\big)\ln{\frac{1}{\big(Z\alpha\big)^2}}+C_0
\right]\nonumber\\
&+&\mbox{\rm higher order terms}\Biggr\}\,.
\end{eqnarray}
We remind here that $R_\infty$ is related to the wave number
($\nu/c$), while $c R_\infty$ corresponds to frequency $\nu$.
Factor of $2\pi$ appears because $h\nu=\hbar\omega$ and the
lifetime $\tau$ is related to the line width of the angular
frequency $\omega=2\pi\nu$.

The logarithmic part of $\alpha(Z \alpha)^2$ term was calculated
in \cite{del1,PS98} (see also \cite{iktau1,iktau}). Recently a
non-logarithmic term was also calculated $C_0=6.576\dots$
\cite{newjskp}. Since our result \cite{iktau} already included an
estimation of the constant $C_0=6.2(5)$, that leads to a marginal
shift in the value of $C_0$, but makes the result previously based
on a naive estimation \cite{iktau} more reliable. We also have to
mention that an exact calculation of $C_0$ alone is not sufficient
for any application because the line width is not well defined. As
it is well known, the line shape contributions to the width enter
in fractional order $\hbar/[\tau\cdot \Delta E]$, which in the
case of an $E1$ transition in a hydrogen-like atom is related to
an $\alpha(Z\alpha)^2$ correction. That is of the same order as
under question and thus the non-logarithmic term should depend on
the phenomenon under study. If we like to express the result of
the Sokolov's experiment \cite{Sokolov} in terms of
$\tau(2p_{1/2})\cdot [E_L(2s_{1/2}) - E_L(2p_{1/2})]$ and to
present the effective lifetime $\tau(2p_{1/2})$ in terms of
Eq.~(\ref{tau2p}), we see that the coefficient $C_0$ is
phenomenon-dependent. However, we note that a numerically big
contribution to the effective constant $C_0$ cannot appear in such
a way since $\hbar/[\tau(2p)\cdot\Delta E(2p-1s)]\simeq
0.33\,\alpha^3/\pi$. To be conservative we double the value of
$\hbar/[\tau(2p)\cdot\Delta E_L(2p-1s)]$ and estimate the
phenomenon-dependent contribution to $\tau(2p_{1/2})$ as $\Delta
C_0\simeq \pm 0.7$ which is to be added to $C_0$ for the
interpretation of experiment \cite{Sokolov}.

We have not taken into account the result of \cite{Sokolov} when
calculating the average values in the summary picture
(Fig.~\ref{figLamb}) because there are certain doubts in its
accuracy noted by Hinds \cite{hinds} and it is not really clear if
the result is as accurate as it was stated originally. The result
of \cite{Sokolov} is rather in a not good agreement with both the
theoretical value and a grand average over the other data. The
discrepancy with theory does not exceed two standard deviations
and it is too early to consider that as a real contradiction.

We need, however, to note that a common opinion on the future of
the microwave measurements of the Lamb shift in hydrogen and on
the reliability of the result \cite{Sokolov} with a 2-ppm
uncertainty has assumed two contradicting statements.
\begin{itemize}
\item First, it is generally believed that since any $2s-2p$
intervals, (either $2s_{1/2}-2p_{1/2}$ (the Lamb splitting) or
$2p_{3/2}-2s_{1/2}$ (the fine structure)) have a width of 0.1 GHz
(because of the radiative lifetime of the $2p$ state), the Lamb
splitting (which is about 1 GHz) cannot be measured better than
at the 10 ppm level. Thus, it is expected that the {\em
statistical error\/} of a microwave measurement should be larger
than 10 ppm.
\item Secondly, sharing the Hinds' point of view \cite{hinds}, it
is believed that the accuracy of the Sokolov's experiment
\cite{Sokolov} is not as high as claimed only because of a
possible {\em systematic\/} error in the determination of the
atomic beam velocity, which, in principle, might be somehow fixed
in the next generation of experiments. Meanwhile, there has been
no doubt expressed on the {\em statistical\/} treatment of the
Sokolov's data.
\end{itemize}

Because of this inconsistency in a opinion of experts, the
question on a possibility of a further progress of microwave
measurements of the Lamb shift with an uncertainty below 10 ppm
remains unclear. A measurement of the Lamb shift in deuterium
within the Sokolov's scheme could provide a chance to test partly
systematic effects since the result may be compared with the
hydrogen-deuterium isotopic shift of the $1s-2s$ interval
\cite{huber}. The latter, combined with the theoretical result on
specific difference in Eq.~(\ref {DefDelta}) \cite{del1,del2,rp},
leads to the isotopic shift for the Lamb shift
\begin{eqnarray}
\Delta E(2s_{1/2}-2p_{1/2}) \biggr\vert^{H}_{D} = &-& \Delta E_L(2p_{1/2})\biggr\vert^{H}_{D}-\frac{1}{7} \Bigr\{\Delta E_{\rm exp}
(1s-2s)  \nonumber\\
&-& \Delta E_{\widetilde{\rm Dirac}}(1s-2s)  +\Delta(2) \Bigl\}
\biggr\vert^{H}_{D}
\end{eqnarray}
with an uncertainty below 1 kHz. Here a calculation of three terms
(namely, the Lamb shift of the $2p_{1/2}$ state, the specific
difference of Eq.~(\ref{DefDelta}) at $n=2$ and the energy related
to an effective Dirac equation
\begin{eqnarray}
E_{\widetilde{\rm Dirac}} = mc^2 &+& m_R
c^2\times\biggl[\Bigl(F_D-1\Bigr)
-\frac{m_R^2}{2Mm}\Bigl(F_D-1\Bigr)^2\nonumber\\
 &+&
\frac{(Z\alpha)^4}{2n^3}\frac{m_R^4}{M^2m^2}
\left(\frac{1}{j+1/2}-\frac{1}{l+1/2}\right)(1-\delta_{l0})\biggr]
 \;,
\end{eqnarray}
which takes into account recoil effects up to
$(Z\alpha)^4m_e^3c^2/M^2$ (the dimensionless Dirac energy
$F_D(nl_j)$ is defined in Eq.~(\ref{defFD})), can be performed
with an uncertainty substantially below 1 kHz. The only
experimental value in the right-hand part of the equation (the
isotopic shift of the $1s-2s$ frequency) is also known with an
uncertainty below 1 kHz \cite{huber}.

Essential progress in studies of the hydrogen Lamb shift came some
time ago from the optical two-photon Doppler-free experiments (see
\cite{twophot,602pho} for detail). The Doppler-free measurements
offer a determination of some transition frequency in the gross
structure with an accuracy high enough to use the results to find
the Lamb shift. However, the cancellation of the linear Doppler
effect is not an only essential advantage of those experiments. A
microwave measurement is to be designed to determine an interval
between the $2s$ state and either the $2p_{1/2}$ or $2p_{3/2}$
state which has a broad line width limiting the accuracy of the
measurement. It is not absolutely clear what the ultimate limit of
the measurements involving the $2p$ states is, but indeed an
experiment with more narrow levels should have a good chance for a
higher accuracy. That is the two-photon excitation of the
metastable $2s$ state to higher excited $ns$ and $nd$ levels that
allows to deal with some relatively narrow levels because of the
$n^{-3}$ dependence of their line width. However, with the
expanding number of the involved energy levels, two special
problems as we mentioned before have appeared in the optical
experiments: a determination of the Rydberg constant and needs to
reduce the variety of the QED contributions for a number of levels
to a single quantity, e.g., to the Lamb shift of the $2s$ (or
$1s$) state $E_L(2s)$.

\begin{figure}[hbtp]
\begin{center}
\includegraphics[width=.7\textwidth]{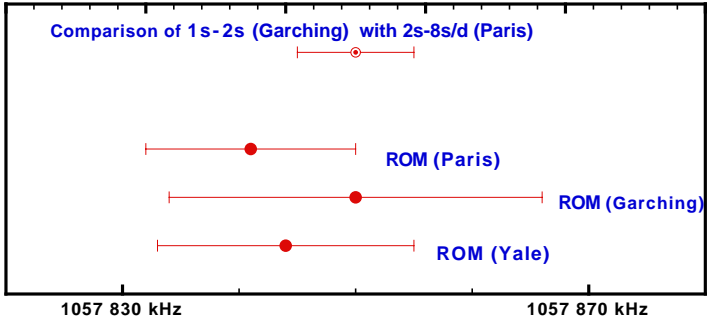}
\end{center}
\caption{An optical determination of the Lamb shift in the
hydrogen atom. The label {\em ROM\/} stands for relative optical
measurements, which dealt with a comparison of two optical
frequencies whithin the same experiment.\label{f:icap4}}
\end{figure}

Two methods were applied to manage the problem with the Rydberg
contribution. The first of them was for a measurement of two
different frequencies within one experiment with the ratio of the
frequencies being approximately an integer number. Extracting a
beat frequency one can avoid the problem of determining the
Rydberg constant at all. Three experiments have been performed in
this way: the Garching experiment dealt with the $1s-2s$
transition and the $2s-4s$ (and $2s-4d$) transition \cite{obfMPQ},
at Yale the $1s-2s$ frequency was compared with the one-photon
$2s-4p$ transition \cite{obfY} and actually that was the only
precision optical experiment with an one-photon transition among
the recent generation of the experiments on hydrogen spectroscopy.
More recent Paris experiment dealt with the $1s-3s$ and $2s-6s$
(and $6d$) intervals \cite{obfP}. The values derived from those
experiments are collected in Fig.~\ref{f:icap4}.

Another way to manage the problem with a determination of the
Rydberg constant is to perform two independent absolute frequency
measurements (i.e., the measurements with respect to a primary
cesium standard). Comparing them afterwards one can determine both
the Rydberg constant and the Lamb shift. Combining results from
Garching on the $1s-2s$ transitions \cite{mpq} and from Paris on
the $2s-8s/8d/12d$ intervals \cite{paris}, the most accurate
optical value of the Lamb shift has been reached
(Fig.~\ref{f:icap4}).

Some of the optical experiments were also performed for deuterium
(see, e.g., \cite{paris}) and that may improve the accuracy in the
determination of the Rydberg constant and, thence, indirectly, of
the hydrogen Lamb shift.

The recent progress with the visible and ultraviolet optical
measurements became in part possible because of use of the
frequency comb generator, an advanced technique which allows to
compare two optical frequencies to each other and an optical
frequency to a microwave frequency \cite{comb}. Because of the
frequency comb generation, a comparison of two transitions within
one experiment gives no advantages anymore. We expect that further
progress will be due to independent absolute measurements of
different transitions.

The optical experiments delivered data which involve a number of
levels ($1s$, $2s$, $3s$ etc.). Therefore, a new problem appeared
in the increasing number of levels and each of these levels has
its own Lamb shift, the value of which should be determined
somehow. A large number of quantities to extract from experimental
data may reduce advantages of the optical approach. A problem with
the number of Lamb shifts was solved with the help of the specific
difference in Eq.~(\ref{DefDelta}) which is found to be of the
form \cite{del1,del2}
\begin{eqnarray}\label{Delta}
\Delta(n) = \frac{\alpha(Z\alpha)^4}{\pi}\frac{m_R^3}{m^2c^2}\times\Biggl\{
&-& \frac{4}{3}\ln{\frac{k_0(1s)}{k_0(ns)}}
\left(1+Z\frac{m}{M}\right)^2 +C_{\rm Rec}\frac{Zm}{M}\nonumber\\
&+&(Z\alpha)^2
\Biggl[A_{61}\ln{\frac{1}{(Z\alpha)^2}}+
A^{VP}_{60}+G^{SE}_{n}(Z\alpha) \nonumber\\ &+&\frac{\alpha}{\pi}
\left(\ln^2{\frac{1}{(Z\alpha)^2}}\,B_{62}+ \dots\right) \Biggr]\Biggr\}\,,
\end{eqnarray}
where the $n$-dependent coefficients $A_{61}$, $ A^{VP}_{60}$, $
C_{\rm Rec}$ and $B_{62}$ are presented for $n=2$ in
Table~\ref{t:2pl}. The coefficients for other $n$ and a table for
the function $ G^{SE}(Z\alpha)$ can be found in Refs.
\cite{del2,rp}. The uncertainty was also discussed there. A
crucial issue is that the difference has a better established
status than that for the $1s$ (or $2s$) Lamb shift. In particular,
various QED corrections vanished for the difference, some other
are of a reduced numerical value. The finite-nuclear-size
contribution, which produces the largest theoretical uncertainty,
is cancelled out after combining the $1s$ and $2s$ contributions.

To analyze the experimental data, we need to use a certain piece
of the bound state QED theory, such as results on the Lamb shift
of $2p_{j}$ states given in Eqs.~(\ref{L2pj}) and (\ref{L2pp}), on
the specific difference $\Delta(n)$ (\ref{Delta}) defined in
Eq.~(\ref{DefDelta}) and a result for the lifetime of the $2p$
state (\ref{tau2p}).

Thus, prior to test the QED {\em theory\/} of the hydrogen Lamb
shift, some {\em theoretical\/} results have to be applied to
extract the {\em experimental\/} values of the Lamb shift from the
data of the measurements. Due to that we have to clarify here the
word `theoretical'. To our mind, a value is a {\em theoretical\/}
one if it is sensitive to {\em theoretical\/} problems (like,
e.g., the problem of the determination of the fundamental constants
and phenomenological parameters needed as an input to the
calculations and the problem of a proper estimation of
uncalculated higher-order QED corrections for the Lamb shift). A
calculated value, which is not sensitive to this kind of problems,
is not theoretical, but rather a {\em mathematical\/} one.

Rigorously speaking, there is neither pure theoretical value nor
experimental value. Theoretical calculations involve
experimentally determined values of certain parameters, or even
functions, while any measurement needs a theory to interpret
measured quantities in terms of collected records of experimental
devices.

The main problem of theoretical calculations for the hydrogen Lamb
shift is a lack of an accurate determination of the proton charge
radius and an insufficient understanding of uncertainty due to the
higher-order QED effects. The former source of theoretical
uncertainty, related to experimental particle physics, dominates.
The proton charge distribution does not affect evaluation of
experimental data via $E_L(2p_j)$, $\Delta(n)$ and $\tau(2p)$,
while the theoretical uncertainty due to the higher-order QED
effects enters the evaluation.

An additional discussion to clarify the problem is needed, since
the auxiliary piece of theory needed to evaluate the experimental
data for the $1s$ Lamb shift is related to the Lamb shift theory,
which in principle we would like to verify measuring the $1s$ Lamb
shift. A question on self-consistency of such a test of the QED
theory of the Lamb shift arises. We emphasize, however, that a
crucial point is that the specific difference $\Delta(n)$ for the
$ns$ states, which are involved in actual experiments, and the
Lamb shift and the lifetime of the $2p_j$ states corresponds to a
part of theory which is understood substantially better than
theory of the $1s$ Lamb shift. The higher-order contribution
enters auxiliary theoretical expressions for  $E_L(2p_j)$,
$\Delta(n)$ and $\tau(2p)$ in an essentially weakened form, since
the dominant contributions are proportional to $\delta_{l0}$.

A successful deduction of the Lamb shift (Fig.~\ref{figLamb}) in
the hydrogen atom provides us with a precision test of bound state
QED and offers an opportunity to learn more about the proton size.
Bound state QED is quite different from QED for free particles.
The bound state problem is complicated by itself even in the case
of classical mechanics. The hydrogen atom is the simplest
atomic system; however, a theoretical result for the energy levels
is expressed as a complicated function (often a perturbation
expansion) of a number of small parameters \cite{icap} (see
review~\cite{report} for a collection of theoretical
contributions):
\begin{itemize}
\item the fine structure constant $\alpha$, which counts the QED
loops;
\item the Coulomb strength $Z\alpha$;
\item the electron-to-proton mass ratio;
\item the ratio of the proton size to the Bohr radius.
\end{itemize}

Indeed, in the hydrogen atom $Z=1$, but it is customary to keep
$Z$ even for hydrogen in order to trace out the origin of the
corrections, especially because the behavior of expansions in
$\alpha$ and $Z\alpha$ differs from each other. In particular, the
latter involves large logarithms ($\ln(1/Z\alpha)\simeq 5$,
$\ln^2(1/Z\alpha)\simeq 24$ and $\ln^3(1/Z\alpha)\simeq 120$ at
$Z=1$) \cite{icap,log1} and big coefficients. Comparing $\alpha$
and $Z\alpha$ we note that in the relativistic case, when the
energy transfer $k_0=\Delta E/c$ is of the same order as the
momentum transfer ${\bf k}$, the parameter is rather $\alpha/\pi$.
In the QED loops, such as for the vacuum polarization and the
self-energy, the relativistic condition $k_0\sim |{\bf k}|$ is
always a case. Sometimes it is also realized for the exchange loop
and in particular in the case of some recoil corrections. However,
most of the Coulomb exchanges correspond to another kinematics,
which is close to the external field case with a low energy
transfer $k_0\ll |{\bf k}|$ and a real parameter is even not
$Z\alpha$ but rather $\pi Z\alpha$ at least for a few first terms
of expansion. One of sources of the factor $\pi$ is a fact that
the imaginary part of logarithm is typically $\pi$, not unity, and
with multilogarithmic contributions some factors $\pi$ in
non-logarithmic terms must be expected.

Because of a presence of at least three small parameters it is not
possible to do any exact calculations and one must use expansions
at least in some parameters. The $Z\alpha$ expansion with all its
enhancements has a very bad convergency and only in a few cases
(some one-loop QED contributions \cite{LSoneloop} and recoil
\cite{LSrecoil}) calculations exact in $Z\alpha$ have been
performed. In such a case the hardest theoretical problem is to
estimate systematically all uncalculated terms related to the
higher-order corrections of the $Z\alpha$ expansion.

References to recent theoretical results on the $1s$ Lamb shift
theory are collected in Table~\ref{17LambTh}. Calculation of these
orders has been completed. Some other corrections are under study
(e.g., $\alpha^3(Z\alpha)^5mc^2$ \cite{egs_in}).

We note that accuracy of the most recent calculations of the
two-loop self-energy \cite{morelogs,alltwoloop} is unclear and we
ignore it in our analysis. In particular we doubt that the higher
order terms may neglected in extrapolation to low $Z$
\cite{alltwoloop}. On contrary we expect they can produce
corrections bigger than claimed in \cite{alltwoloop} uncertainty
of the calculation of the non-logarithmic term. We also do not
think that a reference to a similarity with the one-loop
correction is enough to claim a 15\% uncertainty for the
non-logarithmic term in \cite{alltwoloop}. The one-loop and
two-loop self energies are quite different and in particular they
show a very different $n$ dependence when comparing contributions
for various $ns$ states. The $n$ dependence is strongly related to
splitting of the complete result to partial contributions of
specific areas of integration. Besides we were recently informed
that the single-logarithm two-loop calculation needs corrections
\cite{KPprivate}.

\begin{table}[hbtp]
\begin{center}
\begin{tabular}{lc|lc}
\hline
Correction & Reference(s) & Correction & Reference(s)\\
\hline
$\alpha(Z\alpha)^6mc^2$ &  \protect\cite{pach_lamb1} & $(Z\alpha)^6m^2c^2/M$ & \protect\cite{pach_lamb3,LSrecoil,eides_lamb1}\\
$\alpha^2(Z\alpha)^5mc^2$ &  \protect\cite{pach_lamb2,eides_ry} & $\alpha(Z\alpha)^5m^2c^2/M$ & \protect\cite{pach_lamb4,eides_lamb2,meln_lamb2}\\
$\alpha^2(Z\alpha)^6mc^2$ &
\cite{log1,morelogs,alltwoloop,numtwoloop}
 & $\alpha(Z\alpha)^7mc^2$ & \protect\cite{LSoneloop,sgk_lamb} \\
$\alpha^3(Z\alpha)^4mc^2$ &\protect\cite{meln_lamb1}& & \\
\hline
\end{tabular}
\end{center}
\caption{References to recent theoretical results on calculations of the $1s$ Lamb shift.\label{17LambTh}}
\end{table}

A pure QED calculation for the hydrogen atom is not enough and
before performing any comparison of the QED theory and experiment
we need to take into account effects of the internal structure of
a proton.

\section{Lamb shift and effects of the proton charge distribution}

Let us consider a problem related to the distribution of the
proton charge in more detail. The leading correction due to the
finite size of the proton has a simple expression
\begin{equation}
\label{RadP}
\Delta E_{\rm finite~size}(nl) = \frac{2(Z \alpha)^4m c^2}{3n^3}\,\left(\frac{mcR_p}{\hbar}\right)^2\,\delta_{l0}\;,
\end{equation}
where $R_p$ is the mean-squared proton charge radius. However, to
reach any accurate numerical value one needs first to determine
the value for the proton charge radius with a proper precision.
Unfortunately, data available at the moment, if understood
literally with their claimed uncertainties, are not reliable (we
follow here a consideration in \cite{rp}). The dominant
contribution to the uncertainty of theory of the Lamb shift of the
$1s$ state is due to the contribution (\ref{RadP}) \cite{rp}. It
appears that currently the most accurate value of $R_p$ can be
obtained by reading a comparison of the experimental value of the
Lamb shift and theory in a reversed way: not testing the bound state
QED, but determining the proton charge radius \cite{rp}
\begin{equation}
R_p ({\rm Lamb~shift}) \simeq 0.89(2)\;{\rm fm}\;.
\end{equation}

\begin{figure}[hbtp]
\begin{center}
\includegraphics[width=.7\textwidth]{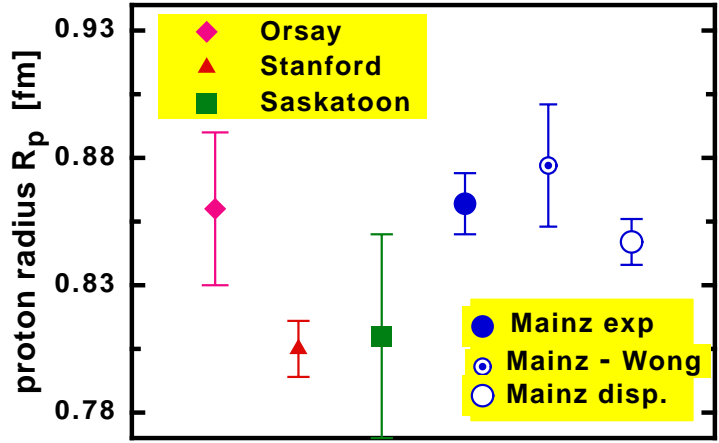}
\end{center}
\caption{Proton charge radius determined from the scattering
experiments. The presented results are phenomenologically
extracted from the scattering at Orsay \cite{orsay}, Stanford
\cite{stanford}, Saskatoon \cite{saskatoon} and Mainz \cite{mainz}
or found with more sophisticated analysis from Mainz data by Wong
\cite{wong} and by Mainz theoretical group from a multi-parameter
dispersion fit of all available data \cite{disp}.\label{f:icap7}}
\end{figure}

The prediction for the theoretical value of the Lamb shift strongly depends on what
value for the proton charge radius is accepted. A crucial level
of accuracy of the radius is 1\%, which is related to the case
when all three uncertainties (experimental, QED and due to the
proton radius) are approximately the same. Two values were under a
long discussion inside the atomic and QED community:
\begin{equation}
R_p({\rm Stanford})=0.805(11)~{\rm fm}\;,~~[79]\;,
\end{equation}
and
\begin{equation}\label{rpmainz}
R_p({\rm Mainz})=0.862(12)~{\rm fm}\;,~~[80]\;.
\end{equation}
More data for the proton radius are collected in Fig.~\ref{f:icap7}.

\begin{figure}[hbtp]
\begin{center}
\includegraphics[width=.8\textwidth]{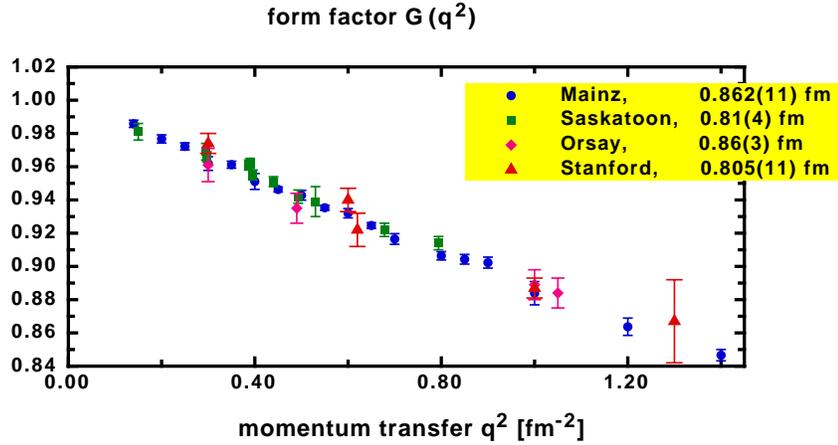}
\end{center}
\caption{Data for the electric form factor of the proton from the
electron-proton elastic scattering experiments performed at Orsay
\cite{orsay}, Stanford \cite{stanford}, Saskatoon \cite{saskatoon}
and Mainz \cite{mainz}.\label{f:icap6}}
\end{figure}

To discuss the scatter of the data, let us look at the most
important data on the electron-proton elastic scattering presented
in Fig.~\ref{f:icap6}. Even from the first glance one can realize
that the Mainz data are
\begin{itemize}
\item presented by a larger number of points;
\item more accurate;
\item obtained for a more broad range of the momentum transfer
including the lowest momentum transfer ever achieved.
\end{itemize}
The latter is important since the easiest way to extract the
proton charge radius from the data is to extrapolate to zero
momentum transfer
\begin{eqnarray}\label{GqRp}
G(q^2) &=&a_0 + a_1 q^2 + a_2 q^4 + \dots\nonumber\\
&=& a_0\cdot \left(1 - \frac{R^2_p}{6} {\bf q}^2 + ...\right)\;,
\end{eqnarray}
where $q^2=-{\bf q}^2$ is related to the three-dimensional
momentum transfer. There is no chance for other data in
Fig.~\ref{f:icap6} to compete with the Mainz results
\cite{mainz}\footnote{I have learned that from a similar picture
presented by Malcolm Boshier for ICAP 1996 (see Fig. 5 in
\cite{boshier}).}. The Mainz experiment is the most appropriate to
precisely determine the proton radius since it contains more
points at lower momentum transfer and with a higher precision than
the rest of the data. Due to this any compilation containing the
Mainz data has to lead to a result close to that from the Mainz
data only, because the Mainz scattering points must be
statistically responsible for the final result. In particular, we
believe the dispersion analysis performed by Mainz theorists
\cite{disp} led to such a result\footnote{Unfortunately, the
authors of \cite{disp} are neither able to explain which part of
data statistically dominates (if any) nor to prove properly the
estimation of the uncertainty assigned by them to the nucleon
radii since the radii were not main objectives of their fitting
procedure and appeared in the very end of the evaluation as a
complicated function of numerous correlated fitting parameters.}
\begin{equation}\label{rpdisp}
R_p({\rm disp})=0.847(9) ~{\rm fm}\;,
\end{equation}
which, however, differs from the empirical value of $R_p=0.862(12)$ fm from Eq.~(\ref{rpmainz}) \cite{mainz}.

\begin{figure}[hbtp] 
\centerline{{\includegraphics[width=0.8\textwidth]{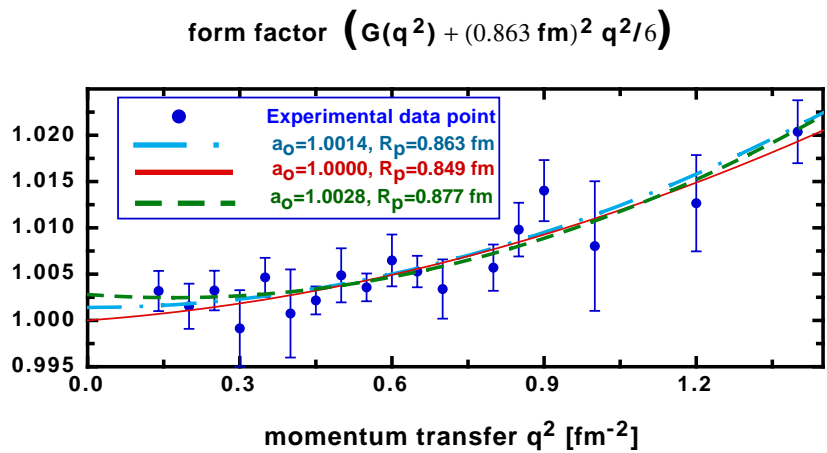}}}
\vspace{10pt}
\caption{Fitting the electric form factor of the
proton from the Mainz experimental data \cite{mainz}. Since
details of fitting by Wong \cite{wong} and a particular result on
the coefficient $a_2$ are not available, we present here similar
fits from \cite{rp}.\label{f:icap8}}
\end{figure}

One of the problems in evaluating the data is their normalization
and actually a different treatment of this problem is in part
responsible for the discrepancy between results from Eqs.
(\ref{rpmainz}) and (\ref{rpdisp}). One can write a low-momentum
expansion of the form factor from Eq.~(\ref{GqRp}) and expect that
$G(0)=1$. However, a real situation is more complicated and
somewhat confusing. The confusion arises from use of the words
`form factor'. The definition of the form factor is clearly
understood theoretically, however, the {\em true\/} form factor
cannot be measured accurately. In the case of scattering one has
to deal with the Rosenbluth formula which describes the cross
section. To make an absolute measurement of the cross section, one
needs to measure parameters of the incoming electron beam with a
high accuracy, as well as determine the efficiency of the
detectors etc. For a relative measurement one has only to ensure
that the details of the beam and detections are the same during
the entire experiment. The normalization is very complicated and
sometimes involves much more efforts than the relative
measurements. However, the normalization measurement used to be
not enough accurate. In particular, in the Mainz case the
uncertainty is about 0.5\%. That means that a value of $G(q^2)$
tabulated from the scattering data, as being the form factor,
actually differs from the {\em true\/} form factor with some
normalization factor, which is consistent with but not equal to
unity.

There is an alternative wording to explain the situation and it
may be helpful for some readers. When the experimentally
determined form factor is presented properly it contains a
statistical error (mainly related to relative measurements) and a
systematic error (mainly due to the normalization). The measured
form factor in such a case directly related to the `theoretical
form factor' and its value at zero momentum transfer is equal to
unity. However, as we mentioned the systematic error is often
higher than the statistical error (at least eventually, when the
statistical error for the fitting parameters is suppressed by the
number of data points). That means that if we at the first
approximation would ignore the systematic error we still should be
able to obtain a result almost without loosing in accurate. If we
drop the systematic error, the central values of the form factor
remain the same, however, it physical meaning is changed
drastically. The measured value $G_{\rm meas}(q)$ can be presented
as
\[
G_{\rm meas}(q) = G_{\rm meas}^\prime(q)\cdot \bigl(1\pm
\delta\bigr)\;,
\]
where $\delta$ is the systematic error of the normalization. While
the measured form factor is defined in such a way that $G_{\rm
meas}(0)=1$; the value with only a statistical uncertainly
satisfies a condition $G_{\rm meas}^\prime(0)=1/(1\pm \delta)$,
i.e., it must be consistent with unity at zero momentum transfer,
but not just equal to it. That is to what we refer above as to a
conventional normalization.

The confusion comes in part from the common attempt to normalize
data as close to the form factor as possible to make their
physical meaning more clear.  However, from a pure practical point
of view that has no sense. The lack of accurate data on the
normalization makes $G(0)$ consistent with unity within the
experimental uncertainty of the normalization, which is nearly
always larger than the uncertainty of relative measurements. The
normalization uncertainty is a systematic error, while the data
are often tabulated with only statistical component of the
uncertainty.

A few different fits were performed by Wong \cite{wong} (see
Fig.~\ref{f:icap8}). Two of them were related to the normalization
used in \cite{mainz} and \cite{disp}\footnote{The paper
\cite{wong} was published prior to \cite{disp}, however, it was
indeed natural to anticipate the use of the `theoretical'
normalization $G(0)=1$.} and the achieved results were close to
published there. A fit with a free value of $a_0$ led to a larger
uncertainty $R_p=0.88(2)\;$fm (see the Wong-Mainz value in
Fig.~\ref{f:icap7}). To our mind, even this result must be treated
with a certain caution. It is necessary to take into account some
higher-order corrections and that is not possible because of the
absence of any complete description of the experiment
\cite{mainz}. Our conservative estimation of the theoretical
uncertainty in Fig.~\ref{figLamb} is related to \cite{rp,icap}
\begin{equation} \label{RpCons}
R_p({\rm conservative}) = 0.88(3) ~{\rm fm}\;.
\end{equation}
The present status of the Lamb shift of the $1s$ ($2s$) state is that
\begin{itemize}
\item the QED computation uncertainty is about 2 ppm,
\item the measurement uncertainty of the grand average value is 3 ppm,
\item while the uncertainty due to the proton size is about 10 ppm.
\end{itemize}

Some other analysis of the proton radius can be found in
\cite{rosenfelder}. The result $R_p=0.880(15)\;$fm is consistent
with (\ref{RpCons}). It is likely that a proper normalization
still is the most important source of correcting the central value
of $R_p$ and the increase of its uncertainty. A very different
treatment by Sick \cite{sick1} which takes into account Coulomb
effects and various systematic sources led to a result
$R_p=0.895(18)\;$fm also consistent with our value in
(\ref{RpCons}). Our uncertainty is somewhat bigger than that of
\cite{wong,rosenfelder,sick1} because we consider possible
systematic errors due to higher order QED corrections \cite{rp} to
scattering which are beyond evaluation in \cite{mainz}.

\begin{figure}[hbtp] 
\centerline{{\includegraphics[width=0.4\textwidth]{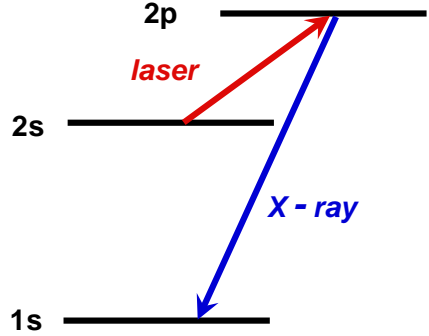}}}
\vspace{10pt}
\caption{Level scheme of the PSI experiment on the Lamb shift
in a muonic hydrogen \cite{psi} (not to scale). The hyperfine structure is not
shown.\label{f:icap10}}
\end{figure}

Improvement of the determination of the proton charge radius is
expected from a study of the Lamb shift in muonic hydrogen. As one
can see from discussion in Sect. 2, the line width scales linearly
with the mass of the orbiting particle, while the
finite-nuclear-size contribution is proportional to $m^3$ and that
offers a good chance for an accurate determination of the proton
charge radius. The level scheme of the experiment which is in
progress at Paul Scherrer Institut \cite{psi} is presented in
Fig.~\ref{f:icap10}. The metastable $2s$ state is to be excited by
the laser light to the $2p$ state which then decays to the ground
state producing an X-ray Lyman-$\alpha$ photon. The intensity of
the X-ray fluorescence is studied as a function of the laser
frequency.

The problem of the nuclear size is not only a problem of the
hydrogen Lamb shift: a similar situation arises with the helium-4
ion Lamb shift, where uncertainties resulting from the QED
computation and the nuclear size are about the same. Theory is in
agreement with experiment \cite{HeLamb} and the leading
theoretical uncertainty is related to the nuclear size. The
theoretical prediction for the Lamb shift in deuterium has a
bigger uncertainty because of the nuclear polarizability effects
\cite{polar-d}. The latter are significant since a deuteron is
known as a very loosely nuclear bound system. For more detail on
the nuclear physics of light nuclei and their effects on atomic
energy levels, see \cite{friartalk}.

\section{Hyperfine splitting in light hydrogen-like atoms and the nuclear structure \label{s:hfs}}

\begin{figure}[hbtp] 
\centerline{{\includegraphics[width=0.7\textwidth]{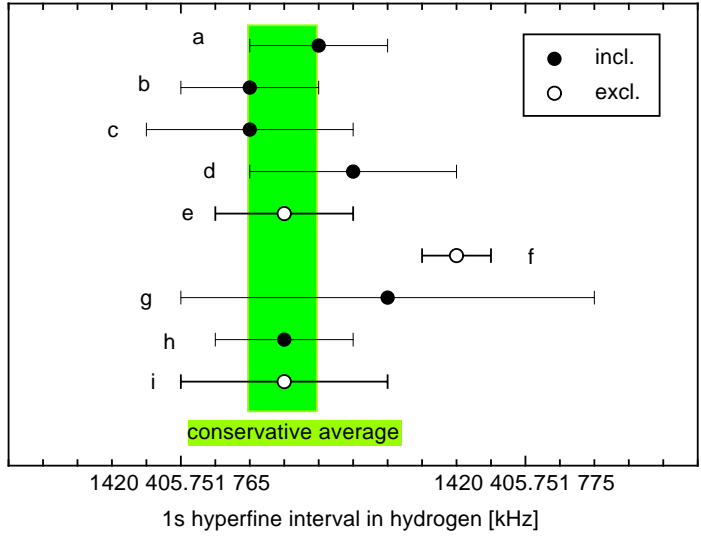}}}
\vspace{10pt} \caption{Measurement of the $1s$ hyperfine splitting
in the hydrogen atom. The references are: $a$ -- \cite{exph1sab}
(experiment \#1 with the wall-shift determined in \cite{exph1sa}),
$b$ -- \cite{exph1sab} (experiment \#2), $c$ -- \cite{exph1sc},
$d$ -- \cite{exph1sd}, $e$ -- \cite{exph1se}, $f$ --
\cite{exph1sf}, $g$ -- \cite{exph1sg}, $h$ -- \cite{exph1sh}, $i$
-- \cite{exph1si}. The conservative average,
1\,420\,405.751\,768(1) kHz, is presented according to
\cite{cjp2000}. It is an average over the results denoted by
filled circles, while the results presented by open circles are
excluded from evaluation (see \cite{cjp2000} for
detail).\label{f:h1shfs}}
\end{figure}

A similar problem of significant interference of the nuclear
structure and QED effects exists for the hyperfine structure (HFS)
in the hydrogen atom and actually the case of the hyperfine
structure is in a sense even much less favorable for the QED
tests. The hyperfine interval in the ground state was for a while
the most accurately measured quantity being known with an
uncertainty at the level of a part in $10^{12}$. The results of
the most accurate experiments are summarized in
Fig.~\ref{f:h1shfs}. It so happened that most results were
published in metrological journals and were missing in various
`physical' compilations. This figure includes only results
published in refereed journals, but not those from conference
proceedings. We also note that the most frequently quoted result
\cite{npl_nature}
\[
f_{\rm HFS}(1s) = 1\,420\,405.751\,767(1)\;\mbox{kHz}
\]
was later corrected \cite{exph1sc} and the uncertainty was
increased by a factor of 3.

\begin{table}[hbtp]
\begin{center}
\begin{tabular}{lcccc}
\hline
Atom, & $E_{\rm HFS} ({\rm exp})$ & Ref. & $E_{\rm HFS} ({\rm QED})$& $\Delta E$ \\
state~~~~~~~~~~& [kHz]&&[kHz]&[ppm]\\
\hline
Hydrogen, $1s$ & 1\,420\,405.751\,768(1) & \protect\cite{exph1sab,exph1sa,exph1sc,exph1sd,exph1sg,exph1sh,cjp2000} &1\,420\,452& - 33\\
Deuterium, $1s$ & ~~327\,384.352\,522(2) & \protect\cite{expd1s} & ~~~327\,339&138\\
Tritium, $1s$ & 1\,516\,701.470\,773(8)  &\protect\cite{mathur} & 1\,516\,760& - 38\\
$^3$He$^+$ ion, $1s$ & - 8\,665\,649.867(10)~~~~~~ &\protect\cite{exphe1s}  & - 8\,667\,494~~& - 213\\
\hline
Hydrogen, $2s$ & 177\,556.860(16)~~~ & \protect\cite{2shydr} & ~~~~177\,562.7 &- 32\\
Hydrogen, $2s$ & ~177\,556.785(29)~~~ & \protect\cite{rothery} & & - 33\\
Hydrogen, $2s$ & ~177\,556.860(50)~~~ & \protect\cite{exph2s} & &- 32\\
Deuterium, $2s$ & ~40\,924.454(7)~ & \protect\cite{mpqd2s} & ~~~~~~~40\,918.82 & 137\\
Deuterium, $2s$ & ~40\,924.439(20)~ & \protect\cite{expd2s} &  &137\\
$^3$He$^+$ ion, $2s$ & ~- 1083\,354.980\,7(88)~~~~ & \protect\cite{prior} & ~- 1083\,585.7& - 213\\
$^3$He$^+$ ion, $2s$ & ~- 1083\,354.99(20)~~~~~~~ & \protect\cite{exphe2s}  &&- 213\\
\hline
\end{tabular}
\end{center}
\caption{Hyperfine structure in light hydrogen-like atoms. The
numerical results are presented for the frequency $E/h$. The
difference ($\Delta E$) between the actual value ($E_{\rm HFS}
({\rm exp})$) and a result of the pure QED calculation ($E_{\rm
HFS} ({\rm QED})$) is caused by the nuclear structure. The
negative sign for the $^3$He$^+$ ion reflects the fact that the
nuclear magnetic moment is negative, i.e., in contrast to other
nuclei in the Table, its direction is antiparallel to the nuclear
spin.\label{17tab1}}
\end{table}

We summarize in Tables~\ref{17tab1} all accurate data available
for the hyperfine intervals of the $1s$ and $2s$ states in
conventional light hydrogen-like atoms (hydrogen, deuterium and
tritium and in the helium-3 ion). Pure leptonic atoms are
considered separately in Sects.~\ref{s:muhfs} and \ref{s:ps}.

A value of the hyperfine interval in the $ns$ state is determined
in the leading order by the so-called Fermi energy $E_F$ as
following
\begin{eqnarray}
E_{ns}({\rm leading})&=&E_F\cdot\frac{1+a_e}{n^3}\nonumber\\
&=& - \frac{8\pi \alpha}{3}\frac{\hbar^3}{ m_e^2 c} \frac{\langle
\mbox{\boldmath$\mu$}_e \cdot \mbox{\boldmath$\mu$}_{\rm
Nucl}\rangle}{\mu_B^2 } \big\vert \Psi_{ns}({\bf r}=0)
\big\vert^2\;.\label{EnsLead}
\end{eqnarray}
Here ${\bf \mu}_e $ and ${\mu}_{\rm Nucl}$ stand for the magnetic
moment of a free electron and the nucleus, respectively, and
\begin{equation}\label{psi0}
\big\vert\Psi_{nl}({\bf r}=0)\big\vert^2 = \frac{1}{\pi}
\left(\frac{Z\alpha\,m_R c}{n\hbar}\right)^3\delta_{l0}
\end{equation}
is a squared value of the non-relativistic wave function at the
origin. Note, that we have included a contribution of the
anomalous magnetic moment of the electron into the leading term
$E_{ns}({leading})$, but not into the definition of $E_F$.

The most important corrections to (\ref{EnsLead}) are due to
relativistic effects, bound-state quantum electrodynamics and due
to the nuclear structure. In the case of a point-like infinitely
heavy nucleus, i.e., when only the relativistic and bound state
QED contributions are taken into account, the correction to
(\ref{EnsLead}) is of the form
\begin{eqnarray}\label{QED1s}
\Delta E_{1s}({\rm QED}) =
E_F&\times
&\left\{\frac{3}{2}(Z\alpha)^2+ \alpha(Z\alpha)\left(\ln2-\frac{5}{2}\right)\right.\nonumber\\
&+&{\alpha (Z \alpha )^2\over \pi}\left[-\frac{2}{3}\ln{1\over(Z\alpha)^2}
\left(\ln{\frac{1}{(Z\alpha)^2}}\right.\right.\nonumber \\
&+&\left.4\ln2-\frac{281}{240}\right) +17.122\,339\ldots \nonumber \\
&-&\left.\left.\frac{8}{15}\ln{2}+\frac{34}{225}\right]+0.7718(4)\,\frac{\alpha^2(Z\alpha)}{\pi}\right\}\,.
\end{eqnarray}
This term is in fact smaller than the nuclear-structure correction (see
Table~\ref{17tabQEDHFS}). The contributions induced by the nuclear
structure are determined there by subtracting the results of the
pure QED calculations
\[
E_{\rm HFS}({\rm QED}) =E_{ns}({\rm leading}) + \Delta E_{ns}({\rm QED})
\]
from the actual values of the hyperfine intervals (see Table~\ref{17tab1})
\[
E_{\rm HFS}({\rm Nucl}) =E_{\rm HFS}({\rm exp}) - E_{\rm HFS}({\rm
QED}) \,.
\]
The result for the $1s$ state is presented in Eq.~(\ref{QED1s})
and the $2s$ result is of a similar form and discussed in detail
in the next section. We mainly follow here \cite{d21}, but the QED
helium result in Table~\ref{17tab1} and related data in
Tables~\ref{17tabQEDHFS} and \ref{17tabD21} are corrected as
explained in \cite{new_rep} and \cite{2sZETF,2seprint} (cf.
\cite{d21}).

\begin{table}[hbtp]
\begin{center}
\begin{tabular}{lcccc}
\hline
Atom & $\Delta E({\rm QED})$& $(Z\alpha)^2E_F$ & $\alpha(Z\alpha)E_F$& $\Delta E({\rm Nucl})$ \\
&[ppm]&[ppm]&[ppm]&[ppm]\\
\hline
Hydrogen& $-23$& 80 & $-96$ & $-33$\\
Deuterium& $-23$& 80 & $-96$ &138\\
Tritium & $-23$ & 80 & $-96$ & $- 38$\\
$^3$He$^+$ ion& 108 & 319& $-192$  & $- 213$\\
\hline
\end{tabular}
\end{center}
\caption{A comparison of the bound QED and nuclear-structure
corrections to the $1s$ hyperfine interval. The QED term $\Delta
E({\rm QED})$ defined in (\ref{QED1s}) contains only the
bound-electron effects while the anomalous magnetic of free
electron $a_e=\alpha/2\pi+...\simeq 1.159\,652\dots\times10^{-3}$
is excluded. Additionally, to the whole QED correction we also
present separately two major contributions into it: the leading
relativistic term in order $(Z\alpha)^2E_F$ and the leading bound
state QED correction in order $\alpha(Z\alpha)E_F$. The nuclear
contribution $\Delta E({\rm Nucl})$ has been found via a
comparison of the experimental results with the pure QED values
(see Table~\ref{17tab1}).\label{17tabQEDHFS}}
\end{table}

There are three kinds of corrections which involve the nuclear-structure effects.
\begin{itemize}
\item The dominant contribution for the hydrogen HFS interval is
due to a distribution of the charge and magnetic moment inside the
nucleus (so-called {\em Zemach correction\/})
\begin{equation}\label{e:zem}
\Delta E({\rm Zemach}) =  \frac{2Z\alpha }{\pi^2}\,E_F\times
mc\,\int{\frac{d^3{\bf q}}
{{\bf q}^4}}\left[\frac{ G_E(-{\bf q}^2) G_M(-{\bf q}^2)}{1+\kappa}-1\right]
\;,
\end{equation}
where the electric/magnetic form factor $G_{E/M}(-{\bf q}^2)$ is a
Fourier transform of the space distribution of the proton electric
charge/magnetic moment $\rho_{E/M}({\bf r})$ and $\kappa$ is the
proton anomalous magnetic moment\footnote{It is necessary to
mention that a straightforward consideration of the form factor as
a Fourier transform of the space distribution of the related value
(such as the electric charge) is a transparent but rough
approximation, which fails for the relativistic momentum transfer
(see, e.g., \cite{Geqrho}).}. In the coordinate space the equation
takes the form
\begin{equation}\label{e:zem1}
\Delta E({\rm Zemach}) = -\frac{2(Z\alpha)mc}{\hbar}\,E_F\times
\int{ d^3{\bf r}d^3{\bf r}^\prime\, \rho_E({\bf r}) \rho_M({\bf
r}^\prime)\, \vert {\bf r}-{\bf r}^\prime\vert } \;.
\end{equation}
For light atoms the uncertainty in (\ref{e:zem}) comes from the
distribution of the magnetic moment which involves the magnetic
form factor $G_{M}(-{\bf q}^2)$ at low momentum transfer, which is
difficult to study experimentally in such kinematic region (in
contrast to the electric form factor $G_{E}$).
\item A different effect dominates for a deuteron, which is a
loosely bound system and the electron can see it not only as the
whole but can also recognize its composites, namely a proton and a
neutron. This is a nuclear polarizability contribution.
\item A theory of a point-like particle with a non-vanishing value
of the anomalous magnetic moment is {\em
inconsistent\/}\footnote{We need to mention a certain
inconsistency in the {\em terminology\/} related to the leptons.
They are referred as ``point-like'' since all effects of the
distribution of the electric charge and the magnetic moment can be
derived from QED. However, they ``have a non-vanishing anomalous
magnetic moment'' which comes from the same QED effects. Indeed,
both effects are to be considered on the same ground. The
inconsistency historically comes from the fact that the anomalous
magnetic moments was first discovered experimentally and later
explained, while the ``electron charge radius'' is hardly to be
seen in a straightforward experiment but can be easily considered
theoretically. The Lamb shift is a result of the electron charge
distribution related to the electron structure, however, it is
understood only through the QED interpretation. So the internal
electron structure effects look rather as a theoretical
construction, while the anomalous magnetic moment is a direct
experimental fact.}. In particular, calculating recoil corrections
to the hyperfine structure in atoms with such a nucleus, certain
divergencies have to appear at high momentum. Those are not
ultraviolet QED divergencies. They are related to an inconsistency
of the consideration of the operators of the non-relativistic
expansion as fundamental operators. The operator
$\sigma_{\mu\nu}k_{\mu}/Mc$ is such an operator and it gives
reasonable results only when $k/Mc\ll 1$. A complete theory with
the finite nuclear size removes all divergencies of this kind and
thus the nuclear size serves as an effective cut-off for them
entering the equations originally designed for a point-like
particle. Since the divergence in the related recoil contribution
to the HFS interval is logarithmic, the recoil effects only
slightly depend on this cut-off and through it on detail of the
nuclear structure.
\end{itemize}

The magnitude of the nuclear corrections to the hyperfine interval
in light hydrogen-like atoms lies at the level from 30 to 200~ppm
(depending on the atom) and theoretical understanding of such
effects is unfortunately by far not sufficient \cite{rp,khr,d21}.
The correction is compatible with the bound-state QED term (note
that the anomalous magnetic moment is related to the free QED
consideration and it is included into the leading term, but not
into the QED correction). The results for the $1s$ state are
summarized in Table~\ref{17tabQEDHFS} (see Ref.~\cite{d21} for
detail). The results for the $2s$ state are slightly different for
the relative value of the QED contribution. A certain progress
\cite{friar_Ze} is on the way, however, it should more contribute
to nuclear physics since a gap between accuracy of a pure QED
theory and that of the nuclear effects is of a few orders of
magnitude and cannot be avoided.

From Table~\ref{17tab1} one can learn that in the relative units
the effects of the nuclear structure are nearly the same for the
$1s$ and $2s$ intervals (33~ppm for hydrogen, 138~ppm for
deuterium and 213~ppm for helium-3 ion). A reason for that is the
factorized form of the nuclear contributions in the leading
approximation (cf.~(\ref{RadP}))
\begin{equation}\label{NuclPsi}
\Delta E({\rm Nucl}) = A({\rm Nucl}) \times \big\vert\Psi_{nl}
({\bf r}=0)\big\vert^2\;.
\end{equation}
In other words, the correction is a product of the
nuclear-structure parameter $A({\rm Nucl})$ and the squared wave
function at the origin (\ref{psi0}), which is a result of a pure
atomic problem (a nonrelativistic electron bound by the Coulomb
field). The nuclear parameter $A({\rm Nucl})$ depends on the
nucleus (a proton, a deuteron etc.) and on the effect under study
(the hyperfine structure, the Lamb shift), but does not depend on
the atomic state. We remind, that, similarly, the leading term for
the HFS interval for an $ns$ state in Eq.~(\ref{EnsLead}) is also
proportional to squared wave function at the origin
$\big\vert\Psi_{nl} ({\bf r}=0)\big\vert^2$ and thus a fractional
value of the nuclear contribution is kept $n$ independent in the
leading approximation.

Two parameters in the hydrogen-like wave function (\ref{psi0}) can
be varied for the same nuclei:
\begin{itemize}
\item the reduced mass of a bound particle for conventional
(electronic) atoms ($m_R\simeq m_e$) and muonic atoms ($m_R\simeq
m_\mu$);
\item the principle quantum number $n=1,2$ for the $1s$
and $2s$ states.
\end{itemize}
We note that the nuclear charge $Z$ is indeed fixed if we like to
deal with the same parameter $A({\rm Nucl})$ (i.e., with the same
nucleus).

The former option was already mentioned when we considered the
determination of the proton charge radius via the measurement of
the Lamb shift in muonic hydrogen \cite{psi}. In the next section
we consider the latter option, a comparison of the $1s$ and $2s$
hyperfine intervals in hydrogen, deuterium and the $^3$He$^+$ ion.
The $1s$ state is the ground state and the $2s$ state is
metastable. Other levels are not accessible for the precision
experiments being broad because of their radiative decay.

\section{Hyperfine structure, free of the nuclear effects: comparison
of the HFS intervals for the $1s$ and $2s$ states\label{s:d21}}

As we have seen in the previous section, both the $1s$ and $2s$
hyperfine intervals in various light hydrogen-like atoms are
measured more accurately than theory can predict. Theory, although
limited by the uncertainty in the calculation of the nuclear
effects, can still proceed to a higher accuracy dealing with a
specific difference
\begin{equation}
\label{d21diff}
D_{21}=8\cdot E_{\rm HFS}(2s)-E_{\rm HFS}(1s)\;,
\end{equation}
where any contribution (and, in particular, related to the leading
nuclear effects) which has the form of Eq.~(\ref{NuclPsi}) should
vanish. Removing the main obstacle on the way of theory, we may
study effects related to higher-order QED contributions.

The difference (\ref{d21diff}) was studied theoretically in
several papers long time ago
\cite{17breit,zwanziger,sternheim,pmohr}. A result of the former
study reads
\begin{eqnarray}\label{d21qed3}
D_{21}({\rm QED3})&=& (Z\alpha)^2 \,E_F \times \left\{ {5 \over 8}
+ {\alpha\over \pi} \left[ \left( {16 \over 3} \,\ln2-7\right)\,
\ln(Z\alpha) -5.551\,55\dots \right] \right.
\nonumber\\
&+& \left.{m\over M} \left[ \left({ \ln2 \over 2}-{7 \over
32}\right) \left(1-{1\over \eta} \right) -{9 \over 8} -\left({145
\over 128} - { 7 \over 8} \,\ln2 \right) \eta\right] \right\} \;,
\end{eqnarray}
where
\begin{equation}\label{defeta}
  \eta = {\mu \over \mu_B} \,{M \over m}\,{1 \over {Z\, I}}\;.
\end{equation}
We remind here about a difference in the notation in particle and
nuclear physics. If the $g$ factor of a nucleus would be defined
in the particle notation, one has to arrive at $\eta$.

The HFS difference $D_{21}$ is similar to $\Delta(n)$ in
Eq.~\ref{DefDelta}, however, there are a number of very different
features. The HFS difference is designed to be compared with
experimental data and it is of interest only for the $1s$ and $2s$
states. The Lamb difference $\Delta(n)$ is to be used for the
interpretation of the data and those involves various $n$, still
$n=2$ is the most important for application case. The leading term
$E_{ns}({leading})$, the leading free QED (due to the anomalous
magnetic moment of the electron) and the leading bound state QED
corrections cancel for the HFS difference. On the contrary, the
leading contribution to the Lamb shift does not cancel for
$\Delta(n)$. Still, there is a similarity in calculations of
various bound state QED contributions.

\begin{table}[hbtp]
\begin{center}
\begin{tabular}{lccc}
\hline
Contribution to HFS in & Hydrogen & Deuterium & $^3$He$^+$ ion\\
& [kHz] & [kHz] & [kHz] \\
\hline
$D_{21}({\rm QED3})$ & 48.937 &  11.305\,6 & -1\,189.253\\
$D_{21}({\rm QED4})$ & {0.018(5)} & {0.004\,4(10)}  &-1.13(14)\\
$D_{21}({\rm Nucl})$ & {-0.002} & {0.002\,6(2)} & 0.307(35)\\
\hline
$D_{21}({\rm theo})$ & 48.953(5) &  11.312\,5(10) & -1\,190.08(15)  \\
\hline
\end{tabular}
\end{center}
\caption{Theory of the specific difference $D_{21}=8E_{\rm
HFS}(2s)-E_{\rm HFS}(1s)$ in light hydrogen-like atoms (see
\cite{2sZETF,2seprint,d21,new_rep} for detail). The numerical
results are presented for the related frequency
$D_{21}/h$.\label{17tabD21}}
\end{table}

Recent studies of $D_{21}$ \cite{sgkH2s} showed that certain
higher-order QED and nuclear corrections have to be taken into
account for a proper comparison of theory and experiment. The
theory has been essentially improved \cite{d21,yero2001} and it is
summarized in Table~\ref{17tabD21}. The QED corrections up to the
third order ($D_{21}({\rm QED3})$) and the fourth-order
contribution of the order $(Z\alpha)^4$ (in units of the $1s$
hyperfine interval) have been known for a while
\cite{zwanziger,sternheim,pmohr,17breit}. The new issue here is a
complete result for the fourth-order QED contributions
($D_{21}({\rm QED4})$) and nuclear corrections ($D_{21}({\rm
nucl})$).

The QED contributions in the fourth order are summarized in
Table~\ref{17epj1}. The new QED contributions are of the order
$\alpha(Z\alpha)^3$, $\alpha^2(Z\alpha)^4$,
$\alpha(Z\alpha)^2{m/M}$ and $(Z\alpha)^3{m/M}$. The difference
with \cite{d21} is a more realistic estimation of the
$\alpha(Z\alpha)^3$  uncertainty~\cite{2sZETF,2seprint}. In the
former paper we accepted result from \cite{yero2001}.

Both results for $\alpha(Z\alpha)^3$ at $Z=1,2$, in
\cite{2sZETF,2seprint} and \cite{yero2001}, are based on
extrapolations of numerical data obtained in \cite{yero2001} for
higher values of $Z$. The procedure leads to a calculation of
central values and its uncertainty. While the calculation of the
central value is approximately the same, the uncertainty is quite
different. Systematic uncertainty of the fitting pro cedure is due
to a possible perturbation of the fit by certain higher-order
terms. As we shown in \cite{2sZETF,2seprint}, one should expect
certain relatively large numerical coefficients for these terms
and that increases the uncertainty.

\begin{table}[hbtp]
\begin{center}
\def\arraystretch{1.4}
\setlength\tabcolsep{5pt}
\begin{tabular}{lcccc}
\hline
Contribution                        &  H~~      & D~~        & $^3$He$^+$~~ & Ref. \\
&[kHz]~~&[kHz]~~&[kHz]~~& \\
\hline
$(Z\alpha)^4 E_F $                 & ~0.005\,6~~~~~ &  ~0.001\,3~~~~~&-0.543~~~~~ &\cite{17breit} \\
$\alpha^2(Z\alpha)^2 E_F $         & ~0.003\,3(16)  &  ~0.000\,8(4) & -0.069(35) & \cite{sgkH2s,d21} \\
$\alpha(Z\alpha)^2{m\over M} E_F $ & -0.003\,1(15)  &  -0.000\,4(2) & ~0.022(11) & \cite{sgkH2s,d21}\\
$\alpha(Z\alpha)^3 E_F $ (SE)      & ~0.008(4)  &  ~0.001\,9(9) & -0.39(14) & \cite{2sZETF,2seprint,yero2001}\\
$\alpha(Z\alpha)^3 E_F $ (VP)      & ~0.003\,0~~~~~ &  ~0.000\,7~~~~& -0.145~~~~~ & \cite{sgkH2s,d21}\\
$(Z\alpha)^3{m\over M} E_F $       & ~0.000\,5(5)~  &  ~0.000\,1~~~~& -0.007(10) & \cite{sgkH2s,d21}\\
\hline
$D_{21}({\rm QED4})$ & ~0.018(5) & ~0.004\,4(10)~ & -1.13(14) & \\
\hline
\end{tabular}
\end{center}
\caption{The fourth order QED contributions to the $D_{21}$ in
hydrogen, deuterium and helium-3 ion (see
\cite{2sZETF,2seprint,d21} for detail).\label{17epj1}}
\end{table}

The higher-order nuclear effects survive the cancellation of the
leading term but they are substantially smaller than the leading
term and can be successfully calculated phenomenologically
\cite{d21}
\begin{eqnarray}\label{d21nuc}
D_{21}&=&\left(\ln2+{3\over16}\right)\cdot(Z\alpha)^2\cdot E^{\rm Nucl}_{\rm hfs}(1s)\nonumber\\
&+&\left(2-{4\over3}\ln2\right)\cdot(Z\alpha)^2\left(\frac{mcR_E}{\hbar}\right)^2E_F
-{1\over 4}\cdot(Z\alpha)^2 \left(\frac{mcR_M}{\hbar}\right)^2
E_F\label{cd21} \;.
\end{eqnarray}
The results for the light atoms are summarized in
Table~\ref{17tabD21}. A value for the nuclear correction in the
helium-3 ion is slightly different here from the result in
\cite{d21} due to different nuclear parameters we used (see
Table~\ref{T:par}).

\begin{table}[hbtp]
\begin{center}
\def\arraystretch{1.4}
\setlength\tabcolsep{5pt}
\begin{tabular}{lccccccc}
\hline
Atom       & $Z$& $I$& $M/m$  & $\mu/\mu_B$ & $E_F$ &$R_E$ &$R_M$ \\
  & & & &  [$10^{-3}$]& [kHz]& [fm] &[fm] \\
\hline
Hydrogen   & 1& 1/2 &1\,836.153 & ~1.521\,032  &~1\,418\,840  &0.88(3)& 0.86(4)\\
Deuterium  & 1& 1   &3\,670.483 & ~0.466\,9755 & ~~~326\,968  &  2.13(1)& 2.07(2)\\
Tritium    & 1& 1/2 &5\,496.922 & ~1.622\,3936 &  1\,515\,038      & 1.76(9)&1.84(18)\\
$^3$He$^+$ & 2& 1/2 &5\,495.885 &  -1.158\,741 & -8\,656\,527 &1.96(3)&1.97(15) \\
\hline
\end{tabular}
\end{center}
\caption{Parameters for the calculations of the HFS interval in
hydrogen, deuterium and helium-3 ion. The proton charge radius is
presented according to Ref.~\protect\cite{rp} while the magnetic
radius is from \protect\cite{sick2}. The radii for other nuclei
are taken from \protect\cite{sick}, while the masses and magnetic
moments are from \cite{newcodata,ame,plamudp,2sZETF}. One should
use the values for the radii with caution. The results in
\protect\cite{sick,sick1,sick2} are obtained from the scattering
data as they were published. However, in a number of papers
published decades ago certain radiative corrections were not taken
into account (cf. our analysis in \protect\cite{rp}) and even a
perfect evaluation with uncorrected data underestimates the
uncertainty.} \label{T:par}
\end{table}

\begin{figure}[hbtp]
\begin{center}
\includegraphics[width=.6\textwidth]{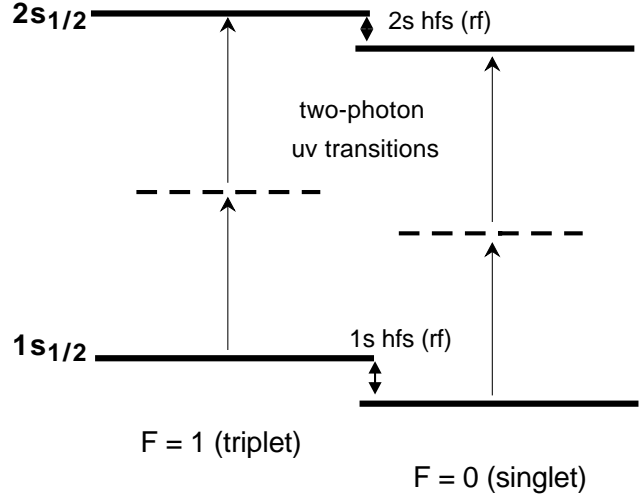}
\end{center}
\caption{Level scheme for the optical measurement of the $2s$
hyperfine structure in the hydrogen atom (not to scale)
\cite{2shydr,2sZETF}. The label {\em rf\/} stands here for
radiofrequency intervals, while {\em uv\/} is for the ultraviolet
transitions.\label{figD21Opt}}
\end{figure}

For all atoms in Table~\ref{17tab1} the hyperfine splitting in the
ground state was measured more accurately than for the $2s$ state.
All experimental results but two were obtained by a direct
measurement of the microwave transitions for the $1s$ and $2s$
hyperfine intervals. However, two of the most recent results
obtained for hydrogen \cite{2shydr,2sZETF} and deuterium
\cite{mpqd2s,2sZETF} have been achieved by means of laser
spectroscopy and the measured transitions lie in the ultraviolet
range \cite{2shydr}. The hydrogen level scheme is depicted in
Fig.~\ref{figD21Opt}. The measured transitions were the
singlet--singlet ($F=0$) and triplet--triplet ($F=1$) two-photon
$1s-2s$ ultraviolet transitions. The eventual uncertainty for the
hyperfine structure is related to 5 parts in $10^{15}$ of the
measured $1s-2s$ interval. A similar experiment for deuterium
achieved an accuracy of the hyperfine interval related to even a
smaller portion of the `big' $1s-2s$ transition, namely 3 parts in
$10^{15}$.

\begin{figure}[hbtp]
\begin{center}
\includegraphics[width=.7\textwidth]{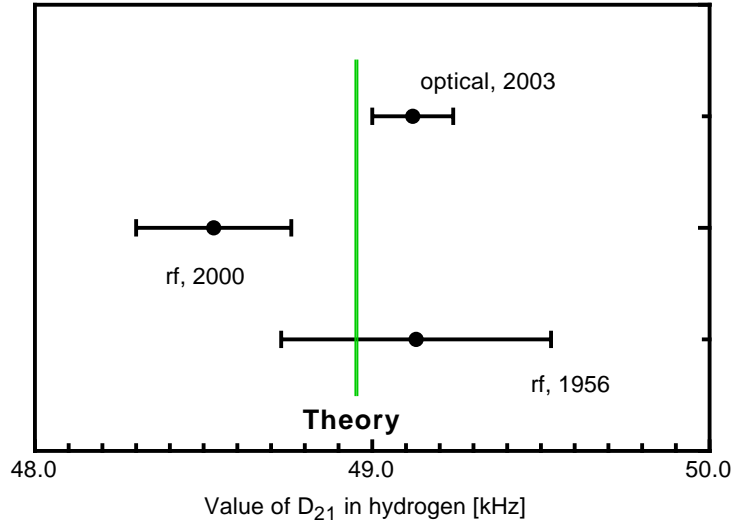}
\end{center}
\caption{Present status of experiment and theory of $D_{21}$ in
the hydrogen atom. The results are labelled with the date of the
measurement of the $2s$ hyperfine structure. See
Table~\ref{17tab1} for references.\label{figD21h}}
\end{figure}

\begin{figure}[hbtp]
\begin{center}
\includegraphics[width=.7\textwidth]{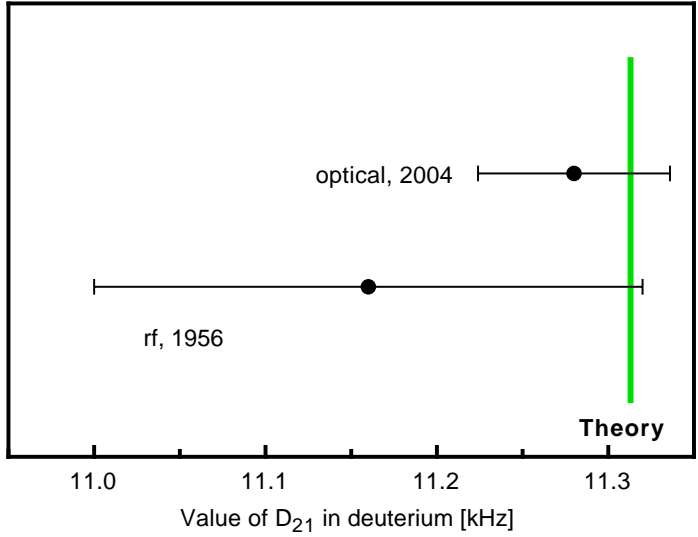}
\end{center}
\caption{Present status of experiment and theory of $D_{21}$ in
the deuterium atom. The results are labelled with the date of the
measurement of the $2s$ hyperfine structure. See
Table~\ref{17tab1} for references.\label{figD21d}}
\end{figure}

\begin{figure}[hbtp]
\begin{center}
\includegraphics[width=.7\textwidth]{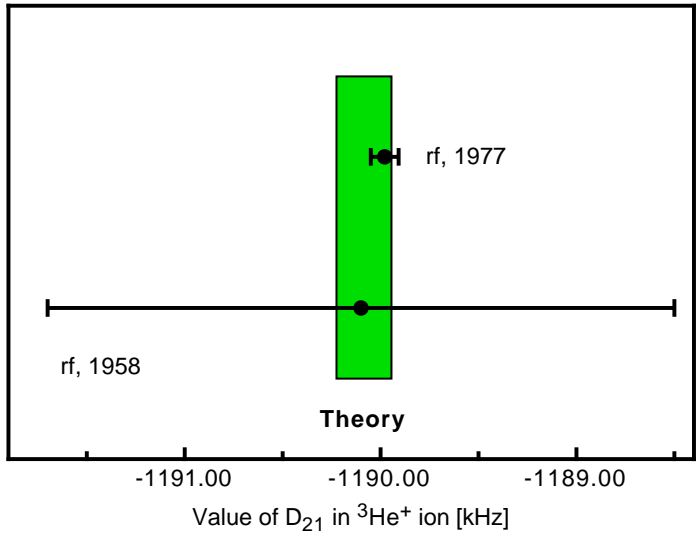}
\end{center}
\caption{Present status of experiment and theory of $D_{21}$ in
the helium ion $^3$He$^+$. The results are labelled with the date
of the measurement of the $2s$ hyperfine structure. See
Table~\ref{17tab1} for references.\label{figD21he}}
\end{figure}

The comparison of theory to experiment for hydrogen, deuterium and
helium-3 ion is summarized in Figs.~\ref{figD21h}, \ref{figD21d}
and \ref{figD21he}. We acknowledge a substantial recent progress
in experimental studies of neutral hydrogen and deuterium atoms,
however, the helium ion data are still of the most interest since
theory and experiment are compatible.

\section{Hyperfine splitting in the ground state of muonium\label{s:muhfs}}

Another possibility to eliminate the nuclear structure effects is
based on studies of nucleon-free atoms. Such an atomic system can
be formed of two leptons. Two atoms of the sort have been produced
and studied for a while with high accuracy, namely, muonium and
positronium. We consider the former below in this section, while
the next section is devoted to the latter.

Muonium is a bound system of a positive muon and an electron. It
can be produced with the help of accelerators. The muon lifetime
is $2.2\cdot 10^{-6}$ sec. However, it is sufficient to perform
accurate spectroscopic measurements. The most accurately measured
transition is the two-photon $1s-2s$ transition with an uncertainty
of 4~ppb \cite{Maas}. The other accurately measured interval is
the $1s$ hyperfine splitting (12~ppb) \cite{MuExp}. A detailed
review of muonium physics can be found in \cite{jungmann}.

\begin{table}[hbtp]
\begin{center}
\begin{tabular}{lcc}
\hline
Term  &  Fractional~~~~ & ~~~~~~~~~~~~~~$\Delta E$~~~~~~~~~~~~~~ \\
&  contribution~~ &  [kHz] \\
\hline$E_F$ & ~~1.000\,000\,000~~~~~~ & 4\,459\,031.88(50)(3)~~ \\ $a_e$ & ~~0.001\,159\,652~~~~~~ & 5\,170.925(1)~\\
QED2 & - 0.000\,195\,815~~~~~~ &- 873.145~~~~\\
QED3 &- 0.000\,005\,923~~~~~~&- 26.411~~\\
QED4 &- 0.000\,000\,123(49)&~~~~~~- 0.548(218)\\
Hadronic~~~&~~0.000\,000\,054(1)~&~~~~~0.241(4)\\
Weak  &- 0.000\,000\,015~~~~~~&- 0.067~~\\
\hline
Total &~~1.000\,957\,830(49)& ~~~4\,463\,302.88(51)(3)(22)\\
\hline
\end{tabular} \end{center}
\caption{Theory of the $1s$ hyperfine splitting in muonium. The
calculations \cite{hamu1} have been performed for $\alpha^{-1}=
137.035\,998\,76(52)$ \cite{kinoalpha} and
$\mu_\mu/\mu_p=3.183\,345\,17(36)$ which was obtained from the
analysis of the data on the Breit-Rabi levels in muonium
\cite{MuExp,MuExp1} (see Sect.~\ref{s:gmu}) and muon precession
\cite{klempt}. The numerical results are presented for the
frequency $E/h$.\label{17tabMu}}
\end{table}

Despite that the fractional accuracy of the measurement of the
hyperfine interval is three times smaller than that of the $1s-2s$
transition, it is of a bigger interest. The uncertainty of the
hyperfine measurement in muonium is indeed much smaller in
absolute units (53~Hz against 9.8~MHz for the $1s-2s$ interval)
and thus more sensitive to higher-order corrections. The Lamb
shift of the $1s$ state in muonium can be extracted from the
optical measurement of the $1s-2s$ transition with fractional
uncertainty of approximately 1\%, which is indeed not enough for
any interesting QED test (it is larger, e.g., than the
proton-finite-size contribution in the case of hydrogen). A real
importance of the $1s-2s$ experiment is due to the determination
of the muon mass which can be used in theoretical calculations for
other QED tests such as the muonium hyperfine structure. We
discuss the muon mass and magnetic moment in Sect.~\ref{s:gmu}.

Let us consider now the hyperfine interval of the ground state in
muonium. The theoretical status is presented in
Table~\ref{17tabMu}. The budget of theoretical uncertainties in
the case of muonium involves quite different sources and they are
in part of experimental origin. The biggest source is in the
calculation of the Fermi energy, accuracy of which is limited by
the knowledge of the muon magnetic moment or the muon mass in
proper units. To determine the muon magnetic moment
$(g_\mu/2)\,(e\hbar/2m_\mu)$ and its mass is essentially the same
because the $g$ factor of a free muon is known well enough
\cite{redin}. A smaller uncertainty in determination of the Fermi
energy (see the number in the second brackets) is due to the fine
structure constant. The theory up to the third order (i.e., $a_e$,
$QED2$ and $QED3$ terms) is well known (see references in
Table~\ref{17MuQED3}). The nonrecoil part up to the third order is
determined for the ground state by (\ref{EnsLead}) and
(\ref{QED1s}), while the recoil effects lead to an additional
contribution
\begin{eqnarray}\label{mu1srec}
\Delta E_{1s}({\rm rec}) &=& (Z\alpha) {m\over M} \frac{E_F}{1 + a_{\mu}}\times
\left\{- \frac{3}{\pi}\ln {\frac{M}{m}}
\right.
+(Z\alpha) \left[ \ln {\frac{1}{(Z\alpha)^2}}
- 8 \ln 2 + \frac{65}{18}\right]
\nonumber \\
&+& \left.
\frac{\alpha}{\pi^2}
\left[ - 2 \ln^2 {\frac{M}{m}} + \frac{13}{12} \ln {\frac{M}{m}}
+ \frac{21}{2} \zeta (3) + \frac{\pi^2}{6} + \frac{35}{9}
\right]\right\}\,.
\end{eqnarray}

\begin{table}[hbtp]
\begin{center}
\begin{tabular}{lclc}
\hline
Correction & References & Correction & Reference(s)\\
\hline
$\alpha(Z\alpha)^2E_F$ & \protect\cite{Pachfs,KNioII,schneider} & $(Z\alpha)^2(m/M)E_F$  & \protect\cite{BYG}\\
$\alpha^2(Z\alpha)E_F$ & \protect\cite{KNio,eides_ry} & $\alpha(Z\alpha)(m/M)E_F$  & \protect\cite{hamuold1,EKSe,EKSmu}\\
\hline
\end{tabular}
\end{center}
\caption{Recent references on the third order QED corrections for
the ground state muonium hyperfine splitting. See
\protect\cite{report} for detail.\label{17MuQED3}}
\end{table}

The fourth-order corrections ($\Delta E({ QED4})$) have not yet
been included either into a non-recoil term $E_{1s}({ QED})$ (see
(\ref{EnsLead}) and (\ref{QED1s})) or into a recoil term $\Delta
E_{1s}({recoil})$. They are the corrections that determine the
uncertainty related to QED theory. These corrections are related
to essentially the same diagrams as the $D_{21}({ QED4})$
contribution in the previous section. The results are summarized
in Table~\ref{T:fourth}. Unknown corrections are estimated
according to \cite{hamu1}. Meantime some of them and in particular
the corrections of relative order $\alpha^3(Z\alpha)$
\cite{egs_in} and $\alpha^2(Z\alpha)(m/M)$ \cite{egs_new} are
under study and a few gauge invariant sets of diagrams have been
already calculated. The muonium QED uncertainty is mainly due to
the calculation of the recoil corrections of order
$\alpha(Z\alpha)^2m/M$ and $(Z\alpha)^3m/M$ \cite{log2}.

\begin{table}[hbtp]
\renewcommand{\arraystretch}{1.4}
\begin{center}
\begin{tabular}{ccc}
\hline
Contribution & Numerical result & Reference \\
\hline
$(Z\alpha)^4$ &~0.03 kHz & \cite{17breit} \\ 
$(Z\alpha)^3\frac{m}{M}$ & $-$0.29(13) kHz & \cite{log1,egas,KNio,log2,logps1,logps2}\\ 
$(Z\alpha)^2\left(\frac{m}{M}\right)^2$ & $-$0.02(1) kHz & \cite{lepage,bodwin,pac2m}\\ 
$(Z\alpha)\left(\frac{m}{M}\right)^3$ & $-$0.02 kHz & \protect{\cite{arno}} \\ 
$\alpha(Z\alpha)^3$ & $-$0.52(3) kHz & \cite{onehfs,yero2001,newnio}, \cite{log2}, \cite{vphfs} \\ 
$\alpha(Z\alpha)^2\frac{m}{M}$ & ~0.39(17) kHz & \cite{log1,logps1,logps2}\\ 
$\alpha(Z\alpha)\left(\frac{m}{M}\right)^2$ & $-$0.04 kHz & \cite{eides2}  \\ 
$\alpha^2(Z\alpha)^2$ & $-$0.04(2) kHz & \cite{log1}  \\ 
$\alpha^2(Z\alpha)\frac{m}{M}$ & $-$0.04(3) kHz &\cite{eslog,ekslog} \\ 
$\alpha^3(Z\alpha)$ & $\pm 0.01$ kHz & \cite{hamu1} \\ 
$\tau$ lepton & $ 0.002$ kHz & \cite{hamu2,hamu3} \\ 
\hline
$\Delta E({\rm QED4})$ & $-0.55(22)$ kHz & \\ 
\hline
\end{tabular}
\end{center}
\caption{The fourth order corrections to the muonium hyperfine
structure. The uncertainty is an rms sum of the partial
uncertainties. The essential part of the contributions (see
\cite{hamu1} for detail) have been found within the leading
logarithmic approximation and their uncertainty is estimated for
each contribution as a half-value of the leading term
\cite{log2,hamu1}. The $\tau$ lepton contribution is also included
into the table and has an order $\alpha(Z\alpha)mM/m_\tau^2$.
\label{T:fourth}}
\end{table}

The muonium theory is not completely free of hadronic
contributions, which appear because of intermediate hadronic
states. That produces one more source of theoretical
uncertainties. The effects of strong interactions in muonium are
discussed in detail in Refs.~\cite{hamu1,hamu2,hamu3}. They are
relatively small, but their understanding is very important
because of the intensive muon sources expected in future
\cite{sources} which can allow a dramatic increase of the accuracy
of muonium experiments.

A comparison of theory and experiment for the muonium HFS interval is
not an isolated problem, since it involves determination of the
muon mass (magnetic moment) in proper units. We discuss problems
of the determination of the muon mass and magnetic moment in
Sect.~\ref{s:gmu} and a comparison of theory versus experiment for
muonium hyperfine interval is presented in the Sect.~\ref{S:log}
along with the other test of bound state QED for the hyperfine
structure (see Table~\ref{17tabHFS}).

\section{Spectrum and annihilation of positronium and recoil
effects\label{s:ps}}

Another pure leptonic atom is positronium. It can be produced at
accelerators or using various radioactive positron sources. The
lifetime of positronium depends on its state, on its orbital and
spin ($S$) quantum numbers. The lifetime for the $1S$ state of
parapositronium (this state with $S=0$ annihilates mainly into two
photons) is $1.25\cdot 10^{-10}$ sec, while orthopositronium (the
$S=1$ state) in the $1S$ state has a lifetime of $1.4\cdot
10^{-7}$ sec because of the three-photon decays. A list of
accurately measured positronium quantities contains the $1S$
hyperfine splitting, the $1S-2S$ interval, the $2S-2P$ fine
structure intervals, the lifetime of the $1S$ state of para- and
orthopositronium and several branching ratios of their decays. We
note that since the nuclear spin effects are not suppressed,
positronium has a structure of energy levels (in respect to their
spin and angular momentum) rather similar to a two-electron system
(such as the neutral helium atom) and thence we use capital letter
for its orbital momentum.

Theoretical predictions for most of the values of interest and
references to their comparison to experimental data are summarized
in Table~\ref{TPsTh}.

\begin{table}[hbtp]
\begin{center}
\begin{tabular}{lcccc}
\hline
Quantity & Leading term & Prediction & Figure \\
\hline
$\Delta\nu(1S-2S)$  & $\frac{3}{8}\, \alpha^2 mc^2$ & 1\,233\,607\,222.2(6) MHz& \protect\ref{F1s} \\
$\Delta\nu_{HFS}(1S)$ & $\frac{7}{12}\, \alpha^4 mc^2$ & 203\,391.7(6) MHz& \protect\ref{figPo}\\
\hline
$\Delta\nu(2^3S_1-2^3P_0)$  & $\frac{5}{96}\, \alpha^4 mc^2$ & 18\,498.25(9) MHz & \protect\ref{figPo2p} \\
$\Delta\nu(2^3S_1-2^3P_1)$  & $\frac{7}{192}\, \alpha^4 mc^2$ & 13\,012.41(9) MHz & \protect\ref{figPo2p} \\
$\Delta\nu(2^3S_1-2^3P_2)$  & $\frac{23}{720}\, \alpha^4 mc^2$ & 8\, 625.70(9) MHz & \protect\ref{figPo2p} \\
$\Delta\nu(2^3S_1-2^1P_1)$  & $\frac{1}{24}\, \alpha^4 mc^2$ & 11\,185.37(9) MHz & \protect\ref{figPo2p} \\
\hline
$\Gamma ({\rm p\!-\!Ps})$ & $\frac{1}{2} \alpha^5 mc^2$ & 7989.62(4)  $\mu$s$^{-1}$ &\protect\ref{Fdpp}\\
$\Gamma ({\rm o\!-\!Ps})$ & $\frac{2(\pi^2-9)}{9\pi} \alpha^6 mc^2$ & 7.039\,96(2) $\mu$s$^{-1}$ & \protect\ref{Fdop}\\
${\rm Br}_{4\gamma/2\gamma}({\rm pPs})$ &$0.0278\,\alpha^2$&
$1.439(2)\cdot10^{-6}$ & \protect\ref{Fbr4g}\\
\hline
\end{tabular}
\end{center}
\caption{\label{TPsTh}Theoretical predictions for positronium.
Comparison to experimental data is presented in figures quoted in
the table. The details and references for experiment can be found
in \protect\cite{conti,ley} while for theory in
\protect\cite{pos,pos1} and below in Table~\protect\ref{TPsRef}.
The leading term for each value is presented in energy units.}
\end{table}

\begin{figure}[hbtp]
\begin{center}
\includegraphics[width=.5\textwidth]{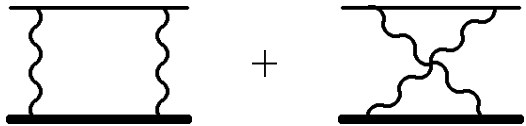}
\end{center}
\caption{Two-photon basic diagrams for the calculation of recoil
contributions.\label{FigSkel}}
\end{figure}

In the case of the atoms with a `heavy' nucleus there is a
significant difference between contributions to the Lamb shift and
the hyperfine splitting. The hyperfine effects used to affect the
Lamb shift and related intervals (such as the gross and fine
structure) only slightly because the hyperfine splitting is of the order
$(Z\alpha)^4m^2c^2/M$, while the Lamb shift is of the order
$\alpha(Z\alpha)^4 mc^2 \ln(1/Z\alpha)$. Since the effects due to
the hyperfine shift are small, the higher order corrections to the
HFS interval are not important for a study of the Lamb shift and
related transitions. Entering into detail, we note that a crucial
problem for theory of the hyperfine structure is a calculation of
the recoil corrections, while for the Lamb shift they are not so
important. The difference originates from the fact that if the
momentum of integration is significantly below $Mc$, the nuclear
recoil corrections are rather of a kinematic origin similar (but
much more complicated) to the reduced mass effects in the
relativistic case (for the electron), since the nucleus can still
be treated as a non-relativistic particle. If the momentum of
integration is of the order of $Mc$, the diagram for the basic block
(see Fig.~\ref{FigSkel}) has order $(Z\alpha)^5(m/M)^2 mc^2$ and
higher-order corrections to this block deliver some extra factors
of $\alpha$ and $Z\alpha$. In other words, a recoil correction
with a relativistic nucleus is of the second order of the
electron-to-nucleus mass ratio. That fact used to be `hidden' in
the standard notation. When one discusses the Lamb shift, the
contributions used to be presented in the absolute units with all
$(m/M)$ factors explicitly presented. Meantime, the relative units
are commonly used for the corrections to the hyperfine structure
and since the Fermi energy
\begin{equation}
E_F \sim (Z\alpha)^4 \frac{m}{M} mc^2
\end{equation}
includes a factor of $m/M$, a linear correction of relative order
$m/M$ is actually the second order contribution in the absolute
units.

Summarizing, we note that:
\begin{itemize}
\item studying the gross and fine structure and the Lamb shift in
a conventional atom such as hydrogen, one needs to take into
account the hyperfine effects with a relatively low accuracy;
\item calculating the gross and fine structure and the Lamb shift,
one does not need to learn much detail on theory of the
second-order recoil corrections and to approach higher-order
recoil effects;
\item accuracy of theoretical predictions suffers from an
uncertainty in the determination of the fundamental constants such
as Rydberg constant $R_\infty$, the fine structure constant
$\alpha$ and the nuclear mass and magnetic moment.
\end{itemize}

Positronium is an exceptional atom since $m/M=1$ and the situation
strongly differs from the conventional atomic systems:
\begin{itemize}
\item studying the gross and fine structure of positronium, one
needs to take into account the hyperfine effects with a high
accuracy;
\item the calculation of truly recoil corrections with
relativistic nucleus is equally important for the gross and fine
structure, as well as for the hyperfine structure and the lifetime
calculations. Actually, the uncertainty of the calculations of the
recoil effects determines the entire theoretical uncertainty for
the most wanted spectroscopic characteristics in positronium.
\end{itemize}

The equality of $m/M$ to unity leads to another important feature
of positronium tests for the bound state QED. Since the
corrections of interest are enhanced ($m/M$ is not a suppressing
factor any longer), the fractional accuracy for successful
high-precision tests is now relatively low. As a result, in
contrast to hydrogen, the interpretation of the measurements of
the $1S-2S$ interval does not crucially involve knowledge of the
Rydberg constant with high accuracy. A study of the hyperfine
interval does not require a value of the fine structure constant
$\alpha$ with high accuracy as it is in muonium. Since $m/M=1$, it
is unnecessary to determine a mass ratio $m/M$ and a ratio of
magnetic moments of the electron and nucleus, i.e.,
positronium\footnote{One may wonder what should happen if the CPT
symmetry is broken and $m_e/m_{\overline{e}}$ and
$\mu_e/\mu_{\overline{e}}$ are not exactly equal to unity. In
principle, certain features of positronium might be sensitive to
such a violation, however, special experiments with selective
sensitivity to this effect would be needed. It is also clear, that
such a CPT violation should show itself not just as a small
mismatch in masses and magnetic moments, but as incorrectness of
the basic equations (e.g., the Dirac equation).}, in an additional
experiment. In other words, positronium offers several high
precision tests of bound state QED without any need to determine
values of any fundamental or phenomenological constant with high
accuracy.

\begin{figure}[hbtp]
\begin{center}
\includegraphics[width=.6\textwidth]{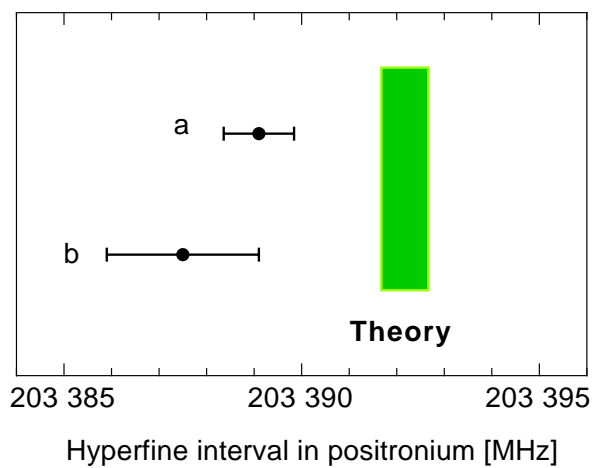}
\end{center}
\caption{The hyperfine splitting in the ground state of positronium. The Yale experiment ($a$) was
performed in 1984 \cite{yale} and the Brandeis one ($b$) in 1975 \cite{brandeis}.\label{figPo}}
\end{figure}

The HFS interval has been determined experimentally with the
highest absolute accuracy among other positronium transitions.
Present experimental data for the positronium hyperfine interval
together with the theoretical result are depicted in
Fig.~\ref{figPo}. The theoretical status for the positronium
hyperfine interval in the $1S$ state is presented in detail in
Table~\ref{17tabPs}. The dominant source of the uncertainty is the
third-order corrections for positronium ($\Delta E({QED3})$),
which are known only in the logarithmic approximation. The
diagrams under question are related to radiative-recoil and recoil
corrections, and they are the same as those responsible for the
uncertainty of muonium hyperfine interval ($\alpha(Z\alpha)^2m/M$
and $(Z\alpha)^3m/M$) \cite{log1,pos,new_rep} (see the previous
section for detail) . The fourth-order recoil terms of theory of
heavy atoms become the third order corrections in positronium.

\begin{table}[hbtp]
\begin{center}
\begin{tabular}{lrrr}
\hline
Term~~~~~~&Fractional~~~~~~&$\Delta E$~~~~~~~~& References \\
&contribution~~~~&[MHz]~~~~~~ \\
\hline
$E_F$~~~~~~~$\alpha^4mc^2$  & 1.000\,000\,0~~~~~~& 204\,386.6~~~~~&\\
QED1~~~$\alpha^5mc^2$   & - 0.004\,919\,6~~~~~~&-1\,005.5~~~~~\\
QED2~~~$\alpha^6mc^2$        & 0.000\,057\,7~~~~~~&11.8~~~~ &\protect\cite{posqed2} ~~~~~\\
QED3~~~$\alpha^7mc^2$        &- 0.000\,006\,1(22)&- 1.2(6) & \protect\cite{log1,logps1,logps2,logps3}\\
 \hline
Total &0.995\,132\,1(22) & ~~~~203\,391.7(6)\\
\hline
\end{tabular}
\end{center}
\caption{Theory of the $1S$ hyperfine interval in positronium. The
numerical results are presented for the frequency $E/h$. The
calculation of the second order terms was completed in Ref.
\cite{posqed2}, the leading logarithmic contributions were found
in \cite{log1}, while the next-to-leading logarithmic terms were
achieved in \cite{logps1,logps2,logps3}. The uncertainty is
presented following \cite{pos,pos1}.\label{17tabPs}}
\end{table}

\begin{figure}[hbtp]
\begin{center}
\includegraphics[width=.7\textwidth]{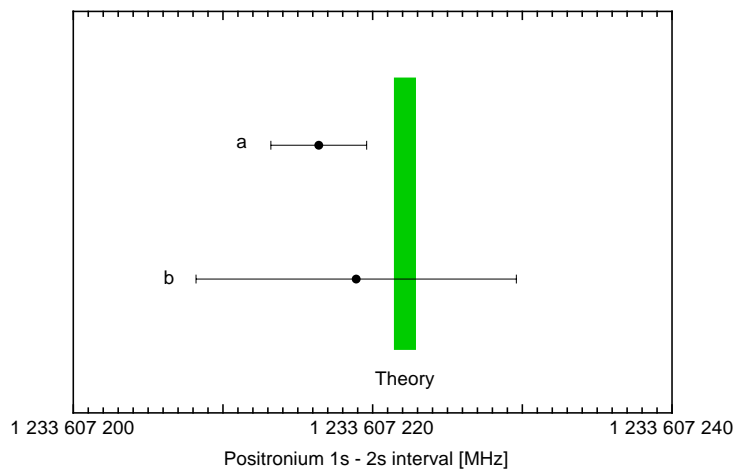}
\end{center}
\caption{\label{F1s} Determination of the $1^3S_1-2^3S_1$ interval
in positronium. The references here are: $a$ is for
\cite{pos1s2sa} and $b$ is for \cite{pos1s2sb}.}
\end{figure}

Studies of the spectrum and decay rates of positronium provide us
with a number of strong tests of bound state QED, some of which
are among the most accurate. Some theoretical predictions from
Table~\ref{TPsTh} can be compared with accurate experimental data,
a review of which can be found in Ref.~\cite{conti}. The most
accurately measured spectroscopic data are related to the ground
state HFS (see Fig.~\ref{figPo}) and to the $1S-2S$ interval (see
Fig.~\ref{F1s} and Table~\ref{t:1s2s}). There are some minor
discrepancies between experimental and theoretical data.

\begin{table}[hbtp]
\begin{center}
\begin{tabular}{lr}
\hline
~~Term~~~~~~&$\Delta E$~~~~~~~~~~~~ \\
&[MHz]~~~~~~~~~~ \\
\hline
~~$\alpha^2mc^2$ &  1\,233\,690\,735.1~~~~~~~~\\
~~$\alpha^4mc^2$ &  -82\,005.6~~~~~~~~\\
~~$\alpha^5mc^2$ &  -1\,501.4~~~~~~~~\\
~~$\alpha^6mc^2$ &  -7.1, \protect\cite{pk1}~~~~~\\
~~$\alpha^7mc^2$ &  1.2(6), \protect\cite{pk2,my} \\
\hline
~~Total & 1\,233\,607\,222.2(6)~~~~\\
\hline
\end{tabular}
\end{center}
\caption{Theory of the $1^3S_1-2^3S_1$ interval in positronium.
The corrections $\Delta E$ are presented in the energy units,
while their numerical values are given in the frequency units for
$E/h$. \label{t:1s2s}}
\end{table}

\begin{figure}[hbtp]
\begin{center}
\includegraphics[width=.7\textwidth]{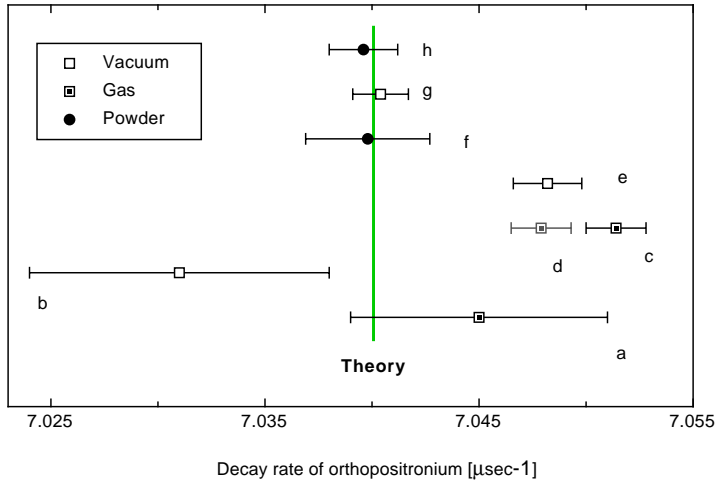}
\end{center}
\caption{\label{Fdop}Measurements of the annihilation decay rate
of orthopositronium: $a$ -- \cite{gopsa}, $b$ -- \cite{gopsb}, $c$
-- \cite{gopsc}, $d$ -- \cite{conti,AAin}, $e$ -- \cite{gopse},
$f$ -- \cite{gopsf}, $g$ -- \cite{recent_pos}, $h$ --
\cite{eopsj2}.}
\end{figure}

The experimental situation with the orthopositronium decay (see
Fig.~\ref{Fdop}) had been not acceptable for a while but it has
been recently improved \cite{recent_pos}. The problem was a
significant inconsistency of various experiments and long standing
strong discrepancy between theory and the most accurate
experimental data. The data presented in Fig.~\ref{Fdop} include
most recent vacuum results from Tokyo \cite{eopsj2} and Ann Arbor
\cite{recent_pos} which are in a good agreement with theory (see
Table~\ref{t:gam}). The original gas result from Ann Arbor (data
point $c$ in Fig.~\ref{Fdop}) is corrected (data point $d$)
according to the preliminary analysis in Ref.~\cite{conti} but
that is not a final result. Further examination is in progress and
it seems that the final uncertainty will be bigger \cite{AAin}.

\begin{table}[hbtp]
\begin{center}
\begin{tabular}{lll} 
\hline
Contribution & ~~~~Decay rate & ~~~~Decay rate \\
&of orthopositronium &of parapositronium \\
&~~~~~~[$\mu$sec$^{-1}$] & ~~~~~~[$\mu$sec$^{-1}$]  \\
\hline
$\Gamma^{(0)}$                 &~7.211\,17              &8\,032.50\\
QED1 
&-0.172\,30              &~ \,-47.25 \\
QED2 
 &~0.001\,11(1), \cite{afs}  &~~\,~4.43(1), \cite{cmy1}\\
QED3
&-0.000\,01(2), \cite{log1,declog} & ~~\,-0.08(4), \cite{log1,declog} \\
\hline
~~Total &  7.039\,96(2) & 7989.62(4) \\
\hline
\end{tabular}
\end{center}
\caption{Theory of the annihilation decay rate of ortho- and
parapositronium (the $1S$ state). The leading contributions are
defined above in Table~\protect\ref{TPsTh}. The decay rate of
ortho/parapositronium into five/four photons is included into
corresponding QED2 terms. \label{t:gam}}
\end{table}

\begin{figure}[hbtp]
\begin{center}
\includegraphics[width=.45\textwidth]{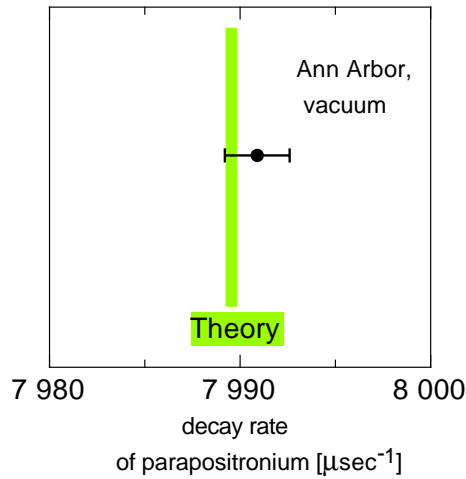}
\end{center}
\caption{\label{Fdpp} Determination of the parapositronium decay
rate. The experiment was performed in Ann Arbor in 1994
\cite{gpsaa}.}
\end{figure}

\begin{figure}[hbtp]
\begin{center}
\includegraphics[width=.4\textwidth]{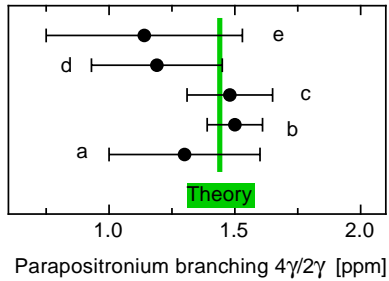}
\end{center}
\caption{\label{Fbr4g}Annihilation of parapositronium: branching
Br($4\gamma/2\gamma$) measured around the world ($a$ --
\cite{4ga}, $b$ -- \cite{4gb}, $c$ -- \cite{4gc}, $d$ --
\cite{4gd}, $e$ -- \cite{4ge}).}
\end{figure}

For the parapositronium decay  theory and experiment are in
perfect agreement (see Fig.~\ref{Fdpp} and Fig.~\ref{Fbr4g}). Most
of positronium decay experiments were reviewed in
Refs.~\cite{conti,ley}.

These two papers also review experiments on the fine structure in
positronium performed at $2^3S_1-2P$ intervals which were less
accurate than experiments at $1S$ HFS and $1S-2S$ intervals. Data
on the fine structure are not as accurate as the results related
to the $1S$ hyperfine splitting and the $1S-2S$ interval but the
progress is possible \cite{conti}.

\begin{figure}[hbtp]
\begin{center}
\includegraphics[width=.6\textwidth]{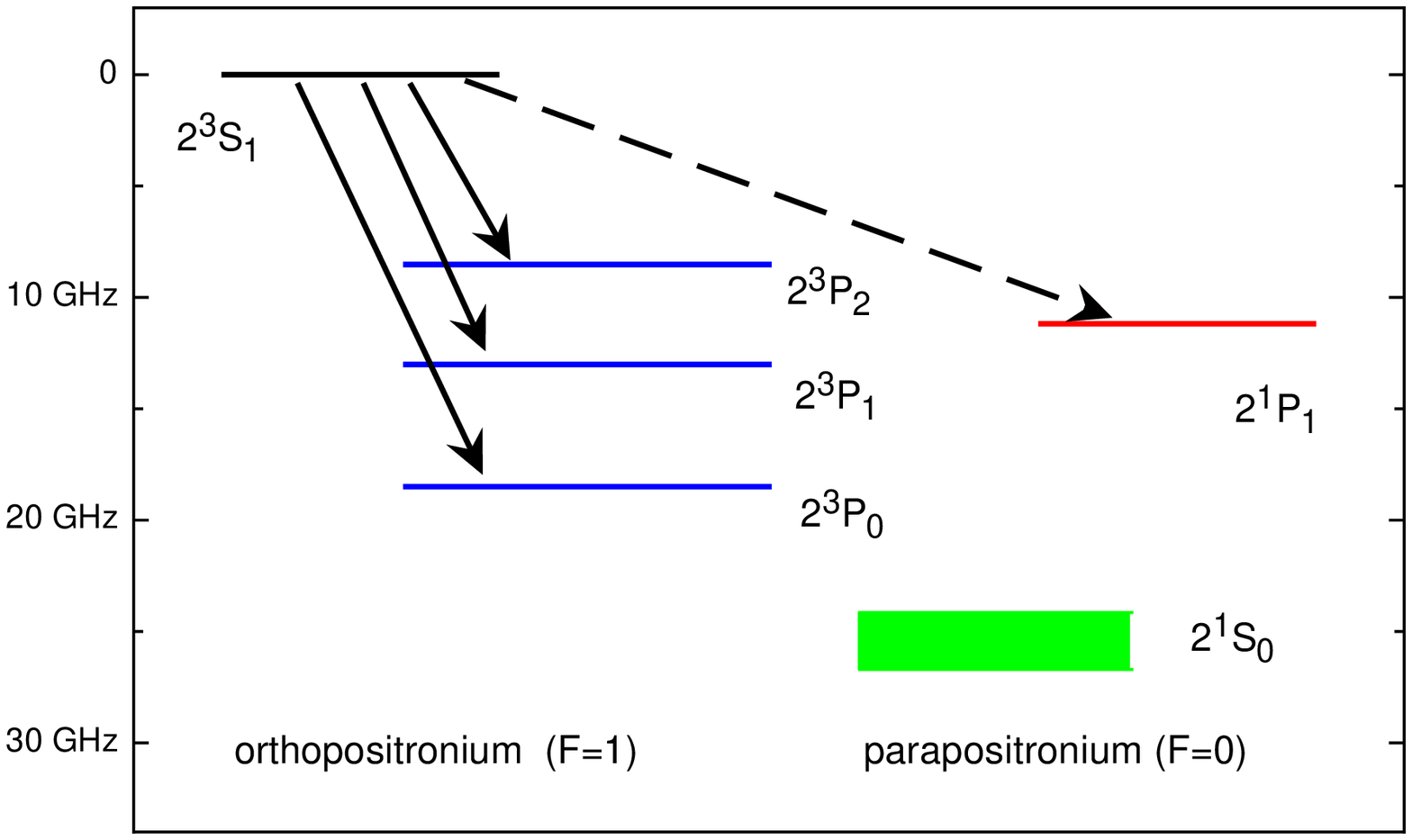}\\
\includegraphics[width=1.0\textwidth]{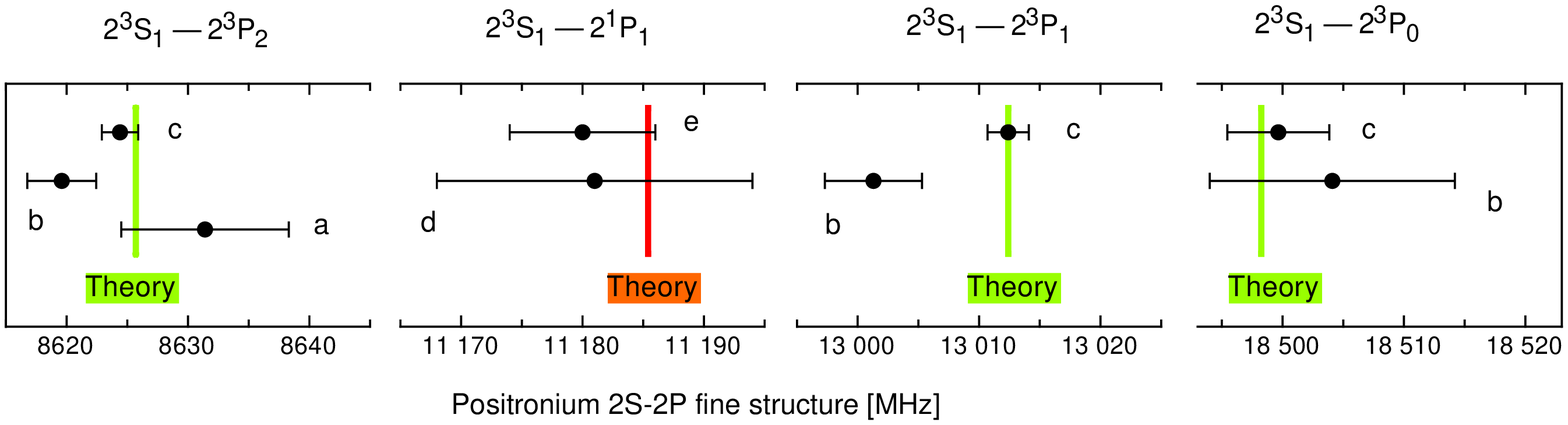}
\end{center}
\caption{The fine structure in positronium at $n=2$: the energy
levels and studied transitions. The $2^3S_1-2^1P_1$ transition is
forbidden for the free positronium atom and it was measured with
an external magnetic field applied. The line width of the $2^3S_1$
state is of 1~MHz and that of all $2P$ states is of 50~MHz and
thus it cannot be seen in the scale of the top picture, while the
width of parapositronium singlet ($2^1S_0$) being of 1.3 GHz is
clearly recognized. The references to the measurements are: $a$ --
\cite{psfsa}, $b$ -- \cite{psfsb}, $c$ -- \cite{psfsc}, $d$ --
\cite{psfsd} and $e$ -- \cite{psfse}.\label{figPo2p}}
\end{figure}

To conclude the section, let us discuss briefly the theory of the
positronium decay rates and energy levels and in particular the
uncertainty of their calculations. We collected all theoretical
predictions in Table~\ref{TPsTh}. The results were published and
presented in different compilations. What we would like to
emphasize here is our {\em estimation of uncertainty\/}.

The uncertainty of any perturbative theoretical calculation is
determined by a plausible estimation of unknown higher-order
corrections which are expected to have large coefficients. There
is a number of corrections enhanced by a big double logarithmic
factor $\ln^2\alpha\simeq 24$ \cite{log1} and the higher-order
terms should be studied to understand better the accuracy of
theory. For the spectroscopy, the crucial order is $\alpha^7mc^2$,
while in the case of the decay the theoretical problems are
related to the corrections of the relative order $\alpha^3$.

\begin{table}[hbtp]
\begin{center}
\begin{tabular}{cccc}
\hline
Value & Ref. to & Ref. to & Ref. to \\
& $\alpha^6mc^2$ & $\alpha^7mc^2\ln^2\alpha $& $\alpha^7mc^2\ln\alpha$\\
& or $\alpha^2\cdot\Gamma^0$ &$\alpha^2\ln^2\alpha\cdot\Gamma^0$ & $\alpha^2\ln\alpha\cdot\Gamma^0$\\
\hline
$1S-2S$ & \cite{pk1} &\cite{pk2,my}  & unknown \\
fine structure &\cite{pk1} &\cite{pk2,my}  & unknown \\
$1S$ HFS &\cite{posqed2} &\cite{log1}  &\cite{logps1,logps2,logps3}
\\
\hline
$\Gamma({\rm p\!-\!Ps})$ &\cite{cmy1} &\cite{log1}  &\cite{declog}
\\
$\Gamma({\rm o\!-\!Ps})$ &\cite{afs} &\cite{log1}  &\cite{declog}
\\
\hline
\end{tabular}
\end{center}
\caption{\label{TPsRef}References to recent progress in the
positronium theory. The contributions to the spectrum are
classified by the absolute energy units (such as $\alpha^6mc^2$),
while those to the decay are presented in units of the leading
contribution $\Gamma^0$ (see Table~\protect\ref{TPsTh}).}
\end{table}

For most of them not only the leading logarithmic corrections
(e.g., in the case of the spectrum that is
$\alpha^7mc^2\ln^2\alpha$) are known, but also the next-to-leading
term ($\alpha^7mc^2\ln\alpha$). We should emphasis, however, that
such a knowledge cannot reduce the uncertainty. The leading term
originates (the most probably, but not always) from a single
source and its magnitude is a natural characteristic of the
correction, while the next-to-leading term used to be a result of
a summation of or a cancellation between different contributions
and can be sometimes small. But that smallness is misleading and
the constant following the single logarithm is not necessary
small. Our estimation of the uncertainty \cite{pos,pos1} is based
on a value of the double logarithmic term in any case.

\section{The $g$ factor of an electron and a nucleus in a light hydrogen-like atom \label{s:glight}}

\subsection{General consideration}

Not only a spectrum of simple atoms can be studied with a high
accuracy. Among other quantities accessible to high precision
measurements are the magnetic moments of bound subatomic
particles. The only way to determine a value of a magnetic moment
precisely (i.e., with uncertainty at the level of 0.01~ppm or
below) is to measure a certain frequency proportional to the
magnetic moment and applied magnetic field. Since the magnetic
field can be neither measured nor calculated with uncertainty
essentially below 0.1 ppm, it is possible to measure accurately
only certain ratios of the magnetic moments. Another option is to
compare a spin precession frequency related to the magnetic moment
with the ion cyclotron frequency. Two-body atomic systems provide
us with an opportunity to study the magnetic moments of their
bound composites and in certain experiments to compare their
magnetic moments to each other.

The interaction of an atom with a weak homogenous magnetic field
${\bf B}$ can be expressed in terms of an effective Hamiltonian.
For an $s$ state in the two-body atom such a Hamiltonian for spin
variables has the form (see, e.g., \cite{00BS})
\begin{equation}
\label{Heff} H_{\rm magn} = g^\prime_e \, {e\hbar\over 2 m}\,
\bigl( {\bf s} \cdot {\bf B} \bigr) -g^\prime_N \, {Ze\hbar\over 2
M} \,\bigl( {\bf I} \cdot {\bf B} \bigr) + J_{\rm HFS}\,\bigl(
{\bf s} \cdot {\bf I}\bigr)\;,
\end{equation}
where ${\bf s}$ stands for the spin of an electron, ${\bf I}$ is for the
nuclear spin, $g^\prime_{e(N)}$ is for the $g$ factor of a bound
electron (nucleus) in the two-body atom and $J_{\rm HFS}$ is the
hyperfine constant\footnote{The notation for the $g$ factors is
different in particle and nuclear physics. For ions one needs to
apply a different notation in which the magnetic Hamiltonian reads
\[
H_{\rm magn} =g^\prime_e \,  {e\hbar\over 2 m} \,\bigl( {\bf s}
\cdot {\bf B} \bigr) -g^\prime_N \, {e\hbar\over 2 m_p}\, \bigl(
{\bf I} \cdot {\bf B} \bigr) + J_{\rm HFS}\,\bigl( {\bf s} \cdot
{\bf I}\bigr)\;,
\]
where $g^\prime_N$ differs indeed from that in Eq.~(\ref{Heff}).
To avoid confusion, in this paper $g$ is used only for the case of
particles (electrons, muons, protons etc.), while if we need the
$g$ factor for a nucleus, we introduce additional notation (see
e.g., (cf. \ref{defeta})).}. We remind here, that $m$ and $M$ are
the masses of the electron and the nucleus, respectively.

We present the $g$ factor of a bound electron in the form \cite{icap,sgkH2,pla}
\begin{equation}\label{gbound}
g^\prime = 2\cdot \Bigl(1+a+b\Bigr)\;,
\end{equation}
where $a$ is the anomalous magnetic moment of a free electron,
while $b$ stands for the binding correction. As it has been known for
a while \cite{00BS,19kin1}, even in absence of QED effects the $g$
factor of a bound electron is not equal to the free Dirac value
(namely, $g_{\rm Dir}^{(0)}=2$). The leading radiative and recoil
effects were later examined in \cite{19kin1a,19kin2}.

Three basic situations with the two-body atomic systems at presence of
a homogeneous magnetic field are of practical interest:
\begin{itemize}
\item a `conventional' low-$Z$ atom with a non-zero nuclear spin
(hydrogen, deuterium, tritium, helium-3 ion);
\item muonium, a
pure leptonic unstable atom with a relatively light nucleus
($m_e/m_\mu\simeq 1/207 \gg m_e/m_p \simeq 1/1836$);
\item a
medium-$Z$ hydrogen-like ion with a spinless nucleus (the ions of
carbon-12 and oxygen-16) with a possible expansion of experimental
activity down to helium-4 and beryllium-10 ions and up to
calcium-40 ion.
\end{itemize}

The bound $g$ factors in light atoms are now known up to the
fourth-order corrections \cite{zee} including the terms of the
order $\alpha^4$, $\alpha^3m/M$ and $\alpha^2(m/M)^2$ and thus the
relative uncertainty is substantially better than a part in
$10^{8}$. All these corrections are of kinematic origin and there
is no need for a higher accuracy and thus for any higher order
corrections for a moment.

For the medium-$Z$ ions some corrections of higher orders are
important for the comparison with experimental data and for
applications to the determination of the fundamental constants.
These corrections involve essential bound state QED effects. The
`kinematic origin' reads that they are a result of a `simple'
quantum mechanical problem of two point-like particles bound by a
Coulomb interaction with their actual values of the masses and
magnetic moments. `Free QED' corrections are a part of kinematic
effects since we consider the actual values of the magnetic
moments which include the anomalous magnetic moment. `Bound state
QED' corrections are related to somewhat beyond this approximation
--- to non-Coulomb additions to the interactions (as, e.g., the
Uehling potential), structure effects (as, e.g., the electron
self-energy). Such higher order corrections were studied and in
particular the linear correction in $\alpha$ has been known
exactly in all orders of $Z\alpha$ as well as some higher order
recoil contributions.

\subsection{Hydrogen and its isotopes: the isotopic effects
for the $g$ factor of a bound electron}

\begin{figure}
\begin{center}
\includegraphics[width=0.6\textwidth]{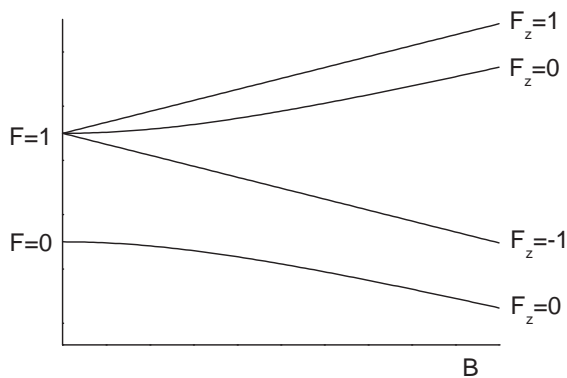}
\end{center}
\caption{Sublevels $E_{\rm magn}(F,F_z)$ of the hyperfine
structure in the $ns$ state of hydrogen or another hydrogen-like
atom with nuclear spin 1/2 at a homogenous magnetic field $B$. The
related equation was first derived by Breit and Rabi (see
\cite{00BS}).} \label{mfFig}
\end{figure}

Few experiments were performed with hydrogen and its isotopes, as
well as with the helium-3 ion. Dependence of energy sublevels on
the magnetic field is presented in Fig.~\ref{mfFig} for the $ns$ state
in a two-body atom with nuclear spin $1/2$ (hydrogen, tritium,
helium-3 ion, muonium).

The hyperfine sublevels were successfully studied by a few
different ways.
\begin{itemize}
\item For example, one is a study of a few splittings related to
the same atom. \item Other experiments on the hydrogen-like atoms
dealt with the $g$ factors of an electron bound in different atoms
studying a kind of isotopic dependence in the electron $g$ factor.
\item One more approach was realized for muonium: the
field-induced energy intervals were measured at the magnetic field
calibrated against the proton spin precession (see
Sect.~\ref{s:gmu}).
\end{itemize}

The isotopic experiments were performed on hydrogen, deuterium and
tritium measured at the same applied magnetic field. The results
are summarized in Table~\ref{19Tab2}. The helium-ion result there
involved a chain of several intermediate comparisons including
experiments with neutral rubidium. All corrections to the electron
$g$ factor up to the fourth order are known, with an exception.
There are two contradicting results for the $(Z\alpha)^2(m/M)^2$
term which can be potentially nuclear-spin-dependent. One group
\cite{gmm21} performed a semi-classical evaluation and reported on
a nuclear-spin-independent result, while the others \cite{gmm22}
claimed a difference between the results for the nuclear spin
$I=1/2$ and $I=1$. For the nuclear spin 1/2 both reproduced the
well known result \cite{19kin2}. Hopefully, for deuterium their
discrepancy is somewhat smaller than the experimental uncertainty,
while the other two-body atoms of interest involve the nuclear
spin $I=1/2$ for which the result in this order is well
established.

\begin{table}
\begin{center}
\def\arraystretch{1.4}
\setlength\tabcolsep{5pt}
\begin{tabular}{c@{\hskip 2em}lc@{\hskip 2em}l}
\hline
Value & Experiment & Reference & Theory\\
\hline
$ {g({\rm H})}/{g_0(e)} $
     & $1- 17.709(13)\cdot10^{-6} $ & \protect{\cite{19h,19d1}}
         & $1- 17.694\cdot10^{-6} $ \\
$ {g({\rm D})}/{g({\rm H})} $
     & $ 1-7.22(3)\cdot10^{-6} $ & \protect{\cite{19d3}}
         & $ 1-7.24\cdot10^{-6} $\\
$ {g({\rm T})}/{g({\rm H})} $
     & $ 1-10.7(15)\cdot10^{-6} $ & \protect{\cite{19t}}
         &$ 1-9.7\cdot10^{-6} $ \\
$ {g({}^4{\rm He}^+)}/{g_0(e)} $
     & $1- 70.88(30)\cdot10^{-6} $ & \protect{\cite{19h,19d1,19he,19rb87,19rb85,198785}}
         & $1- 70.91\cdot10^{-6} $ \\
\hline
\end{tabular}
\end{center}
\caption{The bound electron $g$ factor in light hydrogen-like atoms: a
comparison of theory and experiment. Here, $g_0(e)$ stands for the
magnetic moment of a free electron and it contains the anomalous
magnetic moment $a= \alpha/2\pi+...$} \label{19Tab2}
\end{table}

Setting the nuclear spin equal to 1/2, the equations for the bound $g$ factor in two-body atom read \cite{zee,plb2003}
\begin{eqnarray}\label{getwo}
  g_e^\prime  &=& g_e  \cdot\left\{1-\frac{(Z\alpha)^2}{3}
 \left[   1-\frac{3}{2}\frac{m}{M} \right]
 +\frac{\alpha(Z\alpha)^2}{4\pi}
- \frac{(Z\alpha)^2(1+Z)}{2}\left(\frac{m}{M}\right)^2 \right. \nonumber\\
 &-&\left.    \frac{5\alpha(Z\alpha)^2}{12\pi}\frac{m}{M}
 -\bigl(0.289\dots\bigr)\times\frac{\alpha^2(Z\alpha)^2}{\pi^2}
 -\frac{(Z\alpha)^4}{12}
\right\}
\end{eqnarray}
and
\begin{eqnarray}\label{gntwo}
  \label{gp}
  g_N^\prime &=&  g_N\cdot
  \left\{1- \frac{\alpha(Z\alpha)}{3}\left[ 1 - \frac{m}{2M}
  \frac{3+4a_N}{1+a_N} \right] \right.\nonumber\\
&+&\left.
    \alpha(Z\alpha)\left(\frac{m}{M}\right)^2
    \left(
      -\frac{1}{2} - \frac{Z}{6}\frac{3-4a_N}{1+a_N}
    \right)
    -\frac{97}{108}\,\alpha(Z \alpha)^3
  \right\}
\,,
\end{eqnarray}
where the anomalous magnetic moment of the nucleus $a_N$ is
defined as
\begin{eqnarray}
\mu_{\rm Nucl}\rm &=& 2\,\bigl(1+a_N\bigr)\,\frac{Ze\hbar}{2m}\,I \nonumber\\
&\simeq&2\,\bigl(1+a_N\bigr)\,\frac{Z}{A}\,\mu_{N}\,I\;,
\end{eqnarray}
where, we remind, $M$ is the nuclear mass, $I$ is its spin, while
$\mu_{N}$ in the nuclear magneton. Thus $a_p\simeq 1.793\dots$,
$a_d \simeq - 0.143\dots$, $a_t\simeq 7.916\dots$ and $a_h\simeq
-1.184\dots$ (the latter value is for a {\em helion\/}, the
nucleus of the ${}^3$He). We note that appearance of the factor of
$(1+a_N)$ in the denominator is rather artificial: a more natural
(from the theoretical point of view) expression is of the form
\[
g_N^\prime=2\times(1\cdot g_1 + a_N \cdot g_a)\;,
\]
however, due to applications is it more convenient to present the
result as above (i.e., as a multiplicative correction to $g_N$, a
free value of the $g$ factor).

We remark also that some of the listed values of the nuclear
anomalous magnetic moments are of large numerical values (as,
e.g., $a_t$). That is not an exception. The triton magnetic moment
is essentially a proton magnetic moment (the total spin of two
neutrons is zero) somehow perturbed. However, the magneton related
to the triton is approximately a third part of the proton's since
the charge of the proton and triton is the same, but the triton
mass is threefold higher.

\subsection{Hydrogen and deuterium: determination of the proton
and deuteron magnetic moments}

Studies of a few HFS intervals in the same atom in the presence of
the magnetic field were performed for hydrogen \cite{wink} and
deuterium \cite{phil84} and their results were found to
be\footnote{Here and further in this section we ignore the
direction of the spin and the magnetic moment and thus the sign of
some $g$ factors and ratios of magnetic moments. To simplify
notations, we also drop the prime sign when the bound system is
explicitly specified as, e.g., in the case of $g({\rm H})$.}
\begin{equation}\label{hdatum}
  \frac{\mu_e({\rm H})}{\mu_p({\rm H})}
  = \frac{g_e({\rm H})}{g_p({\rm H})}\;\frac{\mu_B}{\mu_N}=
  658.210\,705\,8(66)
\end{equation}
and
\begin{equation}
  \frac{\mu_e({\rm D} )}{\mu_d({\rm D} )}= 2143.923\,565(23)\;.
\end{equation}
The bound state QED contributions are small and the quoted
experiments were not designed to test them. The purpose was
different, namely, to determine the proton magnetic moment which
is widely used as a probe magnetic moment for other QED tests
(such as the muonium hyperfine structure \cite{MuExp,MuExp1}, the
anomalous magnetic moment of a muon \cite{redin}) or for other
measurements (such as of the deuteron magnetic moment
\cite{plamudp} etc.). In fact, the hydrogen experiment \cite{wink}
with the theoretical evaluation performed in \cite{plb2003} has
delivered the most accurate value of a free ratio of the
electron-to-proton magnetic moments
\begin{equation}
 \frac{\mu_p}{\mu_e} = 658.210\,685\,9(66)
\end{equation}
and the proton $g$ factor. The result of the evaluation in
\cite{plb2003} is somewhat shifted from the CODATA value
\cite{codata} due to recently obtained higher-order corrections
\cite{moor,pyp,zee} and improvement in determination of the
electron-to-proton mass ratio discussed below (see
Sect.~\ref{s:gme}).

\begin{figure}[hbtp]
\begin{center}
\includegraphics[width=0.6\textwidth]{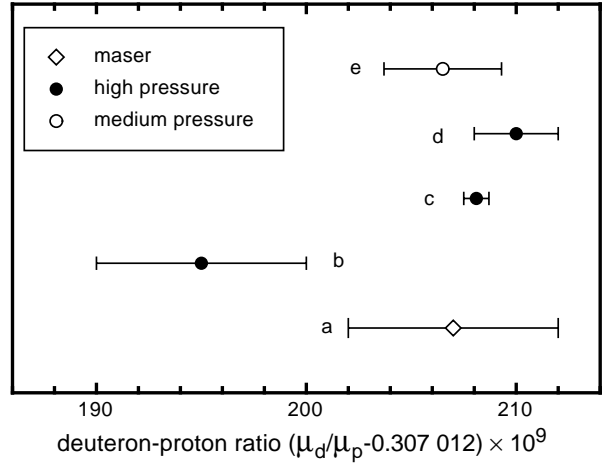}
\end{center}
\caption{The deuteron-to-proton ratio of the magnetic moments:
maser result {\em a\/} is taken from
\protect\cite{wink,phil84,codata}, the other results are from NMR
spectroscopy of hydrogen deuteride HD. Result {\em b\/} is related
to an experiment with an admixture of helium-3 at pressure of
60--80 atm \protect\cite{neronov_he}, while results {\em c\/} and
{\em d\/} correspond to measurements on pure HD gas at pressure of
130~atm with/without rotation of the sample
\protect\cite{neronov_d}. The high-pressure results here have an
uncertainty corrected for an inaccuracy in the calculation of
screening effects (see \cite{plamudp} for detail). The result {\em
e\/} is a recent result obtained at a medium pressure (10~atm) in
order to check effects of the pressure dependence \cite{plamudp}.
The systematic effects due to the pressure shift are under study
and have not been included into the uncertainty
budget.\label{medpres}}
\end{figure}

In the case of deuteron there are competitive NMR experiments on
HD spectroscopy. The experimental results published about twenty
years ago in \cite{neronov_old,neronov_t,neronov_he,neronov_d}
were obtained in NMR high-pressure experiments and they are not
free of possible systematic effects. We start a program of
additional experiments \cite{plamudp} and we hope to study those
systematic effects in more detail and achieve reliable data. The
present situation is summarized in Fig.~\ref{medpres}.

\subsection{The $g$ factors of an electron and a muon
in muonium and the determination of the muon mass\label{s:gmu}}

A difference between theory for muonium and for `conventional'
atoms is due to a higher importance of recoil effects (since the
nucleus, namely a muon, is approximately tenfold lighter than a
proton) and a possibility to expand the muon magnetic moment in
$\alpha$. After expanding $a_\mu$, the result for the bound muon
$g$ factor reads \cite{zee,new_rep}
\begin{eqnarray}
 g_e({\rm Mu}) &=& g_e^{(0)}\cdot
\left\{1-{(Z\alpha)^2\over3}
\left[1-{3\over2}{m\over M}\right] +{\alpha(Z\alpha)^2\over{4\pi}}
- {(Z\alpha)^2(1+Z)\over2}\left({m\over M}\right)^2
\right.
\nonumber\\
&-& {5\alpha(Z\alpha)^2\over{12\pi}}{m\over M}
-\left.{(Z\alpha)^4\over12}
 -\bigl(0.289\dots\bigr)\cdot\frac{\alpha^2(Z\alpha)^2}{\pi^2}
\right\}
\end{eqnarray}
and
\begin{eqnarray}
g_\mu({\rm Mu}) &=&
g_\mu^{(0)}\cdot \left\{
1-{\alpha(Z\alpha)\over3} \left[1-{3\over2}{m\over M}\right]\right.\nonumber\\
&-&\left.{\alpha(Z\alpha)(1+Z)\over2}\left({m\over M}\right)^2+
{\alpha^2(Z\alpha)\over{12\pi}}{m\over M} -{97\over108}\,\alpha(Z
\alpha)^3\right\},
\end{eqnarray}
where the $g$ factor of a free lepton includes the anomalous
magnetic moment $g_{e,\mu}^{(0)}=2\cdot(1+a_{e,\mu})$.

Studies of the Breit-Rabi levels (see Fig.~\ref{mfFig}) in muonium
delivered the best datum on the electron-to-muon mass ratio needed
for a determination of the fine structure constant $\alpha$ from
the muonium hyperfine structure (see Sect.~\ref{s:muhfs}).
Equation~(\ref{Heff}) has been applied \cite{MuExp,MuExp1} to
determine the muon magnetic moment and the muon mass by measuring
the splitting of sublevels in the hyperfine structure of the $1s$
state in muonium in a homogenous magnetic field. Since the
magnetic field was calibrated via the spin precession of a proton,
the muon magnetic moment was measured in units of the proton
magnetic moment, and the muon-to-electron mass ratio was derived
as
\begin{equation}
\frac{m_\mu}{m_e}=\frac{\mu_\mu}{\mu_p}\,\frac{\mu_p}{\mu_B}\,\frac{1}{1+a_\mu}\;.
\end{equation}

\begin{figure}[hbtp]
\begin{center}
\includegraphics[width=.7\textwidth]{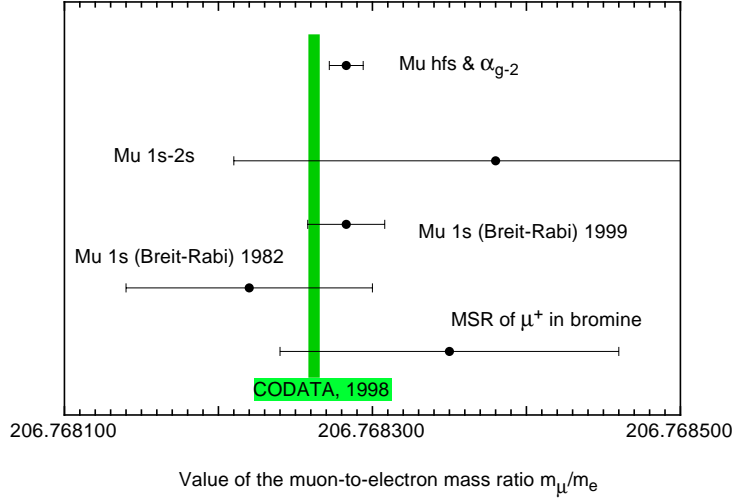}
\end{center}
\caption{The muon-to-electron mass ratio. The most accurate result
 obtained from comparison of the measured hyperfine interval in
muonium \cite{MuExp} to a theoretical calculation \cite{hamu1}
performed with $\alpha_{g-2}^{-1}= 137.035\,998\,76(52)$
\cite{kinoalpha}. The results derived from study of the Breit-Rabi
sublevels are related to two experiments performed at LAMPF in
1982 \cite{MuExp1} and 1999 \cite{MuExp}. The others are taken
from the measurement of the $1s-2s$ interval in muonium
\cite{Maas}, muon-spin-precession-resonance (MSR) study of a muon
in bromine \cite{klempt} and from the CODATA adjustment
\cite{codata}.\label{figMa}}
\end{figure}

The results on the muon mass extracted from the Breit-Rabi formula are
among the most accurate (see Fig.~\ref{figMa}). Still there is a
more precise value available derived from the muonium
hyperfine structure after a comparison of the experimental result
with theoretical calculations. However, the latter is of reduced
interest, since the most important application of the precise
value of the muon-to-election mass is to use it as an {\em
input\/} for calculations of the muonium hyperfine structure while
testing QED or determining the fine structure constants $\alpha$.
The adjusted CODATA result \cite{codata} in Fig.~\ref{figMa} was
extracted from the muonium hyperfine structure studies and in
addition used a somewhat over-optimistic estimation of the
theoretical uncertainty (see for detail \cite{hamu1}).

\section{The $g$ factor of a bound electron in a hydrogen-like ion
with a spinless nucleus and a determination of $m_e/m_p$\label{s:gme}}

A study of the $g$ factor of a bound electron in a medium-$Z$
hydrogen-like ion with a spinless nucleus offers a comparison of
the electron spin precession frequency affected by QED effects and
the ion cyclotron frequency. That can be used to test QED theory
as well as to determine the electron-to-proton mass ratio. An
important feature of such a study is that in contrast to most of
the other tests it is possible to do both without any interference
between these two tasks. That is possible because of an
opportunity to perform experiments with different hydrogen-like
ions.

Since the anomalous magnetic moment of a free electron
is known with high accuracy \cite{trap,codata}
\begin{equation}
a_e=0.001\,159\,652\,187(4)
\end{equation}
to calculate the bound $g$ factor defined in Eq.~(\ref{gbound}),
one has to obtain a bound correction $b$. A summary of the
calculation of the bound corrections is presented in
Table~\ref{17table1}. We mainly follow here our consideration in
\cite{zee}\footnote{Recently a certain progress in the calculation
of the one-loop and two-loop self-energy contribution
\cite{gtwoloop} as well as of the the magnetic part of the
light-by-light scattering contribution \cite{gmagloop} was
achieved. Since presently the experimental uncertainty dominates,
this improvement of theory has no consequences for a moment.}. In
particular, the uncertainty of unknown two-loop contributions is
taken from \cite{gjetp}, where the theoretical uncertainty was
reasonably estimated for the first time. The results of the
one-loop self-energy are achieved in different ways for different
atoms. For lighter elements (helium, beryllium) it is based on
fitting \cite{zee} data of \cite{beier}, while for heavier ions we
use the results of \cite{oneloop}. The other results are taken
from \cite{gjetp} (for the one-loop vacuum polarization; see also
\cite{neweprint}), \cite{pla} (for the nuclear correction and the
electric part of the light-by-light scattering (Wichmann-Kroll)
contribution; see also \cite{neweprint}), \cite{plb2002} (for the
magnetic part of the light-by-light scattering contribution) and
\cite{recoil2} (for the recoil effects).

\begin{table}[hbtp]
\begin{center}
\begin{tabular}{cc}
\hline Ion & $g$ \\
\hline
$^4$He$^+$ & 2.002\,177\,406\,7(1)\\
$^{10}$Be$^{3+}$ & 2.001\,751\,574\,7(2) \\
$^{12}$C$^{5+}$ & 2.001\,041\,590\,0(4)\\
$^{16}$O$^{7+}$ & 2.000\,047\,020\,3(8) \\
$^{18}$O$^{7+}$ & 2.000\,047\,021\,1(8) \\
\hline
\hline
\end{tabular}
\end{center}
\caption{The bound electron $g$ factor in low-$Z$ hydrogen-like
ions with spinless nucleus.\label{17table1}}
\end{table}

Before comparing the theory to experimental data, let us shortly
describe certain details of the experiment crucial for its
interpretation. To determine a quantity like the $g$ factor, one
needs to measure a certain field-induced frequency (such as the
spin precession frequency) at some known magnetic field $B$. It is
clear that there is no way to determine the strength of the
magnetic field with a high accuracy directly. The conventional way
is to measure two frequencies and to compare them. The frequencies
measured in the GSI-Mainz experiment \cite{werth} are the ion
cyclotron frequency
\begin{equation}
{\omega_c}=\frac{{(Z-1)e}}{M_i}B\;,
\end{equation}
where $M_i$ is the ion mass, and the Larmor spin precession
frequency for a hydrogen-like ion with a spinless nucleus
\begin{equation}
{\omega_L}={g_b}\,\frac{{e}}{2m_e}B\;,
\end{equation}
where $g_b$ is the bound electron $g$ factor.

Combining them, one can obtain a result for the $g$ factor of a
bound electron
\begin{equation}\label{g2bound}
\frac{g_b}{2}={\bigl(Z-1\bigr)}\,\frac{m_e}{M_i}\,\frac{\omega_L}{\omega_c}\;
\end{equation}
or for the electron-to-ion mass ratio
\begin{equation}\label{mebound}
\frac{m_e}{M_i}=\frac{1}{Z-1}\,\frac{g_b}{2}\,\frac{\omega_c}{\omega_L}\;.
\end{equation}
To measure the frequencies, one has to apply a certain magnetic
filed and, in principle, it should be as homogenous as possible.
However, the measurement of a frequency consists of two important
parts: an induction of a certain magnetic transition by an
accurately calibrated perturbation and a detection of the
transition. The homogeneity of the field, which is a crucial
requirement of the induction, is in a contradiction with the
detection scheme, which requires a non-homogeneity of the magnetic
field. In particular, a certain $z$-dependence of the field was
applied providing ion oscillations in a vertical direction which
were detected. Since the oscillation frequency depends on the
magnetic state that allows to identify the state and thus to
detect the transition. The breakthrough in a study of the bound
$g$ factor \cite{werth} was a result of a space {\em separation\/}
of the transition and detection areas, which gives an opportunity
to support a very homogenous field for the transitions and
afterwards to measure the response function in a separate area
with $z$ gradients.

\begin{figure}[hbtp]
\begin{center}
\includegraphics[width=.7\textwidth]{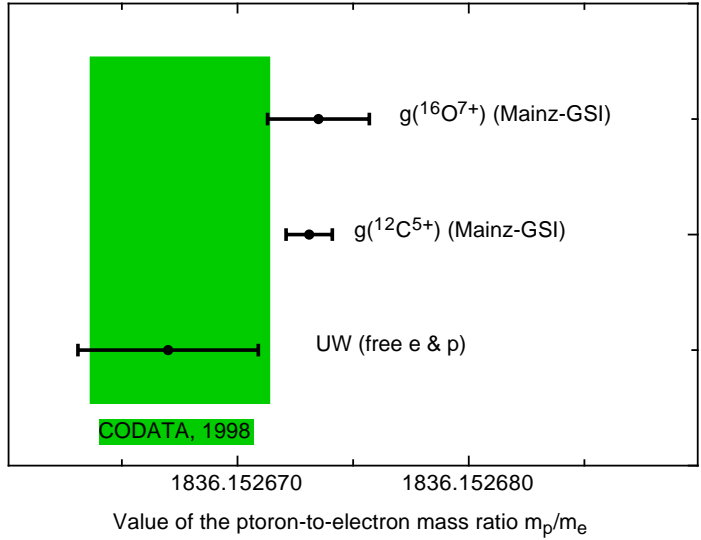}
\end{center}
\caption{The proton-to-electron mass ratio. The theory of the bound
$g$ factor is taken from Table~\ref{17table1}, while the experimental data on the
$g$ factor in carbon and oxygen are from \cite{werth,werthcjp}. The Penning trap result
from University of Washington is from \cite{farnham}.\label{figMe}}
\end{figure}

Today the most accurate value of ${m_e}/{M_i}$ (without using data
from the experiments on the bound $g$ factor) is based on a
measurement of $m_e/m_p$ realized in a Penning trap \cite{farnham}
with a fractional uncertainty of 2~ppm. The accuracy of
measurements of $\omega_c$ and $\omega_L$ as well as of the
calculation of $g_b$ (as shown in \cite{sgkH2}) is substantially
better. That means that it is preferable to apply
Eq.~(\ref{mebound}) to determine the electron-to-ion mass ratio
\cite{me}. Applying the theoretical value for the $g$ factor of
the bound electron and using the experimental results for
$\omega_c$ and $\omega_L$ in hydrogen-like carbon \cite{werth} and
some auxiliary data related to the proton and ion masses from
\cite{codata}, we arrive at the following values
\begin{equation}\label{mpme}
\frac{m_p}{m_e}=1\,836.152\,673\,1(10)
\end{equation}
and
\begin{equation}
m_e= 0.000\,548\,579\,909\,29(31)~{\rm u}\;,
\end{equation}
which slightly differ from those in \cite{me}. The present status of
the determination of the electron-to-proton mass ratio is summarized
in Fig.~\ref{figMe}.

In \cite{sgkH2} it was also suggested in addition to the
determination of the electron mass to check theory by comparing
the $g$ factor for two different ions. In such a case the
uncertainty related to ${m_e}/{M_i}$ in Eq.~(\ref{g2bound})
vanishes. The theoretical uncertainty is chiefly related to
uncalculated higher-order terms. Since various higher-order
corrections are proportional to $Z^4$ or so, even a small change
in $Z$, such as, e.g., a comparison of the carbon result ($Z=6$)
to the oxygen one ($Z=8$) is a quite sensitive probe for the value
of the higher-order terms.

Thus, combining the experimental results for carbon
\cite{werth} and oxygen \cite{werthcjp} ions
\begin{equation}\label{ggexp}
g(^{12}{\rm C} ^{5+})/g(^{16}{\rm O} ^{7+})
=1.000\,497\,273\,1(15)\;,
\end{equation}
one can verify the reliability of the estimation of the theoretical
accuracy. The experiment appears to be in perfect agreement with theory
\begin{equation}\label{ggtheo}
g(^{12}{\rm C} ^{5+})/g(^{16}{\rm O} ^{7+})
=1.000\,497\,273\,3(9)\;.
\end{equation}
Unfortunately, the authors of the experiments did not present this
value directly \cite{werthcjp}. They presented results for the $g$
factors of both ions separately. In both cases the experimental
uncertainties contain systematic and statistical contributions.
The dominant contribution is due to the electron mass and it
should vanish in the ratio. It is also not good to determine
$m_e/M_i$ from one experiment (let us say from carbon's) and use
as an input for the other (oxygen's). In such a way additionally
to the experimental uncertainties the experimental $g$ factor for
the oxygen would contain a theoretical uncertainty for the carbon
$g$ factor.

That is why it is so important to speak in terms of the ratio of
the $g$ factors. The experimental ratio is to be determined by
experimentalists who perfectly know the correlations between the
systematic sources of two measurements, while the theoretical
ratio is to be determined by theorists and that allows to take
into account correlations in the estimations of unknown
higher-order terms.

The estimation in Eq.~(\ref{ggexp}) is based on our understanding
of the papers \cite{werth,werthcjp} and on a private communication
\cite{werthprivate}. The agreement between theory and experiments
means that we have a reasonable estimate of the uncalculated
higher-order terms. Note, however, that for metrological
applications it is preferable to study lower $Z$ ions
(hydrogen-like helium-4 and beryllium-10) to eliminate these
higher-order terms if the experimental accuracy will increase.

As mentioned above, the leading radiative corrections are of
kinematic origin and in particular the contributions of order
$\alpha(Z\alpha)^2$ and $\alpha^2(Z\alpha)^2$ are related to a
relativistic bound-state Coulomb problem for an electron and muon
with non-zero anomalous magnetic moments, which are determined by
free QED. The leading bound-state QED effects are of order
$\alpha(Z\alpha)^4$ (the one-loop contributions) and
$\alpha^2(Z\alpha)^4$ (the two-loop contributions). The former are
known with sufficient accuracy, while lack of accurate results for
the latter determines the uncertainty of the theory. That is
compatible with a test of the Lamb shift theory at the level of a
percent (for the hydrogen-like ions of carbon and oxygen). That is
indeed not accurate enough for any interesting bound-state QED
test. And in fact that is a big advantage of the experiments on
the bound electron $g$ factor for determination of $m_e/m_p$. In a
typical situation (the Lamb shift and the Rydberg constant; the
muonium hyperfine structure and the fine structure constant; the
helium fine structure and $\alpha$) we have to try to verify QED
calculations and to determine a certain fundamental constant
within the same experiment. In the case of the determination of
the electron-to-proton mass ratio the QED theory is relatively
simple. We still need to prove QED calculations, but we have not
yet approached the crucial problem of the Lamb shift theory,
namely, a problem of higher-order two-loop corrections. However,
with sufficient theoretical progress this problem can be studied
with present experimental accuracy if the experiment will be
turned to higher $Z$, in particular, to $Z=20-30$. The
higher-order bound-state QED effects should contribute at a
detectable level and may be successfully studied. Comparing these
higher $Z$ results with $Z=6,8$ and lower $Z$ one can cancel an
uncertainty due to the determination of the electron-to-proton
mass ratio. Thus, the experimental verification of bound-state QED
theory and the determination of $m_e/m_p$ can be clearly
separated.

\section{A determination of the fine structure constant $\alpha$ by means of QED and atomic physics}

\begin{figure}[hbtp]
\begin{center}
\includegraphics[width=.7\textwidth]{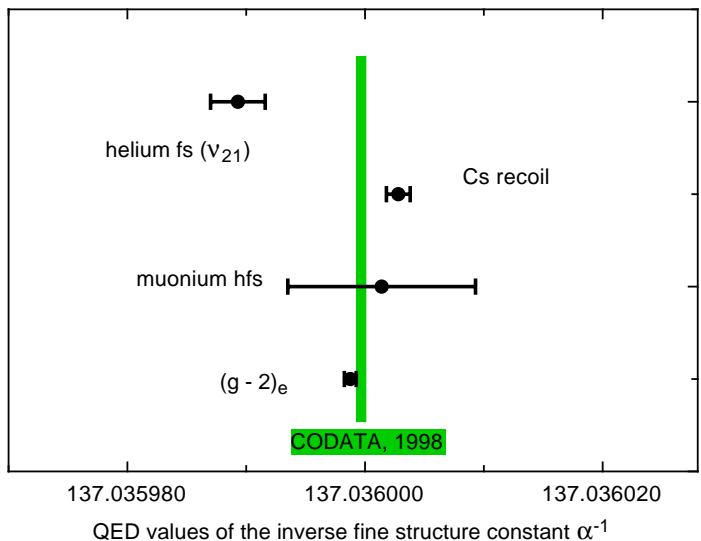}
\end{center}
\caption{The value of the fine structure constant $\alpha$ as
determined by means of atomic physics and QED.\label{figAl}}
\end{figure}

Precision measurements accompanied by accurate theoretical
calculations strongly interfere with high precision determinations
of certain fundamental constants. Those are needed in order to
interpret the theoretical expressions in terms of measurable
quantities. The fine structure constant $\alpha$ plays a basic
role in QED tests. In atomic and particle physics there are
several ways to determine its value. The results are summarized in
Fig.~\ref{figAl}. One method based on the muonium hyperfine
interval was briefly discussed in Sect.~\ref{s:muhfs}. A value of
the fine structure constant can also be extracted from the
neutral-helium fine structure \cite{helium_d,helium_ps,heliumexp}
and from a comparison of theory \cite{kinoalpha} and experiment
\cite{trap} for the anomalous magnetic moment of an electron
($\alpha_{g-2}$). The latter value has been the most accurate one
for a while and there was a long search for another competitive
value. Since the value of $\alpha_{g-2}$ is based on one-group
experiment \cite{trap} and one-group theory \cite{alphag2} there
has been a long standing concern on reliability of this value and
a search for its confirmation, in particular, via an independent
competitive determination of the fine structure constant. At
present, the second best value ($\alpha_{\rm Cs}$) in the list of
the most precise results for the fine structure constant comes
from recoil spectroscopy \cite{chu}. It is only three times less
accurate than $\alpha_{g-2}$. We note, however, that the result
has not yet been published in a refereed journal, but only in
several conference proceedings \cite{chu}. Since results in
proceedings are less reliable than those published in the
journals, we look forward to the eventual publication.

We would like to consider briefly the use and the significance of
the recoil result for the determination of the fine structure
constant. Absorbing and re-emitting a photon, an atom, if it is
initially in rest, gains some kinetic energy which can be
determined through a measurement of the shift of the emitted
frequency in respect to the absorbed one. The recoil shift depends
on relative directions of two photons
\begin{equation}
\Delta f = - C(\theta)\frac{h\,f^2}{2M_ac^2}\;,
\end{equation}
where $M_a$ is the mass of the atom, and under conditions of the
experiment \cite{chu} $C=2$.

That is a measurement of the frequency $\Delta f$ with high
accuracy that was the aim of the photon recoil experiment
\cite{chu}. Combining the absorbed frequency and the shift $\Delta
f$, it is possible to determine a value of atomic mass in
frequency units, i.e., a value of $M_ac^2/h$. In experiment
\cite{chu} the atom was cesium and the transition was the D1 line.
The atomic mass of cesium is known very well in atomic units (or
in units of the proton mass) \cite{Rain} and the D1 line was
accurately measured in \cite{d1mpq}. A value of $M_ac^2/h$ may be
compared to the Rydberg constant $R_\infty=\alpha^2 m_ec/2h$ and
if the electron mass in proper units is known, that eventually
delivers us a value of $\alpha$. A determination of the electron
mass is more complicated than of any atomic mass because of a gap
of three orders of magnitude between masses of an electron and a
proton. At present, the biggest uncertainty of the recoil photon
value of $\alpha_{\rm Cs}$ comes from the experiment \cite{chu},
however, accuracy in the determination of the electron mass is
still significant since that is the second important source of the
uncertainty.

The success of the $\alpha_{\rm Cs}$ determination was sometimes
ascribed to the fact that $\alpha_{g-2}$ is a QED value, derived
with the help of QED theory of the anomalous magnetic moment of an
electron, while the photon recoil result is free of QED. We would
like to emphasize that the situation is not so simple and the
importance of the involvement of QED is rather exaggerated. What is
more important is that the uncertainty of $\alpha_{g-2}$
originates from understanding of the electron behavior in the
Penning trap and it dominates over any QED uncertainty. For this
reason, the value of $\alpha_{\rm Cs}$, based on $m_p/m_e$ from
another Penning trap experiment \cite{farnham} and actually
obtained by the same group, which measured the value of the
anomalous magnetic moment of electron \cite{trap}, can actually be
correlated with $\alpha_{g-2}$. The result
\begin{equation}
\alpha_{\rm Cs}^{-1}=137.036\,0002\,8(10)
\end{equation}
presented in Fig.~\ref{figAl} is obtained using $m_p/m_e$ from
Eq.~(\ref{mpme}). The value of the proton-to-electron mass ratio
found this way is free of the problems with an electron in the
Penning trap, but a certain QED theory is involved. However, one
has to realize that the QED uncertainty for the $g$ factor of a
bound electron and that for the anomalous magnetic moment of a
free electron are completely different. The bound state QED theory
deals with relatively simple Feynman diagrams but in Coulomb field
and in particular to improve theory of the bound $g$ factor, we
need better understanding of Coulomb effects for `simple' two-loop
QED diagrams. In contrast, for the free electron no Coulomb field
is involved, but the uncertainty of the computation arises because
of the accuracy in a numerical calculation of a big number of
complicated four-loop diagrams. There is no correlation between
these two kinds of calculations. The words `simple' and
`complicated' are related to the diagrams, but not to the
simplicity and difficulties of the calculations. Both kinds of
calculations are quite difficult but in a different way. The
crucial free diagrams are complicated diagrams built of simple
blocks. The key bound QED contributions correspond to a simple
diagram constructed of complicated blocks.

\begin{figure}[hbtp]
\begin{center}
\includegraphics[width=.7\textwidth]{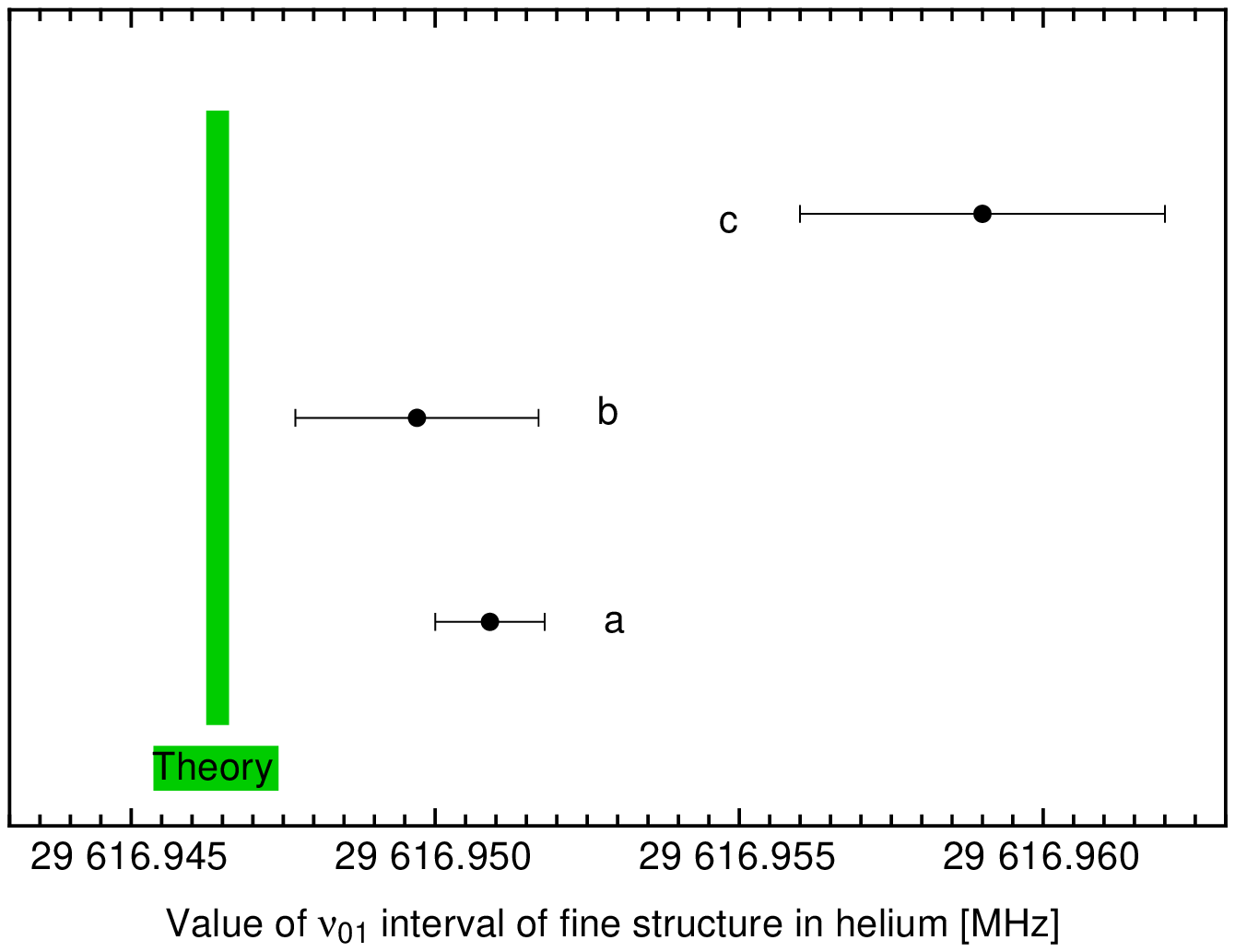}\\
\includegraphics[width=.7\textwidth]{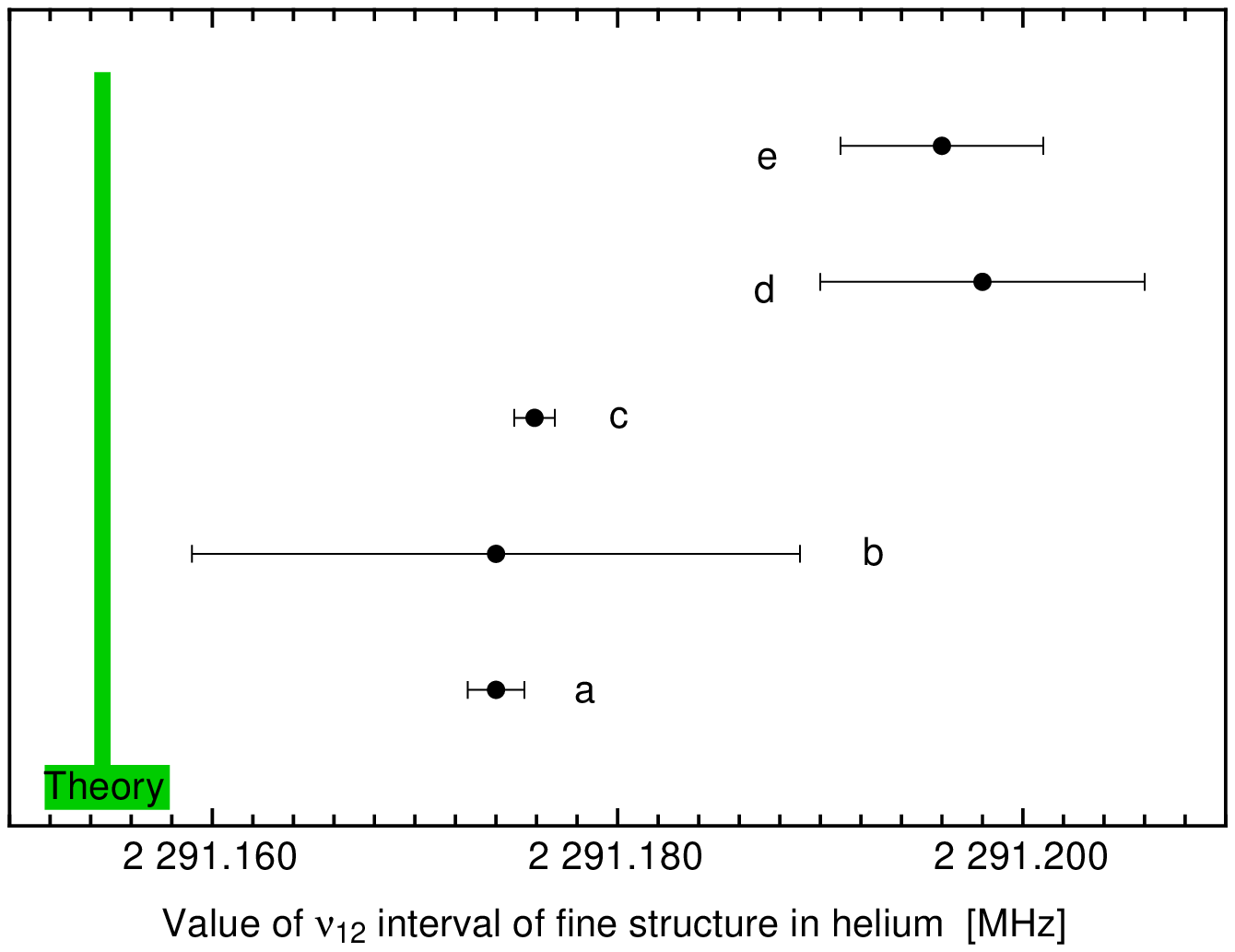}
\end{center}
\caption{The fine structure in neutral helium. $\nu_{01}$ stands
for $1s2p\,^3P_0-{}^3P_1$ and $\nu_{12}$ for
$1s2p\,^3P_1-{}^3P_2$. The theory is presented according to
\cite{helium_d,helium_ps}, the experimental data are taken $a$
from \cite{heliumexp}, $b$ from \cite{iheliumexpb}, $c$ from
\cite{iheliumexpc}, $d$ from \cite{iheliumexpd} and $e$ from
\cite{iheliumexpe}. Less accurate data are not
included.\label{00figHe}}
\end{figure}

The two best values, $\alpha_{g-2}$ and $\alpha_{\rm Cs}$, agree
with the muonium result but contradict to the helium value (see
Fig.~\ref{figAl}). The latter is presented in Fig.~\ref{figAl}
rather to show a potential accuracy of the application of the
helium fine structure method. The helium fine structure of the
$1s2p$ states involves few transitions and in particular so-called
`small interval' ($\nu_{01}$) and `big interval' ($\nu_{12}$). A
comparison of theory against experiment is presented in
Fig.~\ref{00figHe}. One can match a certain scatter of the
experimental data and a bad agreement of theory and experiment. In
principle, one could speculate on a special value of the fine
structure constant which will set an agreement of theory and
experiment. However, the contradiction for $\nu_{01}$ and
$\nu_{12}$ is at the same level in absolute units (from 5 to 20
kHz) and differs in fractional units since the `big interval' is
more than tenfold bigger than the `small interval'. The
contradiction for the `small interval' is too big to be explained
by any acceptable shift in the value of $\alpha$. The situation is
quite unclear but, still, the recent improvement of theoretical
accuracy remains quite promising \cite{helium_d,helium_ps}.

There is a number of other values of the fine structure constant
(see \cite{codata,UFN} for detail). Some of them (based on a
measurement of the gyromagnetic ratio of a proton and a helion and on
a determination of $h/m_n$) involve some QED theory and experiments,
however, the uncertainty of all of them is completely determined
by metrological problems due to electric standards and a
production of a perfect silicon crystal for the Avogadro project.
The other electrical determinations,  based on a measurement of
the quantum Hall resistance in the SI units, needs no QED. Since
the accuracy of those results is substantially below that of
$\alpha_{g-2}$ and since their uncertainty is related to the
electric, magnetic and material standards we do not consider them
here. An updated review of QED and non-QED determination of the
fine structure constant can be found in \cite{newcodata}.

\section{Higher-order logarithmic corrections and
the uncertainty of the bound-state QED calculations\label{S:log}}

\subsection{Hyperfine structure in light hydrogen-like-atoms}

To summarize QED tests related to the hyperfine structure, we
present in Table~\ref{17tabHFS} the data related to the HFS
interval in the $1s$ state of positronium and muonium and to the
specific difference $D_{21}$ in hydrogen, deuterium and helium-3
ion. As we noted in Sects.~\ref{s:d21}, \ref{s:muhfs}
and~\ref{s:ps}, their theoretical uncertainty is related to recoil
effects of relative order $\alpha^3(m/M)$. In the case of helium
$D_{21}$ the higher-order one-loop and two-loop effects are also
important, as well as higher-order nuclear corrections. The
theoretical accuracy supersedes the experimental one (except of
the case of the $D_{21}$ HFS interval in the helium-3 ion, for
which the theoretical uncertainty was recently increased after its
reconsideration \cite{2sZETF,2seprint}). The theory agrees with
experiment very well. A certain expection is the positronium case
where the agreement is not perfect.

\begin{table}[hbtp]
\begin{center}\begin{tabular}{lcccc}
\hline
 Atom  & Experiment & Theory &  $\Delta/\sigma$ & $\sigma/E_F$ \\
 & [kHz] & [kHz]  &   & [ppm] \\
 \hline
Hydrogen, $D_{21}$  &  49.13(13), \protect\cite{2shydr} & 48.953(5)  & 1.3 & 0.09 \\
Hydrogen,  $D_{21}$  &  48.53(23), \protect\cite{rothery}  &  & -1.8 & 0.16 \\
Hydrogen,  $D_{21}$  &  49.13(40), \protect\cite{exph2s} & & 0.4 &  0.28 \\
Deuterium, $D_{21}$  &  11.280(56), \protect\cite{mpqd2s}  & 11.312\,5(10) & -0.6 & 0.17 \\
Deuterium, $D_{21}$  &  11.16(16), \protect\cite{expd2s}  &  & -1.0 & 0.49 \\
$^3$He$^+$ ion, $D_{21}$~~~  &-1\,189.979(71), \protect\cite{prior} &-1\,190.08(15) &0.6 &0.02  \\
$^3$He$^+$, $D_{21}$ & -1\,190.1(16), \protect\cite{exphe2s} &  &  -0.01 & 0.19 \\
\hline
Muonium, $1s$ & 4\,463\,302.78(5)& 4\,463\,302.88(55)& -0.18 &0.11\\
Positronium, $1s$ & 203\,389\,100(740) & 203\,391\,700(500) & -2.9 &4.4\\
Positronium, $1s$ & 203\,397\,500(1600)& & -2.5 &8.2\\
\hline
\end{tabular}
\end{center}
\caption{A comparison of experiment and theory of the hyperfine
structure in light hydrogen-like atoms. The numerical results are
presented for the frequency $E/h$. In the $D_{21}$ case the
reference is given only for the $2s$ hyperfine interval, since for
all the atoms of interest the $1s$ HFS interval was measured more
accurately (see table~\protect\ref{17tab1} for
more detail).\label{17tabHFS}}
\end{table}

\subsection{Crucial higher-order corrections}

The precision physics of light simple atoms provides us with an
opportunity to check higher-order effects of perturbation theory.
The highest-order terms important for the comparison of theory and
experiment are collected in Table~\ref{17tabOrd}. The results for
the energy levels and decay rates of interest. There are also a
few dimensionless quantities of interest, which are mainly the $g$
factors. The uncertainty of the $g$ factor of the bound electron
in carbon and oxygen is related to $\alpha^2(Z\alpha)^5mc^2$
corrections in the energy units, while for calcium the crucial
order is $\alpha^2(Z\alpha)^6mc^2$ if the experiment will reach
the same level of accuracy.

\begin{table}[hbtp]
\begin{center}
\begin{tabular}{lc}
\hline
Value & Order \\
 & [$mc^2$] \\
\hline
Hydrogen, deuterium (gross structure) & $\alpha(Z\alpha)^7$, $\alpha^2(Z\alpha)^6$ \\
Hydrogen, deuterium (fine structure)              & $\alpha(Z\alpha)^7$, $\alpha^2(Z\alpha)^6$ \\
Hydrogen, deuterium (Lamb shift)              & $\alpha(Z\alpha)^7$, $\alpha^2(Z\alpha)^6$ \\
$^3$He$^+$ ion ($2s$ HFS)  & $\alpha(Z\alpha)^7m/M$, $\alpha(Z\alpha)^6m^2/M^2$,\\
& $\alpha^2(Z\alpha)^6m/M$, $(Z\alpha)^7m^2/M^2$\\
$^4$He$^+$ ion (Lamb shift)              & $\alpha(Z\alpha)^7$, $\alpha^2(Z\alpha)^6$ \\
Muonium ($1s$ HFS)      & $(Z\alpha)^7m^2/M^2$, $\alpha(Z\alpha)^6m^2/M^2$,\\
&$\alpha(Z\alpha)^7m/M$ \\
Positronium ($1s$ HFS)  & $\alpha^7$ \\
Positronium (gross structure)      & $\alpha^7$ \\
Positronium (fine structure)       & $\alpha^7$ \\
Para-positronium (decay rate)       & $\alpha^7$ \\
Ortho-positronium (decay rate)      & $\alpha^8$ \\
Para-positronium ($4\gamma$ branching)  & $\alpha^8$ \\
Ortho-positronium ($5\gamma$ branching) & $\alpha^8$ \\
\hline
\end{tabular}
\end{center}
\caption{A comparison of QED theory and experiment: the crucial
orders of magnitude for the energy levels and decay rates in units
of $mc^2$ (see \cite{icap} for more detail).\label{17tabOrd}}
\end{table}

While some of the crucial corrections presented in
Table~\ref{17tabOrd} are completely known, others are not. Many of
the listed corrections and in particular
$\alpha(Z\alpha)^6m^2/M^3$ and $(Z\alpha)^7m^2/M^3$ for the
hyperfine structure in muonium and helium ion, $\alpha^2(Z
\alpha)^6m$ for the Lamb shift in hydrogen and helium ion,
$\alpha^7 m$ for positronium have been known in a so-called
logarithmic approximation. In other words, only the terms with the
highest power of a `large' logarithm (e.g.,
$\ln(1/Z\alpha)\sim\ln(M/m)\sim 5$ in muonium) have been taken
into account. We note that the `large' corrections should have an
essentially non-relativistic origin. Truly relativistic
corrections used to have a few factors $Z\alpha/\pi$ instead of
$Z\alpha$, which is a characteristic non-relativistic factor for
the Coulomb problem, and thus they are numerically suppressed.
First calculations of logarithmically enhanced higher-order terms
were performed for higher-order corrections in muonium HFS,
hydrogen Lamb shift, positronium energy level and decay rates in
our paper \cite{log1} and were successfully developed by various
authors in \cite{log2,del1,log3,pk2,my,sgkH2s,d21}. By now even
some non-leading logarithmic terms have been evaluated by several
groups \cite{logps1,logps2,logps3,morelogs,alltwoloop}.

Such a value of the parameter as $\sim 1/5$ is sufficient to make
a good first approximation. We used to estimate accuracy of the
leading logarithmic approximation by 50\%. However, we cannot
consider such an estimation as an accurate and reliable result. If
the correction happens to have a value significant in comparison
with the overall uncertainty, one has to improve the result. A use
of the leading logarithmic terms is fruitful since they used to
have some `natural' values because they are always (or almost
always) state-independent and not a result of big cancellations.
The latter happens seldom between different gauge-invariant sets
(e.g., in the $\alpha E_F$ term for the positronium hyperfine
splitting) and is easy to recognize. It is clear what to do in the
case of such a cancellation; it is necessary to estimate the
non-leading terms separately by a half-value of the related
leading terms and then to sum these uncertainty contributions as
an rms sum.

``State-independence'' means that the contributions are
proportional to the wave function at the origin squared
$\vert\Psi(0)\vert^2$ or its first derivative
$\vert\Psi(0)\Psi^\prime(0)\vert$, i.e., to $\delta_{l0}/n^3$,
with no other dependence on $n$. In particular, state-independent
effects do not contribute to $\Delta(n)$ and $D_{21}$ considered
in this paper. The non-leading terms, in contrast, depend on a
state and involve sometimes substantial cancellations. They also
may involve various numerical cancellations and their calculation
can hardly improve the reliability of the estimation. However,
their evaluation is necessary as a step in direction towards the
completion of the calculation of the whole correction and to check
whether the non-leading terms are of a reasonable value which
sometimes may be enhanced\footnote{In particular, large
coefficients were discovered for linear logarithmic term for the
Lamb shift \cite{morelogs,alltwoloop}. The coefficients are so
large that the term linear in $\ln(1/(Z\alpha))$ dominates over a
cubic logarithmic term. The result has not yet been confirmed by
an independent group. Neither origin of the large numerical value
of the coefficient is understood, nor possible consequences for
estimation of other higher-order terms.}. We underline that it
could be no rigorous proof for such an empirical estimation --- it
is barely a result of experience based on a high number of known
logarithmic contributions. Such an approach is not much different
from what experimentalists do to estimate possible systematic
errors. The estimation is always a guess, hopefully a plausible
one, and definitely a deeper understanding of the problem is
helpful. Our motivation is not only application of the factor of
1/5 as a parameter, but also an observation that most functions in
physics are quite smooth. The leading logarithmic term for some
value of $Z\alpha$, which may be even much below 1/137 related to
a non-physical value of $Z\ll1$, determines the sign and the
magnitude of the correction. In such a case even if the parameter
approaches unity, the order of magnitude is often still determined
by the leading asymptotics as long as the function is smooth
enough. Most of such smooth functions seldom change their sign and
that is also important for the motivation of our method. A weak
point of this explanation is that we already deal with certain
higher order corrections, i.e., with a function with its first few
terms of expansion subtracted. Even if the function itself is a
`physical' one with the expected smooth behavior, it is not quite
clear how smooth it should be after the subtractions. Also, we
know that some smooth functions change sign etc. and in this case
the asymptotics may be not helpful.

It seems that we have reached a certain numerical limit related to
the logarithmic contribution and the calculation of the
non-logarithmic terms will be much more complicated than anything
else done before. However, it is strongly needed to improve the
reliability and the accuracy of the theoretical predictions.

\subsection{Crucial `soft QED' corrections}

Not only the crucial orders of magnitude offer a cross comparison
of efficiency of different experiments. The modern bound state QED
theory clearly recognizes two kinds of contributions: soft-photon
contributions and hard-photon contributions. The latter are very
similar to free QED, while the former essentially involve binding
effects. There are two crucial soft-QED contributions.
\begin{itemize}
\item One of crucial binding corrections is due to the higher
order of the one-loop and two-loop self-energy. The one-loop self
energy was recently successfully calculated for most of important
applications. The two-loop self-energy terms (see, e.g.,
Fig.~\ref{Ftwoloop}) are partly known and uncertainty of this
knowledge determines the accuracy of the Lamb shift calculations
for hydrogen, helium ion,
the medium- and high-$Z$ Lamb shift and fine structure, while
similar diagrams with a transverse photon contribute to an
uncertainty of $D_{21}$ in the helium-3 ion.
\begin{figure}[hbtp]
\begin{center}
\includegraphics[width=.4\textwidth]{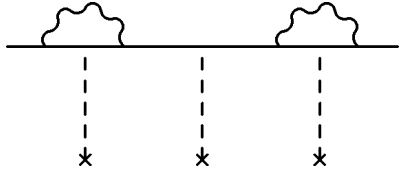}
\end{center}
\caption{One of diagrams for higher-order two-loop corrections
(order of $\alpha^2(Z\alpha)^6mc^2$) to the Lamb shift. The
logarithmic part contains up to the cube of the logarithm.
\label{Ftwoloop}}
\end{figure}
\item The other important corrections are caused by recoil effects
(see Figs.~\ref{Fhighrad} and \ref{Fhighrec}) and they determine
the uncertainty of the muonium HFS, the positronium energy levels
(the hyperfine interval, the $1s-2s$ transition, and the fine
structure) and a part of the uncertainty of the $D_{21}$
difference for the helium-3 ion. The soft-photon nature of these
contributions is clearly seen from the appearance of the bound
logarithm $\ln(1/(Z\alpha))$. Actually this logarithm often (but
not always) arises as $\ln(mc^2/\langle E \rangle)$, where
$\langle E \rangle$ is a characteristic atomic energy. The other
possibility for this logarithm to appear is a logarithmic
integration over coordinate or momentum space which leads to
$\ln(mc \langle r \rangle/\hbar)$, where $\langle r \rangle$ is
the characteristic size of the atomic state. Thus, the
space-logarithm contributions contain a soft part with
integrations over atomic momenta $k\sim (Z\alpha)mc$ and hard part
related to integration over $k\sim mc$. In energy-logarithm cases,
two low-momentum regions ($k\sim (Z\alpha)^2mc$ and $k\sim
(Z\alpha)mc$) can be distinguished.

If one calculates separately the soft- or hard-momentum
contributions, the logarithm of an effective energy or momentum
appears as a logarithm of a certain cut-off. That greatly
simplifies evaluation of logarithmic corrections. Meantime, the
appearance of such logarithm means that both the soft and hard
contributions are divergent and only their sum have sense.

\begin{figure}[hbtp]
\begin{center}
\includegraphics[width=.35\textwidth]{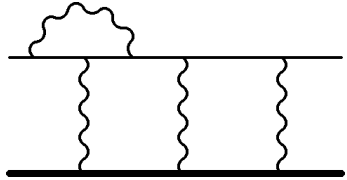}
\end{center}
\caption{An example  of the diagrams for higher-order radiative
recoil corrections to the HFS interval (relative order of
$\alpha(Z\alpha)^2m/M$). The logarithmic part contains up to the
square of the logarithm.\label{Fhighrad}}
\end{figure}
\begin{figure}[hbtp]
\begin{center}
\includegraphics[width=.3\textwidth]{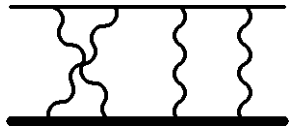}
\end{center}
\caption{An example of the diagrams for higher-order pure recoil
corrections to the HFS interval (relative order of
$(Z\alpha)^3m/M$). The logarithmic part contains double
logarithmic contributions.\label{Fhighrec}}
\end{figure}
\end{itemize}

\subsection{Crucial `hard QED' corrections}

The hard-photon effects are very similar to effects of free QED,
however, the areas of integrations over momentum space are
different. Both the free QED and `hard' contributions to the
bound-state QED  suffer from infrared divergencies. This
problem is also solved in different ways.

Most accurate free QED calculations which may be compared with
experiment are related to the anomalous magnetic moments of an
electron and a muon. The integration for them is performed over
kind of an isotropic area in the Euclidean 4D space\footnote{The
physical expressions to calculate originate indeed from a
description related to the Minkowsky space with the metrics
$(1,-1,-1,-1)$, where, e.g., $k_\mu k^\mu = k_0^2-{\bf k}^2$.
However, while calculating the diagrams it is often convenient to
make the Wick rotation of the integration contour and to arrive at
the Euclidian metric $(1,1,1,1)$ (or rather $(-1,-1,-1,-1)$).},
and the crucial level is a four-loop approximation. For the bound
problems there are two other specific areas of integration.
\begin{itemize}
\item For some problems it is sufficient to apply an
external-field approximation to hard-photon corrections, in which
the exchange photons carry zero energy transfer (see, e.g., a
diagram in Fig.~\ref{Ftwoloop}). Even for recoil effects, the
integration is essentially not covariant in the Euclidian 4D
space. The highest crucial orders are related to four-loop
corrections for the external-field and three-loop corrections for
recoil effects (see, e.g., diagrams in Figs.~\ref{Fhighrad} and
\ref{Fhighrec}). Still, for certain four-dimensional Euclidian
variable one can observe for characteristic momenta $|k_0|\ll
|k_i|$. Note that in contrast to the anomalous magnetic moment,
this is a calculation for two different particles and thus it has
its own simplifications and difficulties.
\item The other specific situation for integration is related to
the positronium annihilation when some photons are real and thus
their momentum satisfy the condition $k^2=k_0^2-{\bf k}^2=0$.
Studies of the orthopositronium decay mode (see, e.g.,
Fig.~\ref{F3phot}) and a branching fraction for five-photon decay
\cite{bran35,4ge} allow us to check calculations of four-loop
corrections in such a non-isotropic area of integration. The
accuracy of determination of the branching of four-photon decay of
parapositronium \cite{4ga,4gb,4gc,4gd,4ge} is approaching a level
where four-loop diagrams are important.
\begin{figure}[hbtp]
\begin{center}
\includegraphics[width=.35\textwidth]{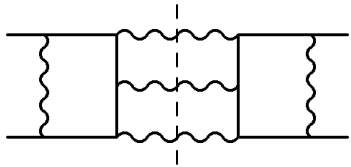}
\end{center}
\caption{An example of the diagrams for two-loop radiative corrections to the
three-photon annihilation of orthopositronium). \label{F3phot}}
\end{figure}
\end{itemize}
The bound state problems supply us with an opportunity to check
modern algorithms for four-loop calculations, competitive to the
anomalous magnetic moment of an electron, and allow us to test
them for different kinematic areas.

\section{What is next?}

Above, we have reviewed the present status on precision theory and
experiment on light two-body atoms. In this section, we will
briefly overview problems and possible experiments in the field.

The Lamb shift was successfully measured in hydrogen, deuterium
and an ion of He$^+$. Experiments in neutral atoms have reached a
certain level of accuracy, which will be difficult to improve very
much without completely new approaches. However, a certain
progress is possible and is expected from optical frequency
measurements.

Nevertheless, microwave experiments may be important. There
remains a question about reliability of a microwave experiment by
Sokolov and Yakovlev. Their results are reproducible within their
experiment, which means that any problem related to possible
systematic effects could be, in principle, resolved. Their
statistical uncertainty is competitive with the present optical
experiments. Unfortunately, the experiment was stopped. We hope it
may be continued or reproduced.

We also note that a microwave study of hydrogen circular states at
MIT is only a project that can deliver a really independent value
of the Rydberg constant, but, unfortunately, the evaluation of the
data remains unfinished.

In the case of the helium ion only microwave measurements have
taken place. Ongoing optical experiments can offer essential
progress.

Tritium is a light neutral hydrogen-like atom whose the Lamb shift
has not yet been measured accurately. We are interested in tritium
not because of QED, but mostly because we hope to learn more about
its nuclear structure and in particular on its charge radius.

Uncertainty in determining the proton charge radius is the main
obstacle in any improvement of the precision theory for the Lamb
shift in hydrogen. The former scattering measurements were
evaluated improperly, while their original data are not available.
A new measurement, even with the same accuracy but with a
transparent and reliable evaluation, would be very helpful. We can
also learn the value of the proton charge radius by measuring the
Lamb shift in muonic hydrogen; and we anticipate news from a
related PSI experiment.

The hyperfine interval of the ground state in hydrogen has been
known for a while with a high experimental accuracy. Any
theoretical progress is possible there only with a better
understanding of the proton structure. In particular it is crucial
to have a reliable description of the proton polarizability
effects. This may be done with improvement of the experimental
data on scattering (muon-proton's or electron-proton's). We
observed a certain interest by high-energy community to
re-evaluate of various experimental data for the scattering
(mainly due to determination of the proton charge radius). We hope
for a reliable re-evaluation of existing data on inelastic
scattering to be obtained soon. More data on spin-dependent
inelastic muon-proton scattering are strongly needed.

An advanced phenomenological description of the nucleon structure
and nucleon-nucleon interaction, based on a constraint from
experiment, is crucial not only for hyperfine splitting in
hydrogen but also in few-nucleon light atoms, such as deuterium,
tritium, helium-3 ion. Accuracy can be improved and we have
observed some progress in this field. We hope that nuclear theory
will help with prediction of the nuclear-structure contributions
to the HFS interval for $A=3$ nucleus.

Successful comparison of the $1s$ and $2s$ hyperfine intervals is
not completely sufficient. Accuracy of the experimental
determination of the $2s$ HFS interval in hydrogen and deuterium
is still not high enough, however, its threefold improvement after
applying optical methods allows to hope for future progress. The
$2s$ measurement in the ${}^3$He$^+$ ion is still competitive with
theory. However, it is not satisfactory when so important
experimental value was never re-measured again. It seems that it
is possible to go to somewhat higher $Z$ and study Li$^{++}$ ions.

The comparison of the $1s$ and $2s$ HFS intervals is not the only
way to get rid of the nuclear effects. Another important option is
spectroscopy of pure leptonic atoms, such as muonium and
positronium. Unfortunately, this kind of experiments was
completely stopped some time ago. There are no running or planned
muonium projects, while all present positronium projects aim only
at the study of decay. We think that a measurement of the $1s$ HFS
interval, $1s-2s$ transitions in muonium and positronium and
$2S-2P$ fine structure intervals in positronium can deliver
results quite competitive with theory. In the case of the
hyperfine splitting in positronium the experiment is necessary
also to resolve a discrepancy of approximately $3\sigma$ which
shows that our understanding of the situation is not completely
satisfactory.

Successful experiments designed to determine the $g$ factor of a
bound electron in a hydrogen-like ion with a spinless nucleus
eventually led to the most accurate determination of the
electron-to-proton mass ratio (or, which is essentially
equivalent, of the electron mass in unified atomic mass units).
The experiment was done with carbon-12 and oxygen-16. Improvement
in accuracy is possible. It is also of crucial importance to
present the data properly. A value $g({}^{12}{\rm
C}^{5+})/g({}^{12}{\rm O}^{7+})$ should be presented by
experimentalists by taking into account correlations between two
measurements. The range of ions should be broadened to include
ions from ${}^{10}$Be$^{3+}$ to ${}^{40}$Ca$^{19+}$. Medium-$Z$
lithium-like ions may be helpful at least to study systematic
effects since they broaden a possible charge-to-mass ratio of ions
and can be reasonably well understood theoretically.

These experiments are more accurate than the MIT study of
Breit-Rabi levels in hydrogen and deuterium which delivered the
best values of $\mu_p/\mu_e$ and $\mu_d/\mu_e$. However, these
experiments were never reproduced and that is not satisfactory
because of the great importance of an accurate value of
$\mu_p/\mu_e$ for QED tests and determination of fundamental
constants. In principle a value of $\mu_p/\mu_e$ can be and should
be determined independently.

One of possibilities can be a determination of the anomalous
magnetic moment\footnote{We follow the common notation which is
somewhat misleading. What is called `anomalous magnetic moment' is
rather `anomalous $g$ factor'. In this paragraph we are discussing
the anomalous contribution to the magnetic moment.} of the
deuteron (similarly to a measurement of the anomalous magnetic
moment of muon at the Brookhaven National Laboratory). That
measurement was done for a value of
$a_\mu\cdot(e\hbar/m_\mu)/\mu_p$. A similar result for a deuteron
would be very helpful since the deuteron anomalous magnetic moment
is small ($a_d \simeq - 0.143$) and its `normal' value, related to
the $g$ factor $g^{(0)}_d=2$, i.e., $g^{(0)}_d\cdot(e\hbar/2m_d)$,
in units of the nuclear magneton is well known (that is simply the
mass ratio for a proton and a deuteron multiplied by two).
Smallness of the magnetic moment provides an enhancement of
accuracy in determination of the magnetic moments
\[
\frac{\mu_d}{\mu_N}=\frac{2m_p}{m_d}+\frac{2a_d\cdot(e\hbar/2m_d)}{\mu_p}\cdot\frac{\mu_p}{\mu_N}
 \;.
\]
Comparing the deuteron
and proton magnetic moments via NMR spectroscopy of HD molecules,
one can find $g_p$.

Determination of the $g$ factor of a proton ($g_p=\mu_p/\mu_N$) is
essentially the same as a determination of $\mu_p/\mu_e$, since
the conversion factor $\mu_N/\mu_e$ is known well enough. W.~Quint
suggested \cite{Quint} to measure the proton $g$ factor directly
using technics developed for the experiment on hydrogen-like ions
with spinless nuclei.

The bound state QED theory is mainly satisfactory for
hydrogen-like atoms for comparison with various present-day
experiments. Taking into account future experiments, it is
necessary to improve its accuracy calculating higher-order
corrections. As one of the crucial problems, let us mention
numerical and analytic calculations of one-loop contributions for
the hyperfine difference $D_{21}$, two-loop corrections for the
Lamb shift, $D_{21}$ and bound electron $g$ factor and various
fourth-order contributions to the positronium energy levels, the
muonium $1s$ hyperfine interval and $D_{21}$.

This review is focussed on light atoms which offer various
precision tests of bound state QED. ``High precision'' is
understood as a high relative precision of various measurements
and calculations and sometimes as a high sensitivity to certain
high-order QED corrections. Certain study of medium- and high-$Z$
ions can be also of interest for QED tests. However, it is
crucially important to clarify uncertainty of the corrections due
to the nuclear charge distribution.

We note also that the high precision is not the only quantity of
interest. One of the motivations to `test' QED is a search for its
violations; another is a test of various approaches for precision
calculations and measurements. It is not necessary that the most
precise tests are the most sensitive to any open questions. The
sensitivity is often quite selective and depends on a problem
under question. Study of tests related to any particular problem
needs a separate study and other experiments may be important.

\section{Summary}

\begin{flushright}
\begin{minipage}{6cm}
When they said, ``Does it fit?"\\
He replied, ``Not a bit!"\\
 \centerline{\em Edward Lear\/}
\end{minipage}
\end{flushright}

The situation concerning precision studies of simple atoms is
fortunately opposite to that in the quoted verse by Lear. The
theory and experiment agree well in general. There may be some
discrepancy in a certain particular case, even a long standing
one, but we have no reason to doubt quantum electrodynamics in its
basic moments. However, the theory of simple atoms is a bound
state QED theory and the crucial question is not whether QED
proper is correct. The question is how successfully QED can be
applied and has been applied to the bound states. The latter is
not a simple issue and progress in QED calculations can be very
fruitful for better understanding of the bound states in nuclear
and particle physics. In early time of quantum physics theoretical
progress established new theories and was of great significance.
Now the most important part of physics of simple atoms lies in
experiment.

Fascinating experimental progress with high-precision
measurements, by bridging microwave, infrared, optical and in part
ultraviolet domains of spectra, with dramatic cooling of atoms,
with access to trapped single atoms and particles, with producing
muonic and exotic atoms and various highly charged ions offers us
a broad range of applications. However, studying simple objects
with predictable properties is still the most natural choice. That
is how simple atoms enter modern physics, most of progress of
which is now rather a technological one and that is how simple
atoms can serve for quite practical applications testing new
experimental methods. Another application of simple atoms of
practical importance is the determination of certain fundamental
constants. These constants have been more and more involved in the
modern system of units and standards.

Can we learn from simple atoms any really new physics? The answer
for most experiments is rather negative. They have pursued other
purposes. Simplicity of the atoms and reliability of the QED
predictions make them  a powerful tool to study effects beyond
atomic physics, such as particle and nuclear properties. Precision
study of two-body atoms delivers us information on structure of
light nuclei (deuteron, triton, helion, and alpha particle), the
structure of a proton, various properties of muons, pions and
other particles. In general, we check consistency of QED theory
and experiment and as long as we see the consistency, the
constraints on new physics are related to the uncertainty of the
experiment and theory. There are still some projects with simple
atoms which are designed to look beyond the established physics: a
search for exotic decay of positronium, a search for variations of
the fundamental constants, a search for CPT violation studying
properties of antihydrogen, a conversion of muonium into
antimuonium etc.

Twenty years ago, when I joined the QED team at Mendeleev
Institute for Metrology and started to work on theory of simple
atoms, the accuracy of the experimental data for most QED tests
was substantially higher than that of the theoretical predictions.
Work of several theoretical groups from around the world during
this twenty-year period has put theory of two-body atomic systems
to a dominant position and we are now looking forward to obtaining
new experimental results. Except the case of the $2s$ hyperfine
interval in the $^3$He$^+$ ion, theory is more accurate than
experiment. However, theory is still not perfect. For some
quantities the gap between the theoretical and experimental
accuracy is not big and we should not think that the theory is
above any doubts. For example, starting my work on the review, I
believed that the theory for the $D_{21}$ difference in helium-3
ion is more accurate than the experimental data, but a
conservative re-estimation of the theoretical uncertainty
\cite{2sZETF,2seprint} has reversed the situation.

Several experiments are in progress or planned and substantial
experimental progress is possible. Since the theoretical
improvement needs some long-term programs, we have to start
working on a further development of theory now. It will be indeed
helpful for difficult theoretical calculations to be motivated by
new experimental projects and new experimental results and I hope
that both will follow.

The obvious success in one-electron atoms turned attention of some
theorists to three-body (such as helium, muonic helium,
antiprotonic helium) and four-body systems and we hope that in
some future it will be possible to state that theory of three- or
maybe even four-body atoms has also superseded experiment.

\section*{Acknowledgements}
I am grateful to G. W. F. Drake, S. I. Eidelman, J. Friar, T. W.
H\"ansch, E. Hinds, V. G. Ivanov, K. Jungmann, A. I. Milstein, P.
Mohr, L. B. Okun, M. H. Prior, J. R. Sapirstein, V. A. Shelyuto,
I. Sick, B. Taylor and G. Werth  for useful and stimulating
discussions. This work was supported in part by RFBR (grants \#\#
02-02-07027, 03-02-16843, 03-02-04029) and DFG (grant GZ 436 RUS
113/769/0-1).


\appendix

\section{Notations\label{s:not}}

\begin{table}[hbtp]
\begin{center}
\begin{tabular}{cl}
\hline
$n$& the principal quantum number\\
$l$& the orbital moment of an electron (in units of $\hbar$) \\
$s$& the electron spin (in units of $\hbar$) \\
$j$& the angular momentum of an electron (in units of $\hbar$) (${\bf j} = {\bf l} + {\bf s}$)\\
$I$ & the nuclear spin (in units of $\hbar$) \\
$F$ & the atomic angular momentum (${\bf F} = {\bf j} + {\bf I}$)\\
$m$ & the mass of the orbiting particle (mainly, an electron) \\
$M$ & the nuclear mass \\
$m_R$ & the reduced mass of the orbiting particle ($m_R=Mm/(M+m)$)\\
$Z$ & the nuclear charge (in units of the proton charge) \\
$A$ & the atomic mass number (i.e., $M\approx A\,m_p$) \\
$R_\infty$ & the Rydberg constant ($R_\infty=\alpha^2m_ec/2h$)\\
$\alpha$ & the fine structure constant ($\alpha=e^2/4\pi\epsilon_0\hbar c$)\\
$c$ & the speed of light\\
$h$ & the Planck constant\\
$\hbar$ & the reduced Planck constant ($\hbar=h/2\pi$)\\
$e$ & the proton (positron) charge \\
$\mu_B$ & the Bohr magneton ($\mu_B = e\hbar/2m_e $) \\
$\mu_N$ & the nuclear magneton ($\mu_N= e\hbar/2m_p $) \\
$m_e$ & the mass of an electron$^\ast$\\
$\mu_h$ & the magnetic moment of a helion$^\ast$ (the nucleus of $^3He$)\\
$a_e$ & the anomalous magnetic moment of an electron$^\star$ ($a_e=(g_e-2)/2$)\\
$g_\mu$ & the $g$ factor of a muon$^\dag$ ($g_\mu=2\mu_\mu/( e\hbar/2m_\mu )$)\\
$R_N$ & the nuclear charge radius$^\ddag$\\
$R_E$ & also the nuclear (rms) charge radius, when it is necessary to distinguish it from $R_M$\\
$R_M$ & the nuclear (rms) magnetic-distribution radius\\
\hline
\end{tabular}
\end{center}
\caption{The most frequently used notation.\protect\newline
$^\ast$ -- similar for other particles and nuclei; $^\star$ --
similar for a muon; $^\dag$ -- similar for an electron and a
proton; $^\ddag$ -- similar for a proton, a deuteron
etc.\label{TNot}}
\end{table}

Unless otherwise stated, we use SI units. However, all numerical
results for energy $E$ are presented in frequency units $E/h$.

Energy levels in a single-electron atom are labelled as $nl_j$ or
if the hyperfine effects are involved as $nl_j(F)$. The orbital
momentum is presented by letters $s$ for $l=0$, $p$ for for $l=1$,
$d$ for $l=2$, $f$ for $l=3$ etc. For few-electron systems (and
positronium) the capital letters for angular momenta are used such
as $n{}^\kappa L_J$ where $\kappa=2S+1$.

\section{Extract from the list of the recommended fundamental
constants (CODATA, 2002)}

Here, we summarize the most recent CODATA recommended values of
fundamental constants related to QED. We need to emphasize that
Table~\ref{Tcodata} contains results obtained in the
2002\footnote{We remind that 1998 and 2002 are years of the
deadline for collecting the data for the evaluation.} adjustment
of the fundamental constants \cite{newcodata}, while through out
the paper we compare original results with the older set of
recommended values \cite{codata}. A reason for that is that a
substantial part of results reviewed here was obtained after 1998
and was not included in the earlier adjustment \cite{codata} and
thus comparing the results we clearly see the recent progress in
the field. The recent adjustment \cite{newcodata} already included
nearly all original results reviewed in this work.

\begin{table}[hbtp]
\begin{center}
\begin{tabular}{lc}
\hline
Constant & CODATA, \\
& 2002 \cite{newcodata} \\
\hline
$R_\infty$& $10\,973\,731.568\,525(73) \;{\rm m}^{-1} $\\
$c\,R_\infty$&$3.289\,841\,960\,360(22)\times 10^{15}\; {\rm Hz} $\\
$R_H$ & $10\,967\,758.340\,642(73)\;{\rm m}^{-1}$ $^\star$\\
$\alpha^{-1}$ & $137.035\,999\,11(46)$ \\
$m_e$ & $5.485\,799\,094\,5(24)\times10^{-4}\;{\rm u}$ \\
$a_e$ & $1.159\,652\,185\,9(38)\times10^{-3}$ \\
$m_p$ &$ 1.007\,276\,466\,88(13)\;{\rm u}$  \\
$m_p/m_e$&$1\,836.152\,672\,61(85)$  \\
$\mu_p/\mu_B$& $1.521\,032\,206(15)\times10^{-3}$ \\
$g_p$& $5.585\,694\,701(56)$ \\
$\mu_d/\mu_B$& $0.466\,975\,456\,7(50)\cdot10^{-3}\,$ \\
$g_d$& $0.857\,438\,232\,9(92)$ \\
$m_\mu$ & $ 0.113\,428\,926\,4(30)\; {\rm u}$ \\
$m_\mu/m_e$&$206.768\,283\,8(54)$  \\
$\mu_\mu/\mu_p$ & $3.183\,345\,118(89)^\ast$  \\
$a_\mu$& $1.165\,919\,81(62)\times10^{-3}$ \\
\hline
\end{tabular}
\end{center}
\caption{New recommended values of the fundamental constants
(CODATA, 2002, \cite{newcodata}) related to atomic physics and
QED. \protect\newline $^\star$ The result for the Rydberg constant
for the hydrogen atom $R_H$ was not presented in \cite{newcodata}
directly and it is derived from values for $R_ \infty$ and
$m_e/m_p$. $^\ast$ We ignore in our paper (except of consideration
of $D_{21}$) the negative sign of the ratio of the magnetic
moments.\label{Tcodata}}
\end{table}

\end{document}